\documentclass[aps,onecolumn,groupedaddress,showpacs,notitlepage,nofootinbib]{revtex4-1}

\usepackage{graphicx}
\usepackage{dcolumn}
\usepackage{bm}
\usepackage{amssymb}
\usepackage{mathrsfs}
\usepackage{amsmath}
\usepackage{epsfig}
\usepackage{color}
\usepackage{hhline}
\usepackage{hyperref}

\allowdisplaybreaks[4]

\begin{document}

\title{Constraining the interaction strength between dark matter and visible matter: I. fermionic dark matter}

\author{Jia-Ming Zheng$^1$}
\author{Zhao-Huan Yu$^{1,2}$}
\author{Jun-Wen Shao$^1$}
\author{Xiao-Jun Bi$^{2}$}
\author{Zhibing Li$^1$}
\author{Hong-Hao Zhang$^1$}
\email[Email:~]{zhh98@mail.sysu.edu.cn}

\affiliation{$^1$School of Physics and Engineering, Sun Yat-Sen University, Guangzhou
510275, China\\
$^2$Key Laboratory of Particle Astrophysics, Institute of High
Energy Physics, Chinese Academy of Sciences, Beijing 100049,
China}

\begin{abstract}
In this work we study the constraints on the dark matter
interaction with the standard model particles, from the observations
of dark matter relic density, the direct detection experiments of
CDMS and XENON, and the indirect detection of the $\bar{p}/p$
ratio by PAMELA. A model independent way is adopted in the study
by constructing the effective interaction operators between dark
matter and standard model particles. The most general 4-fermion
operators are investigated. We find that the constraints from different
observations are complementary with each other. Especially the
spin independent scattering gives very strong constraints for
corresponding operators. In some cases the indirect detection of
$\bar{p}/p$ data can actually be more sensitive than the direct
detection or relic density for light dark matter ($\lesssim 70$
GeV).
\end{abstract}


\maketitle

\section{Introduction}

The existence of a significant component of nonbaryonic dark
matter (DM) in the Universe has been well confirmed by
astrophysical observations
\cite{Tegmark:2006az,Komatsu:2008hk,Komatsu:2010fb} in recent
years, however the nature of this substance remains unclear. Since
there is no candidate for DM in the Standard Model (SM) of
particle physics, it implies the existence of new physics beyond
the SM. Probe of the microscopic identity and properties of DM has
become one of the key problems in particle physics and cosmology
(for reviews of DM, see, for instance,
\cite{kolb-turner,Jungman:1995df,Bertone:2004pz,Murayama:2007ek,Feng:2010gw}).

Among a large amount of theoretical models, a well-motivated
candidate for DM is the weakly interacting massive particle
(WIMP). This WIMP must be stable, nonrelativistic, electrically
neutral, and colorless. If the mass of WIMP is from a few GeV to
TeV and the interaction strength is of the weak scale, they can
naturally yield the observed relic density of DM, which is often
referred to as the WIMP miracle \cite{Feng:2010gw}. A huge variety
of new physics models trying to solve the problems of the SM at
the weak scale can naturally contain WIMP candidates, such as the
supersymmetric models
\cite{Jungman:1995df,Goldberg:1983nd,Ellis:1983ew,Kane:1993td},
extra dimensional models
\cite{Kolb:1983fm,Cheng:2002ej,Hooper:2007qk,Servant:2002aq,Servant:2002hb,Agashe:2004ci,
Agashe:2004bm,Agashe:2007jb}, little Higgs models
\cite{Cheng:2004yc,Low:2004xc,Birkedal:2006fz,Freitas:2009jq,Kim:2009dr},
left-right symmetric models
\cite{Dolle:2007ce,Guo:2008hy,Guo:2008si,Guo:2010vy}, and many
other theoretical scenarios.

The above mentioned specific models are well-motivated, however
they still lack experimental support. We do not know whether
nature really behaves like one of them or some other yet
unconsidered theories. Moreover, in case the DM particle is the
only new particle within the reach of LHC and other new particle
species are much heavier than DM, it will be very difficult to
tell which model the DM particle belongs to. Additionally, it is
possible that the DM may be first observed by direct or indirect
detection experiments. These early observations may only provide
information about some general properties of the DM particle, and
may not be able to distinguish the underlying theories. Therefore,
the model-independent studies of the DM phenomenology are
particularly important for they may avoid theoretical bias
\cite{Birkedal:2004xn,Giuliani:2004uk,Kurylov:2003ra,Beltran:2008xg,Cirelli:2008pk,Shepherd:2009sa}.
Recently there have been quite a few papers following such
consideration and adopting a model-independent way to study
various phenomenologies related with DM
\cite{Cao:2009uv,Cao:2009uw,Beltran:2010ww,Fitzpatrick:2010em,Goodman:2010yf,Bai:2010hh,
Goodman:2010ku,Goodman:2010qn,Bell:2010ei}. Especially the relic density measured
by WMAP \cite{Komatsu:2010fb}, direct detection from CDMS
\cite{Ahmed:2009zw}, XENON \cite{Aprile:2010um} and possible
collider signals from LHC are considered in these
studies.

In this work, we first construct the general effective 4-fermion
interaction operators between DM particles and the SM particles,
which extend the effective fermionic WIMP interactions given in
Ref. \cite{Beltran:2008xg}. Here we focus on Dirac fermionic DM. Discussions on scalar and vector DM will be presented in companion papers. We then give updated constraints from
the DM relic density within the 7-year WMAP data \cite{Komatsu:2010fb} and
the spin-independent WIMP-nucleus elastic scattering searches by
CDMS II \cite{Ahmed:2009zw} and XENON100 \cite{Aprile:2010um}, and
compare our results with those in Ref. \cite{Beltran:2008xg}. In
addition, we present new phenomenological constraints on these
effective models from the spin-dependent WIMP-nucleus elastic
scattering searches by CDMS \cite{Akerib:2005za} and
XENON \cite{Angle:2008we} and the cosmic-ray antiproton-to-proton
ratio by PAMELA \cite{Adriani:2010rc}. We find that the
constraints from different kinds of experiments are rather
comparable. Combination of these constraints provides more
information of the effective models.

This paper is organized as follows. In Sec. \ref{sec-model} we
briefly describe the effective DM models of various 4-fermion
interaction operators. In Sec. \ref{sec-relic}, \ref{sec-direct}
and \ref{sec-indirect} we explore the constraints on these models
from the DM relic density, direct and indirect detection searches,
respectively. In Sec. \ref{sec-combine} we discuss the validity region of effective theory and present the combined
constraints on the effective coupling constants of these models.
Sec. \ref{sec-con} is the conclusion.

\section{Effective Models\label{sec-model}}

We start with the case that DM is a single Dirac fermionic WIMP
($\chi$). Instead of considering a WIMP candidate from a specific
theoretical model, we study the phenomenologies in a
model-independent way by constructing effective interaction
operators between $\chi$ and  the SM particles. These interaction
operators are constrained only by the requirements of Hermiticity,
Lorentz invariance and CPT invariance.

To proceed, we make the following assumptions similar to those in Ref. \cite{Beltran:2008xg}: (1) The WIMP is the only new particle at the electroweak scale, and any new particle species other than the WIMP has a mass much larger than the WIMP. This implies that the WIMP's thermal relic density is not affected by resonances or coannihilations, and this makes it possible to describe the interaction between the WIMPs and the standard model particles in terms of an effective field theory. (2) The WIMP only interacts with the standard model fermions through a 4-fermion effective interaction, but not with other particles like gauge or Higgs bosons. This 4-fermion effective interaction is assumed to be dominated by only one form (scalar, vector, etc.) of the set of 4-fermion operators for simplicity. (3) The WIMP annihilation channels to the standard model fermion-antifermion pairs dominate over other possible channels. In other words, the possible channels to final states that include gauge or Higgs bosons are assumed to be negligible for simplicity.
However, it should be noted that the supersymmetric DM model actually cannot
satisfy the three assumptions above simultaneously in order to give the correct
relic density. Even so these assumptions are still useful for a general
research.

The effective Lagrangian between two fermionic WIMPs ($\chi$ and $\bar{\chi}$) and two standard model fermions ($f$ and $\bar{f}$) is given by only one of the following expressions:
\begin{eqnarray}
\text{Scalar interaction (S)}:&&\qquad \mathcal{L}_{\mathrm{S}}=\sum_{f}\frac{G_{S,f}}{\sqrt{2}}\bar{\chi}\chi\bar{f}f\label{Leff-S}\\
\text{Pseudoscalar interaction (P)}:&&\qquad \mathcal{L}_{\mathrm{P}}=\sum_{f}\frac{G_{P,f}}{\sqrt{2}}\bar{\chi}\gamma_5\chi\bar{f}\gamma_5f
\label{Leff-P}\\
\text{Vector interaction (V)}:&&\qquad \mathcal{L}_{\mathrm{V}}=\sum_{f}\frac{G_{V,f}}{\sqrt{2}}\bar{\chi}\gamma^\mu\chi\bar{f}\gamma_\mu f\label{Leff-V}\\
\text{Axialvector interaction (A)}:&&\qquad \mathcal{L}_{\mathrm{A}}=\sum_{f}\frac{G_{A,f}}{\sqrt{2}}\bar{\chi}\gamma^\mu \gamma_5\chi\bar{f}\gamma_\mu\gamma_5 f\label{Leff-A}\\
\text{Tensor interaction (T)}:&&\qquad \mathcal{L}_{\mathrm{T}}=\sum_{f}\frac{G_{T,f}}{\sqrt{2}}\bar{\chi}\sigma^{\mu\nu}\chi\bar{f}\sigma_{\mu\nu} f\label{Leff-T}\\
\text{Scalar-pseudoscalar interaction (SP)}:&&\qquad \mathcal{L}_{\mathrm{SP}}=\sum_{f}\frac{G_{SP,f}}{\sqrt{2}}\bar{\chi}\chi\bar{f}i\gamma_5f
\label{Leff-SP}\\
\text{Pseudoscalar-scalar interaction (PS)}:&&\qquad \mathcal{L}_{\mathrm{PS}}=\sum_{f}\frac{G_{PS,f}}{\sqrt{2}}\bar{\chi}i\gamma_5\chi\bar{f}f
\label{Leff-PS}\\
\text{Vector-axialvector interaction (VA)}:&&\qquad \mathcal{L}_{\mathrm{VA}}
=\sum_{f}\frac{G_{VA,f}}{\sqrt{2}}\bar{\chi}\gamma^\mu\chi\bar{f}\gamma_\mu\gamma_5f
\label{Leff-VA}\\
\text{Axialvector-vector interaction (AV)}:&&\qquad \mathcal{L}_{\mathrm{AV}}
=\sum_{f}\frac{G_{AV,f}}{\sqrt{2}}\bar{\chi}\gamma^\mu\gamma_5\chi\bar{f}\gamma_\mu f\label{Leff-AV}\\
\text{Alternative tensor interaction ($\tilde{\mathrm{T}}$)}:&&\qquad \mathcal{L}_{\tilde{\mathrm{T}}}
=\sum_{f}\frac{\tilde{G}_{T,f}}{\sqrt{2}}\varepsilon^{\mu\nu\rho\sigma}
\bar{\chi}\sigma_{\mu\nu}\chi\bar{f}\sigma_{\rho\sigma} f\label{Leff-T-tild}\\
\text{Left handed-left handed interaction (LL)}:&&\qquad \mathcal{L}_{\mathrm{LL}}
=\sum_{f}\frac{G_{LL,f}}{\sqrt{2}}
\bar{\chi}\gamma^\mu(1-\gamma_5)\chi\bar{f}\gamma_\mu(1-\gamma_5)f\label{Leff-LL}\\
\text{Right handed-right handed interaction (RR)}:&&\qquad \mathcal{L}_{\mathrm{RR}}
=\sum_{f}\frac{G_{RR,f}}{\sqrt{2}}
\bar{\chi}\gamma^\mu(1+\gamma_5)\chi\bar{f}\gamma_\mu(1+\gamma_5)f\label{Leff-RR}\\
\text{Left handed-right handed interaction (LR)}:&&\qquad \mathcal{L}_{\mathrm{LR}}
=\sum_{f}\frac{G_{LR,f}}{\sqrt{2}}
\bar{\chi}\gamma^\mu(1-\gamma_5)\chi\bar{f}\gamma_\mu(1+\gamma_5)f\label{Leff-LR}\\
\text{Right handed-left handed interaction (RL)}:&&\qquad \mathcal{L}_{\mathrm{RL}}
=\sum_{f}\frac{G_{RL,f}}{\sqrt{2}}
\bar{\chi}\gamma^\mu(1+\gamma_5)\chi\bar{f}\gamma_\mu(1-\gamma_5)f\label{Leff-RL}
\end{eqnarray}
where the sum of $f$ is over all the standard model fermions,
and the effective coupling constants $G$ are real numbers which have mass dimension of $-2$. The 4 chiral interaction operators, $\mathcal{L}_{\mathrm{LL}}$, $\mathcal{L}_{\mathrm{RR}}$, $\mathcal{L}_{\mathrm{LR}}$ and $\mathcal{L}_{\mathrm{RL}}$, are just the combinations of the other operators mentioned above. Here we do not include the interaction operators involving derivative $\partial_\mu$ insertions into fermion bilinears, for they have higher momentum dimensions and may be safely ignored in the small momentum limit. Note that the alternative tensor interaction term in \eqref{Leff-T-tild} has two other equivalent forms, i.e., $\varepsilon^{\mu\nu\rho\sigma}
\bar{\chi}\sigma_{\mu\nu}\chi\bar{f}\sigma_{\rho\sigma} f=-2\bar{\chi}\sigma^{\mu\nu}i\gamma_5\chi\bar{f}\sigma_{\mu\nu} f=-2\bar{\chi}\sigma^{\mu\nu}\chi\bar{f}\sigma_{\mu\nu}i\gamma_5 f$.

Each form of the interaction operators listed above represents an
effective model of the WIMP coupled to the standard model
fermions. For each case, we can calculate the corresponding
annihilation and scattering cross sections, which depend on the
WIMP mass $M_\chi$ and the coupling constants $G_f$. Associated
with the recent results of the DM relic density, direct and
indirect detection experiments, we can obtain the phenomenological
constraints on $G_f$. It would be interesting and meaningful to
compare the constraints derived from different experiments.

It is worthwhile to note the symmetry properties of these operators under discrete C, P, and T transformations. The first 5 forms of the operators, $\mathcal{L}_{\mathrm{S}}$, $\mathcal{L}_{\mathrm{P}}$,
$\mathcal{L}_{\mathrm{V}}$, $\mathcal{L}_{\mathrm{A}}$ and $\mathcal{L}_{\mathrm{T}}$ are
separately  invariant under C, P, and T, while the transformation properties of the other
operators are summarized in Table \ref{Table-Trans-CPT}.
The transformation properties of the operators under CP are the same as those under T,
given that the coupling constants $G$ are real-valued numbers.
Thus, all the operators are actually CPT invariant.
If the future experiments indicate that there were some of the C, P, and T symmetries
in the DM sector, we may use this table to concentrate on or exclude some
interaction operators.
\begin{table}[th]
\caption{The transformation properties of the 4-fermion operators under C, P, and T. Since $\mathcal{L}_{\mathrm{S}}$, $\mathcal{L}_{\mathrm{P}}$, $\mathcal{L}_{\mathrm{V}}$, $\mathcal{L}_{\mathrm{A}}$ and $\mathcal{L}_{\mathrm{T}}$ are separately  invariant under C, P, and T, they are not listed below. The plus '$+$' means being invariant under the transformation, while the minus '$-$' means sign reversal. The transformation properties of the operators under CP are the same as those under T, given that the coupling constants $G$ are real numbers.}
\label{Table-Trans-CPT}
\centering
\renewcommand{\arraystretch}{1.2}
\begin{tabular}{c | c c c c c c c c c}\hline\hline
& $\mathcal{L}_{\mathrm{SP}}$ & $\mathcal{L}_{\mathrm{PS}}$ & $\mathcal{L}_{\mathrm{VA}}$ & $\mathcal{L}_{\mathrm{AV}}$ & $\mathcal{L}_{\tilde{\mathrm{T}}}$ & $\mathcal{L}_{\mathrm{LL}}$ & $\mathcal{L}_{\mathrm{RR}}$ & $\mathcal{L}_{\mathrm{LR}}$ & $\mathcal{L}_{\mathrm{RL}}$\\ \hline
P & $-$ & $-$ & $-$ & $-$ & $-$ & $\mathcal{L}_{\mathrm{RR}}$ & $\mathcal{L}_{\mathrm{LL}}$ & $\mathcal{L}_{\mathrm{RL}}$ & $\mathcal{L}_{\mathrm{LR}}$\\ \hline
C & $+$ & $+$ & $-$ & $-$ & $+$ & $\mathcal{L}_{\mathrm{RR}}$ & $\mathcal{L}_{\mathrm{LL}}$ & $\mathcal{L}_{\mathrm{RL}}$ & $\mathcal{L}_{\mathrm{LR}}$\\ \hline
T & $-$ & $-$ & $+$ & $+$ & $-$ & $+$ & $+$ & $+$ & $+$\\ \hline\hline
\end{tabular}
\end{table}

\section{WIMP annihilation and relic density\label{sec-relic}}

In order to determine the relic density of WIMPs and the source
function of cosmic-ray particles produced by the DM annihilation in the
Galaxy, which is relevant to the DM indirect detection, we need to calculate the cross sections of WIMP-antiWIMP
annihilation  to fermion-antifermion pairs for each case listed in the last section. The result is given by
\begin{eqnarray}
\sigma_{S,\,\mathrm{ann}}&=&\frac{1}{16\pi }\sum_f\bigg(\frac{G_{S,f}}{\sqrt{2}}\bigg)^2c_f\sqrt{\frac{s-4m_f^2}{s-4M_\chi^2}}
\frac{(s-4M_\chi^2)(s-4m_f^2)}{s}\label{diracWIMP-sigma-S}\\
\sigma_{P,\,\mathrm{ann}}&=&\frac{1}{16\pi}
\sum_f\bigg(\frac{G_{P,f}}{\sqrt{2}}\bigg)^2c_f\sqrt{\frac{s-4m_f^2}{s-4M_\chi^2}}s
\label{diracWIMP-sigma-P}\\
\sigma_{V,\,\mathrm{ann}}&=&\frac{1}{12\pi}
\sum_f\bigg(\frac{G_{V,f}}{\sqrt{2}}\bigg)^2c_f\sqrt{\frac{s-4m_f^2}{s-4M_\chi^2}}
\bigg[s+2(M_\chi^2+m_f^2)+4\frac{M_\chi^2m_f^2}{s}\bigg]\label{diracWIMP-sigma-V}\\
\sigma_{A,\,\mathrm{ann}}&=&\frac{1}{12\pi}
\sum_f\bigg(\frac{G_{A,f}}{\sqrt{2}}\bigg)^2c_f\sqrt{\frac{s-4m_f^2}{s-4M_\chi^2}}
\bigg[s-4(M_\chi^2+m_f^2)+28\frac{M_\chi^2m_f^2}{s}\bigg]\label{diracWIMP-sigma-A}\\
\sigma_{T,\,\mathrm{ann}}&=&\frac{1}{6\pi}
\sum_f\bigg(\frac{G_{T,f}}{\sqrt{2}}\bigg)^2c_f\sqrt{\frac{s-4m_f^2}{s-4M_\chi^2}}
\bigg[s+2(M_\chi^2+m_f^2)+40\frac{M_\chi^2m_f^2}{s}\bigg]\label{diracWIMP-sigma-T}\\
\sigma_{SP,\,\mathrm{ann}}&=&\frac{1}{16\pi}
\sum_f\bigg(\frac{G_{SP,f}}{\sqrt{2}}\bigg)^2c_f\sqrt{\frac{s-4m_f^2}{s-4M_\chi^2}}
(s-4M_\chi^2)\\
\sigma_{PS,\,\mathrm{ann}}&=&\frac{1}{16\pi}
\sum_f\bigg(\frac{G_{PS,f}}{\sqrt{2}}\bigg)^2c_f\sqrt{\frac{s-4m_f^2}{s-4M_\chi^2}}
(s-4m_f^2)\\
\sigma_{VA,\,\mathrm{ann}}&=&\frac{1}{12\pi}
\sum_f\bigg(\frac{G_{VA,f}}{\sqrt{2}}\bigg)^2c_f\sqrt{\frac{s-4m_f^2}{s-4M_\chi^2}}
\bigg[s+2(M_\chi^2-2m_f^2)-8\frac{M_\chi^2m_f^2}{s}\bigg]\\
\sigma_{AV,\,\mathrm{ann}}&=&\frac{1}{12\pi}
\sum_f\bigg(\frac{G_{AV,f}}{\sqrt{2}}\bigg)^2c_f\sqrt{\frac{s-4m_f^2}{s-4M_\chi^2}}
\bigg[s+2(m_f^2-2M_\chi^2)-8\frac{M_\chi^2m_f^2}{s}\bigg]\label{diracWIMP-sigma-AV}\\
\sigma_{\tilde{T},\,\mathrm{ann}}&=&\frac{2}{3\pi}
\sum_f\bigg(\frac{\tilde{G}_{T,f}}{\sqrt{2}}\bigg)^2c_f\sqrt{\frac{s-4m_f^2}{s-4M_\chi^2}}
\bigg[s+2(m_f^2+M_\chi^2)-32\frac{M_\chi^2m_f^2}{s}\bigg]\\[0.3cm]
\sigma_{C,\,\mathrm{ann}}&\equiv&\sigma_{LL,\,\mathrm{ann}}=\sigma_{RR,\,\mathrm{ann}}
=\sigma_{LR,\,\mathrm{ann}}=\sigma_{RL,\,\mathrm{ann}}\nonumber\\
&=&\frac{1}{3\pi}\sum_f\bigg(\frac{G_{C,f}}{\sqrt{2}}\bigg)^2c_f\sqrt{\frac{s-4m_f^2}{s-4M_\chi^2}}
\bigg[s-(M_\chi^2+m_f^2)+4\frac{M_\chi^2m_f^2}{s}\bigg]\label{diracWIMP-sigma-chiral}
\end{eqnarray}
where $s$ is the Mandelstam variable, $M_\chi$ is the WIMP mass,
the sum is over the final state fermion species $f$, and $c_f$ are
the color factors, equal to 3 for quarks and 1 for leptons. The
annihilation cross sections of the four chiral interactions,
$\sigma_{LL,\,\mathrm{ann}}$, $\sigma_{RR,\,\mathrm{ann}}$,
$\sigma_{LR,\,\mathrm{ann}}$ and $\sigma_{RL,\,\mathrm{ann}}$,
have the exactly same formula, so they can be denoted by a common
symbol $\sigma_{C,\,\mathrm{ann}}$ with the corresponding coupling
constants denoted by $G_{C,f}$. Note that
Eqs.\eqref{diracWIMP-sigma-S}--\eqref{diracWIMP-sigma-V} agree
exactly with Eqs.(6)--(8) in Ref. \cite{Beltran:2008xg}, while
Eqs.\eqref{diracWIMP-sigma-A},\eqref{diracWIMP-sigma-T} are
slightly different from Eqs.(9),(10) in \cite{Beltran:2008xg}.

In the very early Universe, the WIMPs were in thermal equilibrium.
As the Universe expands, the WIMPs departed from thermal
equilibrium when they were nonrelativistic, and finally froze out
to yield a cold relic roughly when the annihilation rate dropped
below the Hubble rate. This evolution process is described by the
Boltzmann equation
\begin{equation}
\frac{dn_\chi}{dt}+3Hn_\chi =-\langle {\sigma
_{\mathrm{ann}}v_\mathrm{M{\o}l}}\rangle\left[n_\chi n_{\bar{\chi}} -
n_\chi^\mathrm{eq}n_{\bar{\chi}}^\mathrm{eq}\right]=-\langle {\sigma
_{\mathrm{ann}}v_\mathrm{M{\o}l}}\rangle\left[(n_\chi)^2 -
(n_\chi^\mathrm{eq})^2\right]\label{boltz-eq}
\end{equation}
where $H\equiv\dot{a}/a=\sqrt{8\pi\rho/(3 M_\mathrm{Pl}^2)}$ is
the Hubble rate with $M_\mathrm{Pl}$ denoting the Planck mass,
$n_\chi$ ($n_{\bar{\chi}}$) is the number density of WIMPs
(antiWIMPs), and the thermal average
$\langle {\sigma_{\mathrm{ann}}v_\mathrm{M{\o}l}}\rangle$ will be explained below. For Dirac fermions without particle-antiparticle
asymmetry, $n_\chi=n_{\bar{\chi}}$, and the total DM paritcle
number density is $n_{\mathrm{DM}}=2n_\chi$
\cite{Gondolo:1990dk,Jungman:1995df,Srednicki:1988ce}. At the early time, when the
temperature $T\gg M_\chi$, the WIMP number density $n_\chi$ was
very close to its equilibrium value $n_\chi^\mathrm{eq}\propto
T^3$, and the annihilation rate per unit volume $\Gamma = n_\chi
n_{\bar{\chi}}\left\langle {\sigma _{\mathrm{ann}}v_\mathrm{M{\o}l}}
\right\rangle$ was much greater than the Hubble expansion rate per
unit volume $3Hn_\chi$ and enough to maintain the thermal
equilibrium. However, as the temperature $T$ decreased below
$M_\chi$, the equilibrium number density was exponentially
suppressed, $n_\chi^\mathrm{eq}\simeq g [M_\chi T/(2 \pi)]^{3/2}
\exp{(-M_\chi/T)}$, where $g=2$ is the number of degrees of
freedom of a fermionic WIMP. Eventually, the annihilation rate
became smaller than the expansion rate, and the WIMPs froze out of
equilibrium.

The thermally averaged quantity $\langle {\sigma_{\mathrm{ann}}v_\mathrm{M{\o}l}}\rangle$ should be treated
carefully. As pointed out by Ref.~\cite{Gondolo:1990dk}, the M{\o}ller velocity $v_\mathrm{M{\o}l}$ in Eq.~\eqref{boltz-eq} is defined by
$v_\mathrm{M{\o}l} \equiv \sqrt {(p_1 \cdot p_2)^2 - m_1^2m_2^2}/(E_1E_2)=\sqrt {|{{{\mathbf{v}}_1} -
{{\mathbf{v}}_2}} |^2 - |{{{\mathbf{v}}_1} \times {{\mathbf{v}}_2}} |^2}$ with subscripts 1 and 2 labeling
the two initial DM particles and particle velocities $\mathbf{v}_i \equiv \mathbf{p}_i / E_i$ ($i=1,2$).
The M{\o}ller velocity $v_\mathrm{M{\o}l}$ equals the relative velocity $v_\mathrm{rel} \equiv \left|
{{{\mathbf{v}}_1} - {{\mathbf{v}}_2}} \right|$ only when the collision is collinear, ${{{\mathbf{v}}_1} \times {{\mathbf{v}}_2}}=0$. Since the Boltzmann equation, Eq.~\eqref{boltz-eq}, is expressed in the
cosmic comoving frame \cite{Gondolo:1990dk}, in which the gas is at rest as a whole, the thermal average $\langle {\sigma_{\mathrm{ann}}v_\mathrm{M{\o}l}}\rangle$ must be taken
in this frame. Fortunately, even including relativistic effects, it has been shown  \cite{Gondolo:1990dk} that
$\left< \sigma_{\mathrm{ann}} v_\mathrm{M{\o}l} \right> = \left<\sigma_{\mathrm{ann}} v_\mathrm{lab}
\right> ^{\mathrm{lab}}$, where $v_\mathrm{lab}\equiv |\mathbf{v}_{1,\,\mathrm{lab}}-\mathbf{v}_{2,\,\mathrm{lab}}|$ and the right-hand side is computed in the lab frame, in which one of
the two initial particles is at rest. Thus, it is convenient to calculate the thermal average in the lab frame using the method described in \cite{Gondolo:1990dk}.

Cold DM requires that the freeze-out of WIMPs occurred when they were nonrelativistic.
In the nonrelativistic limit, we can parameterize $\sigma _{\mathrm{ann}}v=a+bv^2+\mathcal{O}(v^4)$ \footnote{Note that this expansion in powers of $v_\mathrm{lab}^2$ is equivalent to that in Ref.~\cite{Gondolo:1990dk}: $\sigma _{\mathrm{ann}}v_\mathrm{lab}=a^{(0)}+a^{(1)}\epsilon+\mathcal{O}(\epsilon^2)$ with $\epsilon\equiv (s-4M_\chi^2)/(4M_\chi^2)$. Since $v_\mathrm{lab}=2\sqrt{\epsilon(1+\epsilon)}/(1+2\epsilon)$, one easily obtains the relation between these two expansions: $a^{(0)}=a$, $a^{(1)}=4b$, etc.},
where $v \equiv v_\mathrm{lab}=\sqrt{s(s-4M_\chi^2)}/(s-2M_\chi^2)$.
According to \cite{Srednicki:1988ce,Gondolo:1990dk}, we then obtain $\langle\sigma _{\mathrm{ann}}v\rangle^{\mathrm{lab}}= a+6b/x+\mathcal{O}(1/x^2)$ with $x\equiv
M_\chi/T$. Now let us compute the coefficients $a$ and $b$ in the effective models. Due to the common factor $(s-4M_\chi^2)^{-1/2}$
in Eqs.~\eqref{diracWIMP-sigma-S} -- \eqref{diracWIMP-sigma-chiral}, $s$ must be expanded up to order $v^4$ to get the correct coefficients $b$. In the lab frame, $s=2M_\chi^2(1+1/\sqrt{1-v^2})= 4 M_\chi^2 + M_\chi^2 v^2 + \frac{3}{4}M_\chi^2 v^4+\mathcal{O}(v^6)$.
Substituting this expansion of $s$ into Eqs.~\eqref{diracWIMP-sigma-S} -- \eqref{diracWIMP-sigma-chiral} and
expanding $\sigma _{\mathrm{ann}}v$ in powers of $v$ up to order
$v^2$, we obtain
\begin{eqnarray}
\sigma_{S,\,\mathrm{ann}}v&\simeq&\frac{1}{8\pi}
\sum_f\bigg(\frac{G_{S,f}}{\sqrt{2}}\bigg)^2c_f
\bigg(1-\frac{m_f^2}{M_\chi^2}\bigg)^{3/2}
M_\chi^2v^2\label{sigma_v-S}\\
\sigma_{P,\,\mathrm{ann}}v&\simeq&\frac{1}{2\pi}
\sum_f\bigg(\frac{G_{P,f}}{\sqrt{2}}\bigg)^2c_f
\sqrt{1-\frac{m_f^2}{M_\chi^2}}M_\chi^2 \bigg[1+\frac{{m_f^2/M_\chi
^2}}
{{8( {1 - m_f^2/M_\chi ^2})}}v^2\bigg]\label{sigma_v-P}\\
\sigma_{V,\,\mathrm{ann}}v&\simeq&\frac{1}{2\pi}
\sum_f\bigg(\frac{G_{V,f}}{\sqrt{2}}\bigg)^2c_f\sqrt{1-\frac{m_f^2}{M_\chi^2}}
(2M_\chi^2+m_f^2) \bigg[1+\frac{{ - 4 + 2m_f^2/M_\chi ^2 +
11m_f^4/M_\chi ^4}} {{24( {1 - m_f^2/M_\chi ^2} )( {2 + m_f^2/M_\chi
^2} )}}v^2\bigg]
\label{sigma_v-V}\\
\sigma_{A,\,\mathrm{ann}}v&\simeq&\frac{1}{2\pi}
\sum_f\bigg(\frac{G_{A,f}}{\sqrt{2}}\bigg)^2c_f
\sqrt{1-\frac{m_f^2}{M_\chi^2}}m_f^2 \bigg[1+\frac{{8M_\chi ^2/m_f^2
- 28 + 23m_f^2/M_\chi ^2}} {{24(1 - m_f^2/M_\chi ^2)}}v^2\bigg]
\label{sigma_v-A}\\
\sigma_{T,\,\mathrm{ann}}v&\simeq&\frac{2}{\pi}
\sum_f\bigg(\frac{G_{T,f}}{\sqrt{2}}\bigg)^2c_f
\sqrt{1-\frac{m_f^2}{M_\chi^2}}(M_\chi^2+2m_f^2) \bigg[1+\frac{{ - 2
- 17m_f^2/M_\chi ^2 + 28m_f^4/M_\chi ^4}} {{24(1 - m_f^2/M_\chi
^2)(1 + 2m_f^2/M_\chi ^2)}}v^2\bigg]
\label{sigma_v-T}\\
\sigma_{SP,\,\mathrm{ann}}v&\simeq&\frac{1}{8\pi}
\sum_f\bigg(\frac{G_{SP,f}}{\sqrt{2}}\bigg)^2c_f
\sqrt{1-\frac{m_f^2}{M_\chi^2}}
M_\chi^2v^2\\
\sigma_{PS,\,\mathrm{ann}}v&\simeq&\frac{1}{2\pi}
\sum_f\bigg(\frac{G_{PS,f}}{\sqrt{2}}\bigg)^2c_f
\sqrt{1-\frac{m_f^2}{M_\chi^2}} (M_\chi^2-m_f^2)
\bigg[1+\frac{{3m_f^2/M_\chi ^2}}
{{8(1 - m_f^2/M_\chi ^2)}}v^2\bigg]\\
\sigma_{VA,\,\mathrm{ann}}v&\simeq&\frac{1}{\pi}
\sum_f\bigg(\frac{G_{VA,f}}{\sqrt{2}}\bigg)^2c_f
\sqrt{1-\frac{m_f^2}{M_\chi^2}}(M_\chi^2-m_f^2) \bigg[1+\frac{{ - 2
+ 11m_f^2/M_\chi ^2}}
{{24(1 - m_f^2/M_\chi ^2)}}v^2\bigg]\\
\sigma_{AV,\,\mathrm{ann}}v&\simeq&\frac{1}{6\pi}
\sum_f\bigg(\frac{G_{AV,f}}{\sqrt{2}}\bigg)^2c_f
\sqrt{1-\frac{m_f^2}{M_\chi^2}}
\big(M_\chi^2+\frac{m_f^2}{2}\big)v^2\label{sigma_v-AV}\\
\sigma_{\tilde{T},\,\mathrm{ann}}v&\simeq&\frac{8}{\pi}
\sum_f\bigg(\frac{\tilde{G}_{T,f}}{\sqrt{2}}\bigg)^2c_f
\sqrt{1-\frac{m_f^2}{M_\chi^2}}(M_\chi^2-m_f^2) \bigg[1+\frac{{ - 2
+ 17m_f^2/M_\chi ^2}}
{{24(1 - m_f^2/M_\chi ^2)}}v^2\bigg]\\
\sigma_{C,\,\mathrm{ann}}v&\simeq&\frac{2}{\pi}
\sum_f\bigg(\frac{G_{C,f}}{\sqrt{2}}\bigg)^2c_f
\sqrt{1-\frac{m_f^2}{M_\chi^2}} M_\chi^2\bigg[1+\frac{{2 -
m_f^2/M_\chi ^2 + 2m_f^4/M_\chi ^4}} {{24(1 - m_f^2/M_\chi
^2)}}v^2\bigg] \label{sigma_v-Ch}
\end{eqnarray}
from which, one can easily read off the corresponding thermally averaged quantities $\langle\sigma _{\mathrm{ann}}v\rangle$.
Our results of Eqs.~\eqref{sigma_v-A},~\eqref{sigma_v-AV}, and \eqref{sigma_v-Ch} agree well with Eqs.~(35) and (39) of Ref. \cite{Srednicki:1988ce}.
Note that Eq.\eqref{sigma_v-S} is the same as Eq.(13) in Ref. \cite{Beltran:2008xg}, while the $\mathcal{O}(v^2)$ terms of Eqs.\eqref{sigma_v-P}--\eqref{sigma_v-A}, and both the $\mathcal{O}(v^0)$ and $\mathcal{O}(v^2)$ terms of Eq.\eqref{sigma_v-T} are different from the corresponding terms of Eq.(14)--(17) in \cite{Beltran:2008xg}. In spite of these differences, they do not have much effect on the main results of Ref. \cite{Beltran:2008xg}. This is because the calculated relic density depends  mainly on the leading term of $\langle\sigma _{\mathrm{ann}}v\rangle$, as we will see below.

Using the standard procedure \cite{kolb-turner,Jungman:1995df} to approximately solve the
Boltzmann equation \eqref{boltz-eq}, we obtain a relic density of DM particles as
\begin{equation}
\Omega_{\mathrm{DM}} h^2=2\Omega_\chi h^2\simeq 2\times 1.04\times10^9~\mathrm{GeV}^{-1}\left(\frac{T_0}{2.725~\mathrm{K}}\right)^3 \frac{x_f}{M_{\mathrm{pl}}\sqrt{g_\ast(T_f)}(a+3b/x_f)} \label{relic-d}
\end{equation}
where $x_f\equiv M_\chi/T_f$ with $T_f$ being the freeze-out temperature, $g_\ast(T_f)$ is the total number of effectively relativistic degrees of freedom at freeze-out, $T_0=2.725\pm0.002$~K \cite{Mather:1998gm} is the present CMB temperature. The freeze-out temperature  parameter $x_f$ can be evaluated by numerically solving the following equation:
\begin{equation}
x_f=\ln\left[c(c + 2)\sqrt{\frac{45}{8}}
\frac{gM_\chi M_{\mathrm{Pl}}(a+6b/x_f)}{2\pi^3\sqrt{g_\ast(x_f)}\,x_f^{1/2}}\right]\label{x-f}
\end{equation}
where $c$ is an order one parameter defined by the freeze-out criterion and determined by matching the late-time and early-time solutions. The precise value of $c$ is not so significant for the numerical solution of $x_f$ due to the logarithmic dependence in Eq.~\eqref{x-f}, and we take the usual value $c=1/2$ in the calculation. Noting that $g_\ast$ in Eqs. \eqref{relic-d} and \eqref{x-f} depends on the temperature $T$, we adopt the recent numerical result of $g_\ast(T)$ in Ref.~\cite{Coleman:2003hs} where the quark-hadron transition temperature is taken to be 200 MeV.

\begin{figure}[!htbp]
\centering
\includegraphics[width=0.88\textwidth]{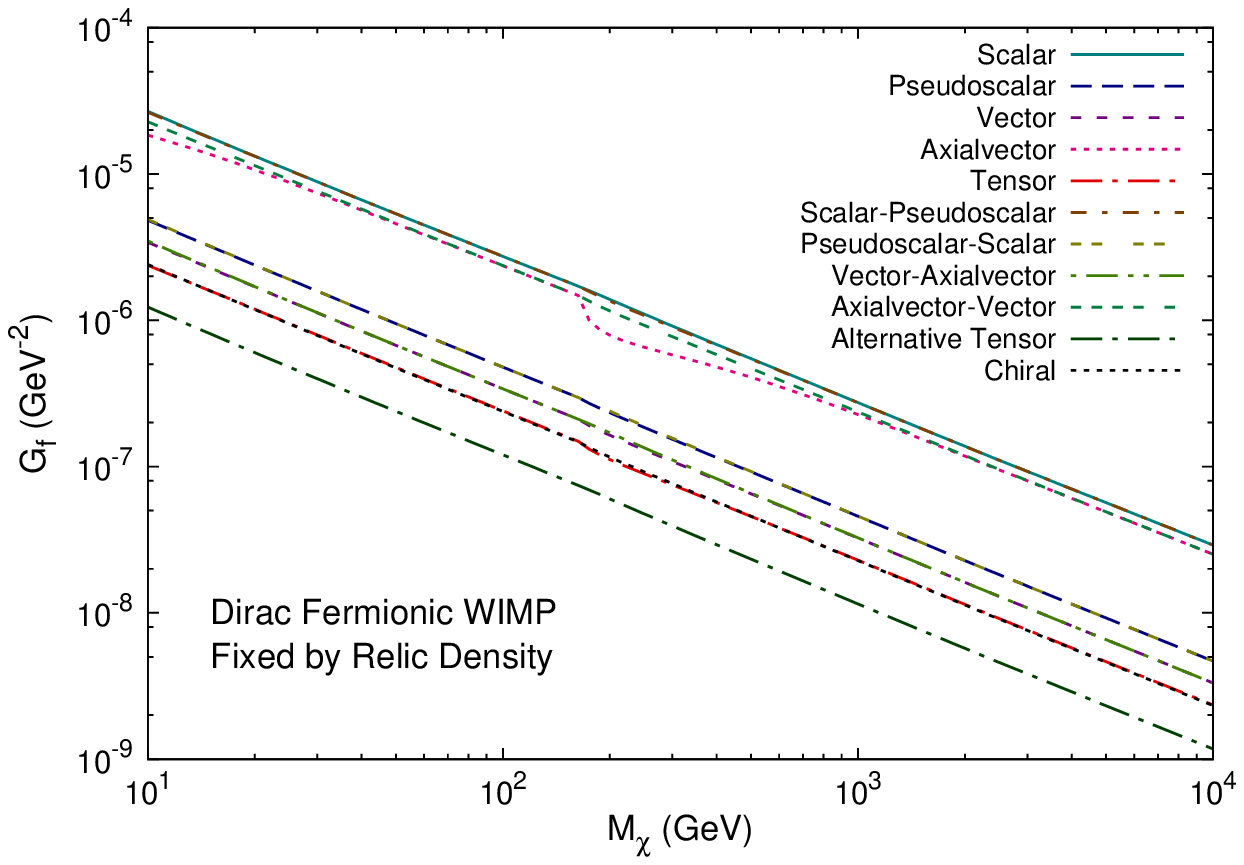}%
\\
\includegraphics[width=0.88\textwidth]{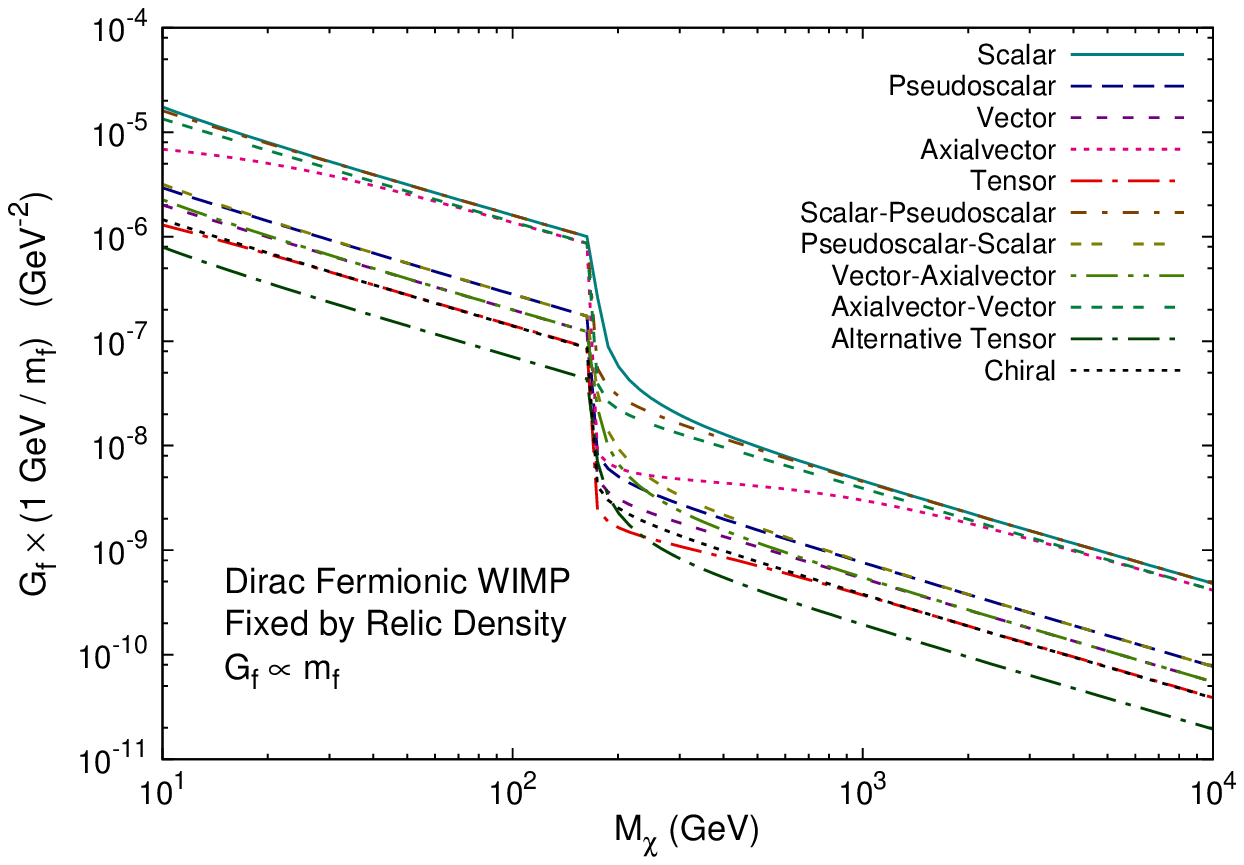}%
\caption{The predicted coupling constant $G_f$ as a function of the WIMP mass $M_\chi$, fixed by the observed relic density, $\Omega_{\mathrm{DM}} h^2=0.1109\pm0.0056$ \cite{Komatsu:2010fb}, in each effective model
of 4-fermion interaction operators. In the upper frame, results are given for the case when the effective couplings to all the standard model fermions are equal (universal couplings). In the lower frame, results are shown for the case when the coupling constants are proportional to the fermion mass $m_f$. In both cases, several pairs of curves are nearly identical. See the text for more details. \label{fig:dirac:rd_coupling}}
\end{figure}
\begin{figure}[!htbp]
\centering
\includegraphics[width=0.44\textwidth]{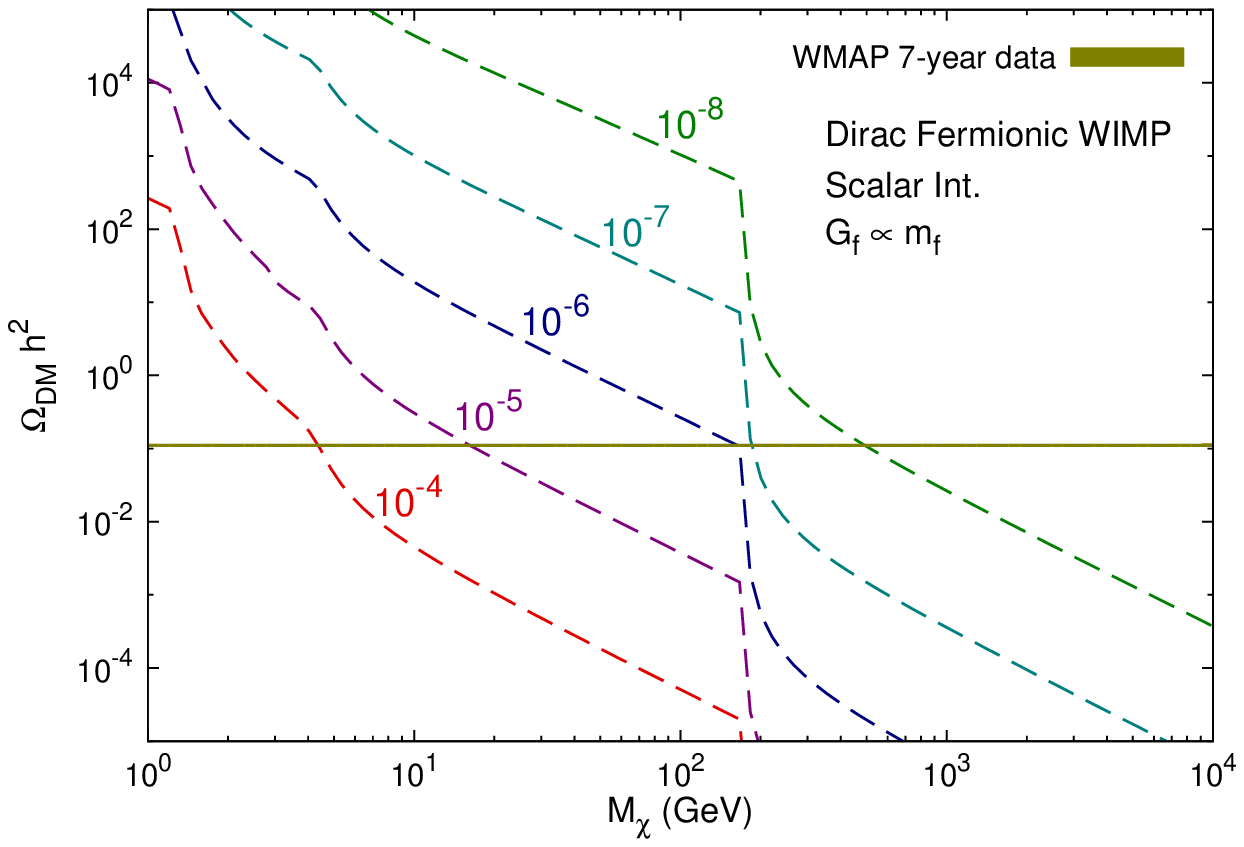}%
\hspace{0.01\textwidth}%
\includegraphics[width=0.44\textwidth]{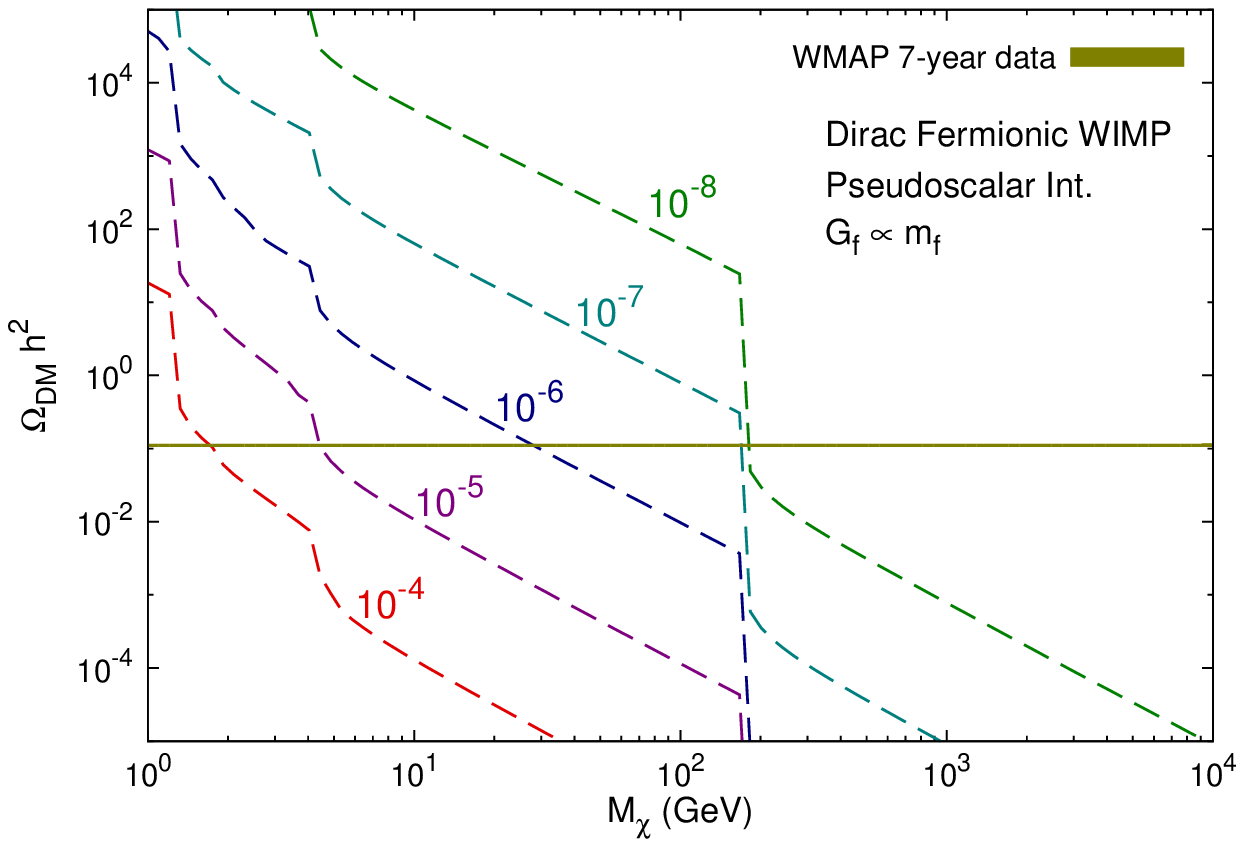}%
\\
\includegraphics[width=0.44\textwidth]{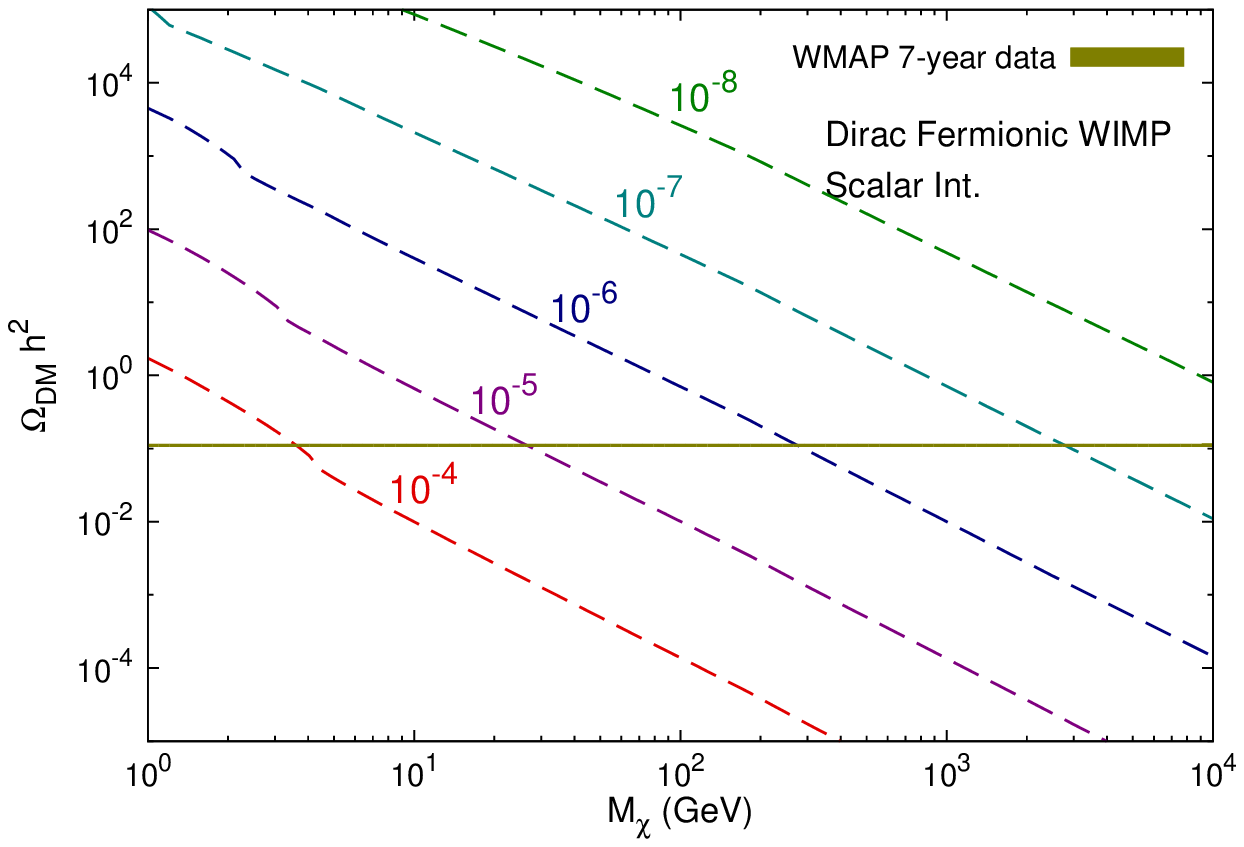}%
\hspace{0.01\textwidth}%
\includegraphics[width=0.44\textwidth]{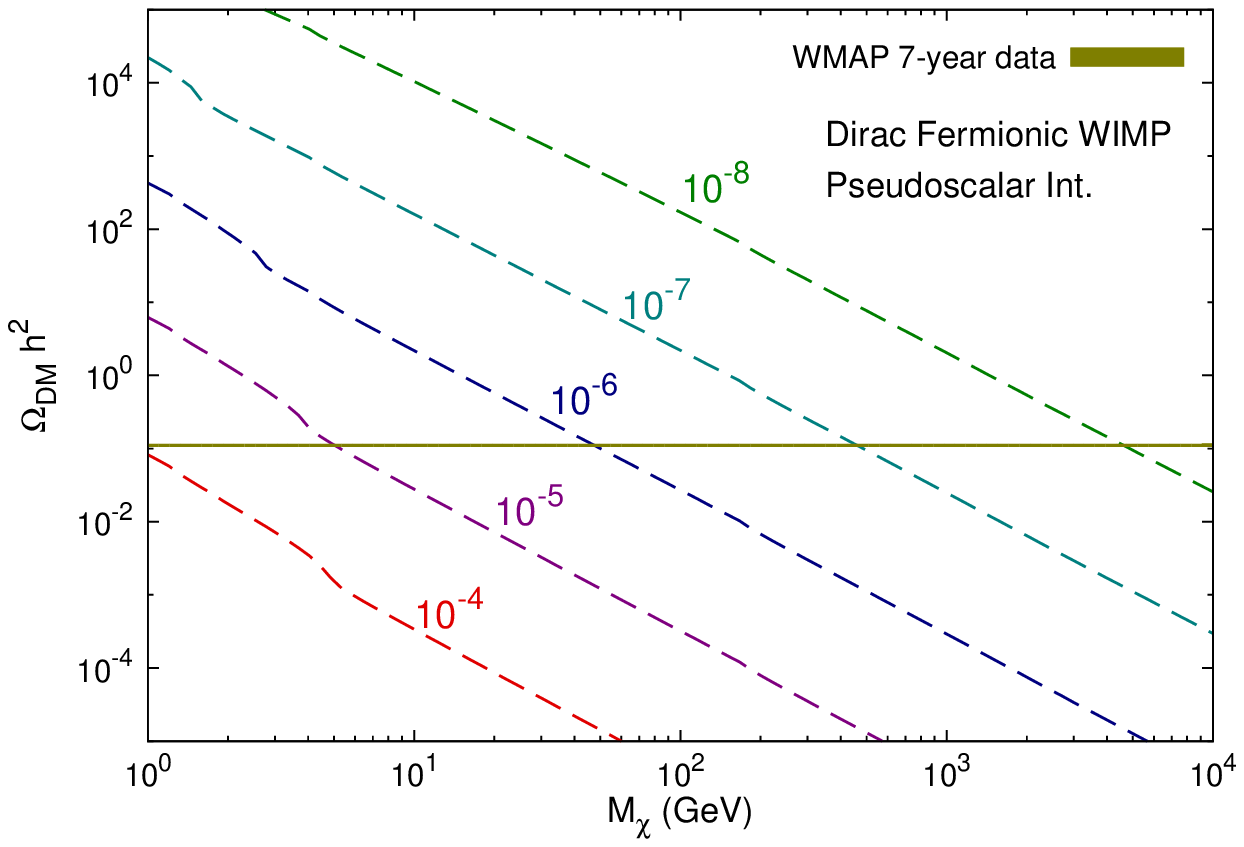}%
\\
\includegraphics[width=0.44\textwidth]{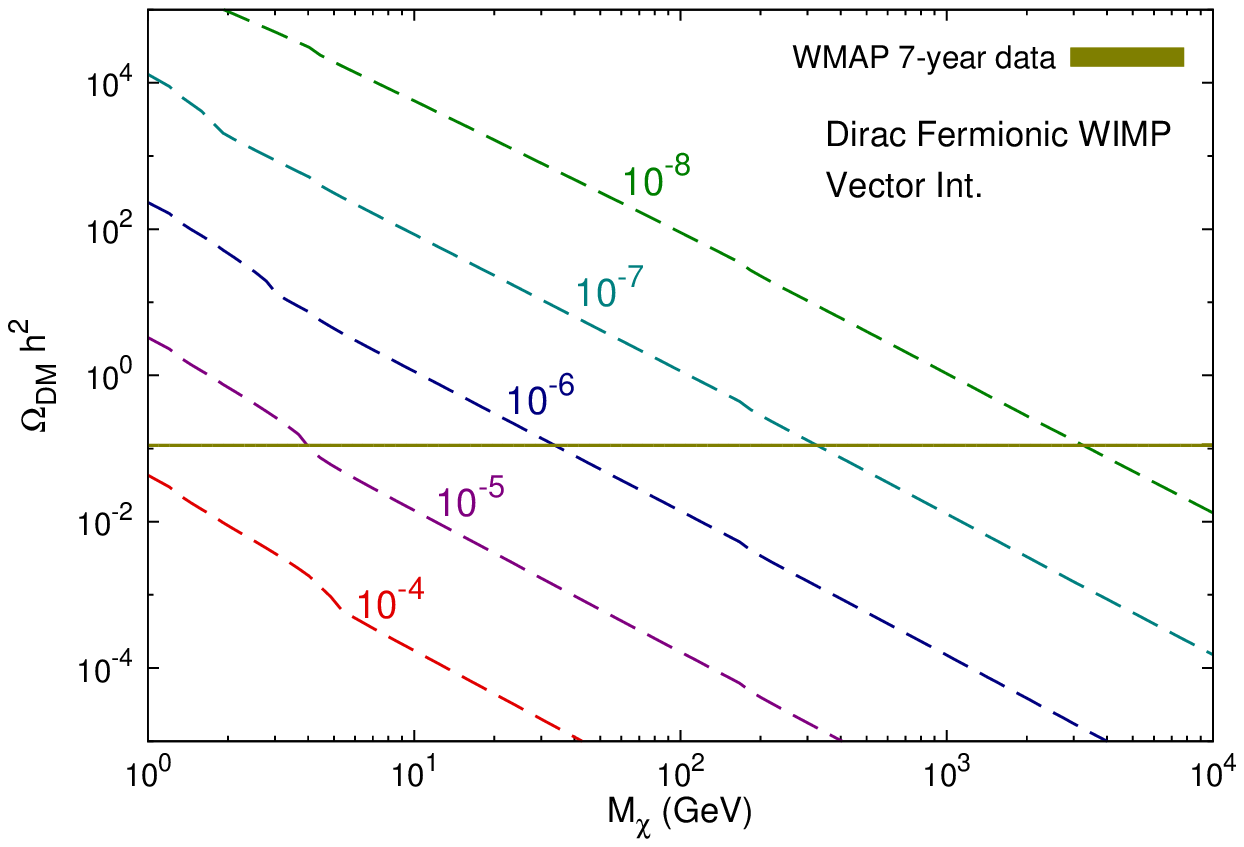}%
\hspace{0.01\textwidth}%
\includegraphics[width=0.44\textwidth]{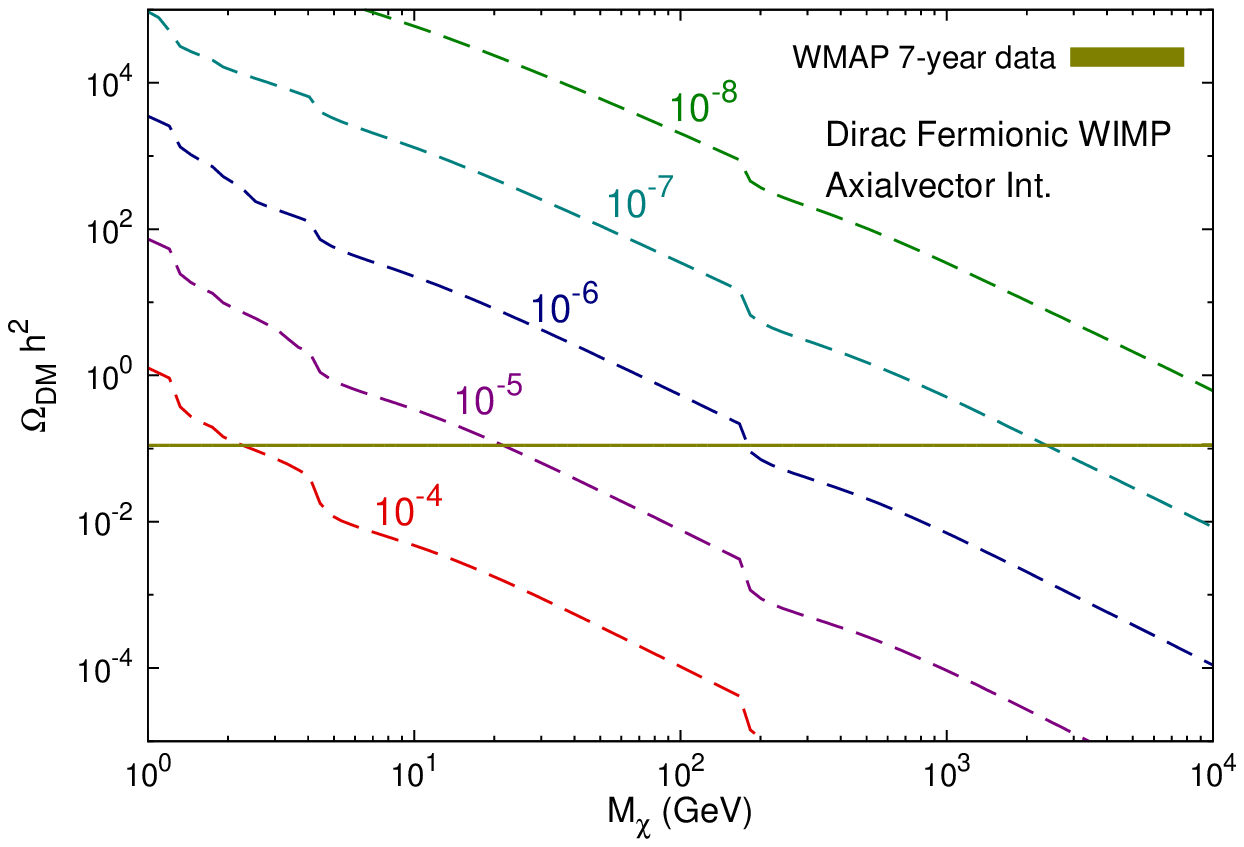}%
\\
\includegraphics[width=0.44\textwidth]{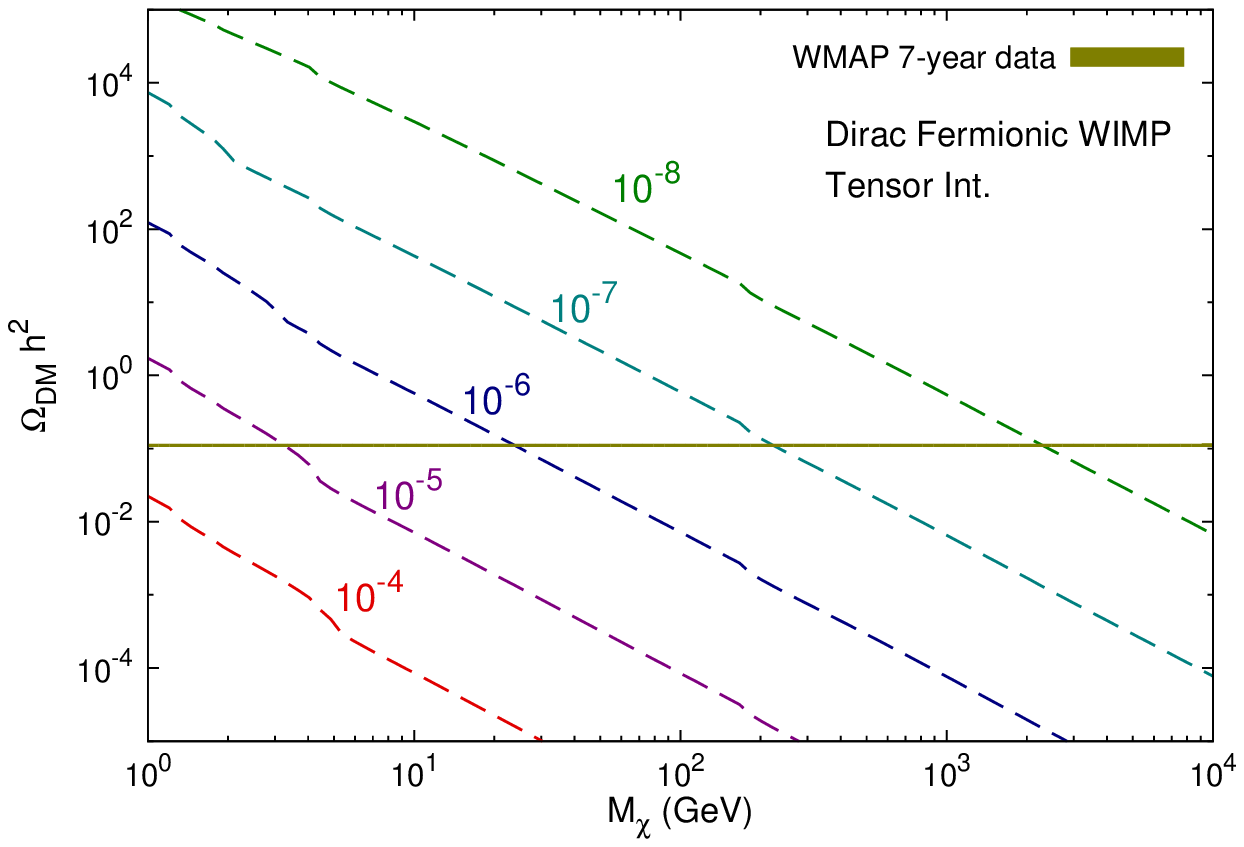}%
\hspace{0.01\textwidth}%
\includegraphics[width=0.44\textwidth]{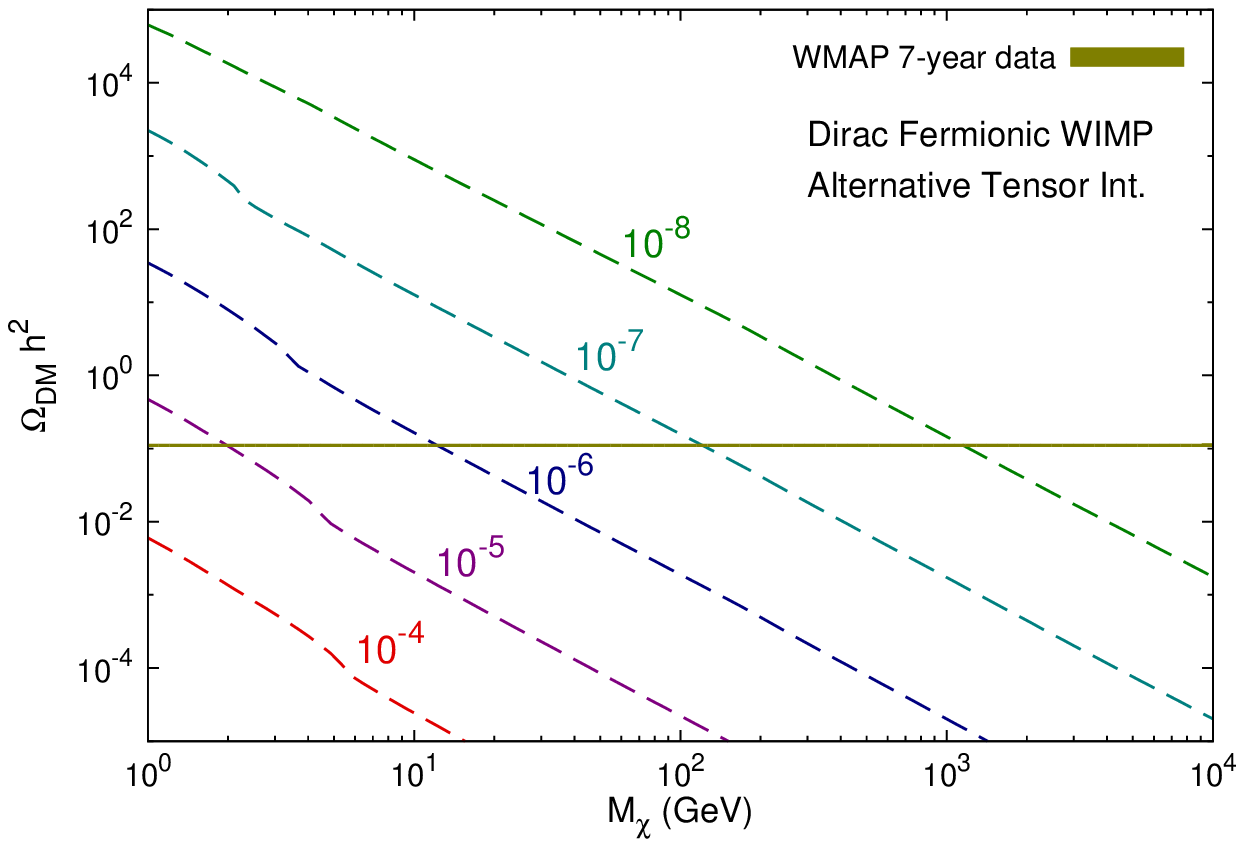}%
\caption{The predicted thermal relic density (dashed lines) of
Dirac fermionic WIMPs with scalar, pseudoscalar, vector,
axialvector, tensor and alternative tensor interactions
respectively. In the upper left and upper right frames, results
are given for the case when the coupling constants are
proportional to the fermion mass $m_f$, $G_f \times \left(
1~\mathrm{GeV} / m_f \right) = 10^{-8}$, $10^{-7}$, $10^{-6}$,
$10^{-5}$ and $10^{-4}~\mathrm{GeV}^{-2}$. In the remaining 6
frames, results are shown for the case when the couplings to all
the standard model fermions are equal (universal couplings), $G_f
= 10^{-8}$, $10^{-7}$, $10^{-6}$, $10^{-5}$ and
$10^{-4}~\mathrm{GeV}^{-2}$. The horizontal solid band shows the
range of the observed DM relic density, $\Omega_{\mathrm{DM}}
h^2=0.1109\pm0.0056$, measured by WMAP \cite{Komatsu:2010fb}.
\label{fig:dirac:rd_density}}
\end{figure}

According to the observed DM relic density, $\Omega_{\mathrm{DM}}
h^2=0.1109\pm0.0056$ \cite{Komatsu:2010fb}, we can estimate the
relation between the effective coupling constants $G_f$ and the
WIMP mass $M_\chi$ in each effective model of 4-fermion
interaction operators, as shown in
Fig.~\ref{fig:dirac:rd_coupling}. Two kinds of coupling constants
are considered here. In the upper frame of
Fig.~\ref{fig:dirac:rd_coupling}, we show the results for the case
when the effective couplings to all the standard model fermions
are equal (universal couplings). In the lower frame of
Fig.~\ref{fig:dirac:rd_coupling}, we show the results for the case
when the coupling constants are proportional to the fermion mass
$m_f$. In both cases, $G_f$ decreases as $M_\chi$ increases for
fixed $\Omega_{\mathrm{DM}} h^2$ in each effective model. Besides,
Fig.~\ref{fig:dirac:rd_coupling} has several interesting features:
\begin{itemize}
  \item In both cases, the 4 curves of $G_f$ vs. $M_\chi$ for the effective models of scalar, scalar-pseudoscalar, axialvector and axialvector-vector interactions lie well above the other curves. This comes from the fact that $\sigma_{S,\,\mathrm{ann}}v$, $\sigma_{SP,\,\mathrm{ann}}v$ and $\sigma_{AV,\,\mathrm{ann}}v$ are of order $\mathcal{O}(v^2)$; although the leading term of $\sigma_{A,\,\mathrm{ann}}v$ is of order $\mathcal{O}(v^0)$, it is smaller by a factor of $m_f^2/M_\chi^2$ than the $\mathcal{O}(v^0)$ terms for other types of interactions.
  \item In the case of $G_f\propto m_f$, there is an obvious downward bend in the curve of $G_f$ vs. $M_\chi$ at about $M_\chi\sim m_t=171.2$~GeV in each effective model. This can be easily explained as follows. In the low velocity limit, the threshold for the annihilation channel $\chi~ \bar{\chi}\to t~\bar{t}$ is about $M_\chi\sim m_t$. Since $G_f\propto m_f$, the WIMP couples much more strongly to the top quark than to other fermions, and the corresponding channel $\chi~ \bar{\chi}\to t~\bar{t}$ gives a tremendous contribution to the total $\langle\sigma _{\mathrm{ann}}v\rangle$. This finally makes the curve to bend down.
  \item In the case of universal couplings, there are 5 pairs of nearly identical curves, because in each pair their corresponding $\sigma_{\mathrm{ann}}v$ differ only by terms of $\mathcal{O}(v^2)$ and/or terms of $m_f^2/M_\chi^2$. This feature is also noted in Ref. \cite{Beltran:2008xg}. These pairs are: (1) the curves for scalar and scalar-pseudoscalar interactions, and we denote this approximate identity of the two curves by S$\;\simeq\;$SP for short here and henceforth; (2) P$\;\simeq\;$PS; (3) V$\;\simeq\;$VA; (4) T$\;\simeq\;$C; (5) A$\;\simeq\;$AV except for some small regions. From these approximate identities for pairs of curves, we see that the predicted relic density relies mainly on the leading term of $\langle\sigma _{\mathrm{ann}}v\rangle$ in each model, especially on the $\mathcal{O}(M_\chi^2)$ term in the leading term.
  \item In the case of $G_f\propto m_f$, there are also 5 pairs of nearly identical curves except for some small regions. These pairs are the same as those in the above case, though deviations of the two nearly identical curves in each pair become large in this case.
\end{itemize}

In Fig.~\ref{fig:dirac:rd_density} we show the curves of
$\Omega_{\mathrm{DM}} h^2$ vs. $M_\chi$ for fixed coupling constants
in the models of scalar, pseudoscalar, vector, axialvector, tensor
and alternative tensor interaction operators. Due to the nearly
identities described above, we do not include the curves for
scalar-pseudoscalar, pseudoscalar-scalar, vector-axialvector,
axialvector-vector and chiral interactions. Here we still consider
the two kinds of coupling constants. In the upper left and upper
right frames, we show the results when the coupling constants are
proportional to the standard model fermion masses, $G_f\propto m_f$,
and various values of the couplings are taken: $G_f \times \left(
1~\mathrm{GeV} / m_f \right) = 10^{-8}$, $10^{-7}$, $10^{-6}$,
$10^{-5}$ and $10^{-4}~\mathrm{GeV}^{-2}$. This proportionality of
the couplings to the fermion masses may come from
Yukawa couplings of a Higgs
mediated interaction or some other unknown underlying mechanism. In
the remaining 6 frames, we show the results for the case when the
effective couplings to all the standard model fermions are equal
(universal couplings), and various values of the couplings are
taken: $G_f = 10^{-8}$, $10^{-7}$, $10^{-6}$, $10^{-5}$ and
$10^{-4}~\mathrm{GeV}^{-2}$. The curves in
Fig.~\ref{fig:dirac:rd_density} bend down to more or less at about
$M_\chi\sim 1.72$~GeV, 4.20~GeV and 171.2~GeV, which exactly
correspond to the masses of charm, bottom, top quarks, respectively.

Comparing with the results in Ref.~\cite{Beltran:2008xg}, we
observe that the curves in the first 6 frames of
Fig.~\ref{fig:dirac:rd_density} are a little higher than those
given by Ref. \cite{Beltran:2008xg}. This slight difference may be
caused by the following reasons: (1) We use
$\Omega_{\mathrm{DM}}=\Omega_\chi+\Omega_{\bar{\chi}}=2\Omega_\chi$
for Dirac fermionic WIMPs with the assumption of no
particle-antiparticle asymmetry. (2) Some formulas of $\sigma
_{\mathrm{ann}}v$ we obtained differ slightly from those given by
\cite{Beltran:2008xg}, as already described at the bottom of
Eq.\eqref{sigma_v-Ch}. (3) We use the effective degrees of freedom
$g_\ast(T)$ given by \cite{Coleman:2003hs}.

It is important to note that the results in
Figs.~\ref{fig:dirac:rd_coupling} and \ref{fig:dirac:rd_density}
are found under the assumptions presented in Sec.~\ref{sec-model}.
If resonances, coannihilations, or annihilations to final states
other than fermion-antifermion pairs are significant, the actual
curves in Figs.~\ref{fig:dirac:rd_coupling} and
\ref{fig:dirac:rd_density} will be significantly lower than these
shown there, as pointed out in \cite{Beltran:2008xg}.

\section{Direct detection\label{sec-direct}}

In this section we discuss the direct detection constraints on the
effective models of Eqs.\eqref{Leff-S}--\eqref{Leff-RL}. Direct
detection experiments are designed to measure the recoil energy of
the atomic nuclei when the WIMPs elastically scatter off them. The
WIMP-quark interactions in the effective models naturally induce
the WIMP-nucleon interactions, and the latter further induce the
WIMP-nucleus interactions. Such interactions may lead to the
elastic scattering of the WIMPs with the nuclei, which may be
detected at the direct detection experiments.

The velocity of the WIMP near the Earth is thought to be of the
same order as the orbital velocity of the Sun, $v\simeq 0.001c$.
Because of this small velocity, the momentum transfer in the
WIMP-nucleus scattering is considerably small compared to the
masses of the WIMP and the nuclei. Thus all the WIMP-nucleus cross
sections can be calculated in the limit of zero momentum tranfer.
In this limit the WIMP-quark interaction operators,
$\mathcal{L}_{\mathrm{P}}$, $\mathcal{L}_{\mathrm{SP}}$,
$\mathcal{L}_{\mathrm{PS}}$, $\mathcal{L}_{\mathrm{VA}}$,
$\mathcal{L}_{\mathrm{AV}}$ and
$\mathcal{L}_{\tilde{\mathrm{T}}}$, and their correspondingly
induced WIMP-nucleon interaction operators have no contribution to
the WIMP-nucleus cross sections and thus they are not sensitive to
direct detection experiments. This is because in the zero momentum
transfer limit some fermion bilinear operators become zero, for
example, the operator $\bar{\psi}\gamma_5\psi$ vanishes, and the
time component of $\bar{\psi}\gamma^\mu\gamma_5\psi$ and the space
components of $\bar{\psi}\gamma^\mu\psi$ vanish as well. For more
details on this issue, see Ref. \cite{Agrawal:2010fh}.

Among the remaining operators relevant to direct detection, the scalar and vector interaction operators, $\mathcal{L}_{\mathrm{S}}$ and $\mathcal{L}_{\mathrm{V}}$, are referred to as spin-independent (SI) interactions, while the axialvector and tensor interaction, $\mathcal{L}_{\mathrm{A}}$ and $\mathcal{L}_{\mathrm{T}}$, belong to spin-dependent (SD) interactions. And the chiral interaction operators are the combinations of SI and SD interactions. The SI interactions of all the nucleons add coherently in the target nucleus, and the corresponding WIMP-nucleus cross section is proportional to the square of the atomic mass number of the nucleus. On the other hand, since the spins of nucleons in a nucleus tend to cancel in pairs, the SD interactions rely mainly on the spin content of one unpaired nucleon and the corresponding cross section is not enhanced for heavy nuclei.

We would like to illustrate the calculation in the effective model
of scalar interaction operators.  Eq.\eqref{Leff-S} can induce the
effective Lagrangian for the WIMP-nucleon couplings, which reads
\begin{eqnarray}
\mathcal{L}_{\mathrm{S},\,\mathrm{induced}}
=\sum_{N=p,n}\frac{G_{S,N}}{\sqrt{2}}\bar{\chi}\chi\bar{N}N\label{Leff-S-N}
\end{eqnarray}
where the WIMP's effective Fermi couplings to the nucleons
(protons and neutrons), $G_{S,N}$ ($N=p,~n$), are related to the
coupling constants to quarks by
\begin{eqnarray}
G_{S,N}
=\sum_{q=u,d,s}G_{S,q}f^N_q\frac{m_N}{m_q}+\sum_{q=c,b,t}G_{S,q}f^N_Q\frac{m_N}{m_q}
\end{eqnarray}
where the nucleon form factors are $f^p_u=0.020\pm0.004$,
$f^p_d=0.026\pm0.005$, $f^p_s=0.118\pm0.062$, $f^n_u=0.014\pm0.003$,
$f^n_d=0.036\pm0.008$, $f^n_s=0.118\pm0.062$ \cite{Ellis:2000ds}, and
\begin{equation}
f^N_Q=\frac{2}{27}\bigg(1-\sum_{q=u,d,s}f^N_q\bigg)
\end{equation}

From Eq.\eqref{Leff-S-N} and its induced WIMP-nucleus interactions, it follows that the cross section for a WIMP ($\chi$) scattering elastically from a nucleus ($A$) in the zero momentum transfer limit is given by \cite{Belanger:2008sj}
\begin{eqnarray}
\text{Scalar Int.}:\qquad\sigma_{S,\,\chi A}=\frac{4}{\pi}
\frac{M^2_{\chi}m^2_A}{(M_\chi+m_A)^2}\left[\frac{1}{2}\bigg(Z\frac{G_{S,p}}{\sqrt{2}}
+(A-Z)\frac{G_{S,n}}{\sqrt{2}}\bigg)\right]^2\label{cross-sec-chi-A-S}
\end{eqnarray}
where $m_A$ is the target nucleus mass. $Z$ and $(A-Z)$ are the
numbers of protons and neutrons in the nucleus.  The factor $1/2$
in the square bracket comes from the fact that a Dirac fermionic
WIMP and its antiparticle are different \cite{Belanger:2008sj}.
However, this factor is missing in Eq.(20) of
Ref.\cite{Beltran:2008xg}. Indeed, such a factor $1/2$ does not
exist in the expression for a self-conjugated WIMP such as a
Majorana fermion, but it seems that a Dirac fermionic WIMP is
considered in Ref.\cite{Beltran:2008xg}, otherwise the fermion
bilinears $\bar{\chi}\gamma^\mu\chi$ and
$\bar{\chi}\sigma^{\mu\nu}\chi$ vanish for a Majorana fermion
$\chi$. Taking the special case when the nucleus is just the
nucleon (proton or neutron) in Eq.\eqref{cross-sec-chi-A-S}, we
obtain the WIMP-nucleon cross section in the zero momentum
transfer limit:
\begin{eqnarray}
\text{Scalar Int.}:\qquad \sigma_{S,\,\chi N}=\frac{M^2_{\chi}m^2_N}
{\pi(M_\chi+m_N)^2}\bigg(\frac{G_{S,N}}{\sqrt{2}}\bigg)^2\label{cross-sec-chi-N-S}
\end{eqnarray}

Likewise, we compute the WIMP-nucleus cross section in the effective models of vector, axialvector, tensor and chiral interaction operators, resulting in
\begin{eqnarray}
\text{Vector Int.}:&&\qquad\sigma_{V,\,\chi A}=
\frac{M^2_{\chi}m^2_A}{\pi(M_\chi+m_A)^2}\left[Z\frac{G_{V,p}}{\sqrt{2}}
+(A-Z)\frac{G_{V,n}}{\sqrt{2}}\right]^2\label{cross-sec-chi-A-V}\\
\text{Axialvector Int.}:&&\qquad\sigma_{A,\,\chi A}=\frac{4M^2_{\chi}m^2_A}{\pi(M_\chi+m_A)^2} \frac{J_A+1}{J_A}\left[\frac{G_{A,p}} {\sqrt{2}}S^A_p +
\frac{G_{A,n}}{\sqrt{2}}S^A_n\right]^2\label{cross-sec-chi-A-A}\\
\text{Tensor Int.}:&&\qquad\sigma_{T,\,\chi A}=\frac{16M^2_{\chi}m^2_A}{\pi(M_\chi+m_A)^2} \frac{J_A+1}{J_A}\left[\frac{G_{T,p}}{\sqrt{2}}S^A_p+\frac{G_{T,n}}{\sqrt{2}}S^A_n\right]^2
\label{cross-sec-chi-A-T}\\
\text{Chiral Int.}:&&\qquad\sigma_{C,\,\chi A}=\frac{M^2_{\chi}m^2_A}{\pi(M_\chi+m_A)^2}\left[Z\frac{G_{V,p}}{\sqrt{2}}
+(A-Z)\frac{G_{V,n}}{\sqrt{2}}\right]^2\nonumber\\
&&\qquad\qquad\qquad+\frac{4M^2_{\chi}m^2_A}{\pi(M_\chi+m_A)^2}\frac{J_A+1}{J_A}
\left[\frac{G_{A,p}}{\sqrt{2}}S^A_p+\frac{G_{A,n}}{\sqrt{2}}S^A_n\right]^2\nonumber\\
&&\qquad\qquad\quad\simeq\frac{M^2_{\chi}m^2_A}{\pi(M_\chi+m_A)^2}\left[Z\frac{G_{V,p}}{\sqrt{2}}
+(A-Z)\frac{G_{V,n}}{\sqrt{2}}\right]^2\quad \text{(when $A\gg 1$)}
\label{cross-sec-chi-A-C}
\end{eqnarray}
where $J_A$ is the nuclear spin, $S^A_N$ is the
expectation value of the total spin of the nucleon ($N$) in the nucleus ($A$), and the WIMP's effective Fermi couplings, $G_N$, to the nucleons are related to those to quarks, $G_q$, by
\begin{eqnarray}
&&G_{V,p}=2G_{V,u}+G_{V,d}\;,\qquad G_{V,n}=G_{V,u}+2G_{V,d}\\
&&G_{A,N}=\sum_{q=u,d,s}G_{A,q}\Delta^N_q\\
&&G_{T,N}=\sum_{q=u,d,s}G_{T,q}\Delta^N_q
\end{eqnarray}
with the form factors $\Delta^p_u=0.842\pm0.012$,
$\Delta^p_d=-0.427\pm0.013$, $\Delta^p_s=-0.085\pm0.018$
\cite{Airapetian:2007mh}, $\Delta^n_u=\Delta^p_d$,
$\Delta^n_d=\Delta^p_u$, $\Delta^n_s=\Delta^p_s$. In
Eq.\eqref{cross-sec-chi-A-C} the WIMP-nucleus cross section for
chiral interactions receives contributions from both SI and SD
interactions. As heavy nuclei are used in CDMS and XENON, the SD
part in Eq.\eqref{cross-sec-chi-A-C} is several orders of
magnitude smaller than the SI part, and thus the SD contribution
may be omitted when comparing the results in those experiments.
So, in this approximation, the WIMP-nucleus cross section for
chiral interactions has the same form as that for vector
interactions. The WIMP-nucleus cross sections in Eqs.
\eqref{cross-sec-chi-A-V} -- \eqref{cross-sec-chi-A-C} can be
normalized to the corresponding WIMP-nucleon cross sections as
follows:
\begin{eqnarray}
\text{Vector Int.}:&&\qquad\sigma_{V,\,\chi N}=\frac{M^2_{\chi}m^2_N}{\pi(M_\chi+m_N)^2}
\left(\frac{G_{V,N}}{\sqrt{2}}\right)^2\label{cross-sec-chi-N-V}\\
\text{Axialvector Int.}:&&\qquad\sigma_{A,\,\chi N}=\frac{3M^2_{\chi}m^2_N}{\pi(M_\chi+m_N)^2}
\left(\frac{G_{A,N}}{\sqrt{2}}\right)^2\label{cross-sec-chi-N-A}\\
\text{Tensor Int.}:&&\qquad\sigma_{T,\,\chi N}=\frac{12M^2_{\chi}m^2_N}{\pi(M_\chi+m_N)^2}\left(\frac{G_{T,N}}{\sqrt{2}}\right)^2
\label{cross-sec-chi-N-T}\\
\text{Chiral Int.}:&&\qquad\tilde{\sigma}_{C,\,\chi N}\simeq\frac{M^2_{\chi}m^2_N}{\pi(M_\chi+m_N)^2}
\left(\frac{G_{V,N}}{\sqrt{2}}\right)^2\label{cross-sec-chi-N-C}
\end{eqnarray}
Note that in Eq.\eqref{cross-sec-chi-N-C}
$\tilde{\sigma}_{C,\,\chi N}$ is not the actual WIMP-nucleon cross
section for chiral interactions, it just means a normalized
quantity of the corresponding WIMP-nucleus cross section for a
heavy nucleus $A$. In other words, a normalization procedure from
Eq.\eqref{cross-sec-chi-A-C} to Eq.\eqref{cross-sec-chi-N-C} is
used: $\tilde{\sigma}_{C,\,\chi N}\equiv
[\mu_N^2/(\mu_A^2A^2)]\cdot\sigma_{C,\,\chi A}$ when $A\gg 1$,
with the reduced masses $\mu_A\equiv M_\chi m_A/(M_\chi+m_A)$ and
$\mu_N\equiv M_\chi m_N/(M_\chi+m_N)$, where we have used the fact
that $G_{V,p}$ and $G_{V,n}$ have the same order of magnitude:
$G_{V,p}\sim G_{V,n}$.

\begin{figure}[!htbp]
\centering
\includegraphics[width=0.49\textwidth]{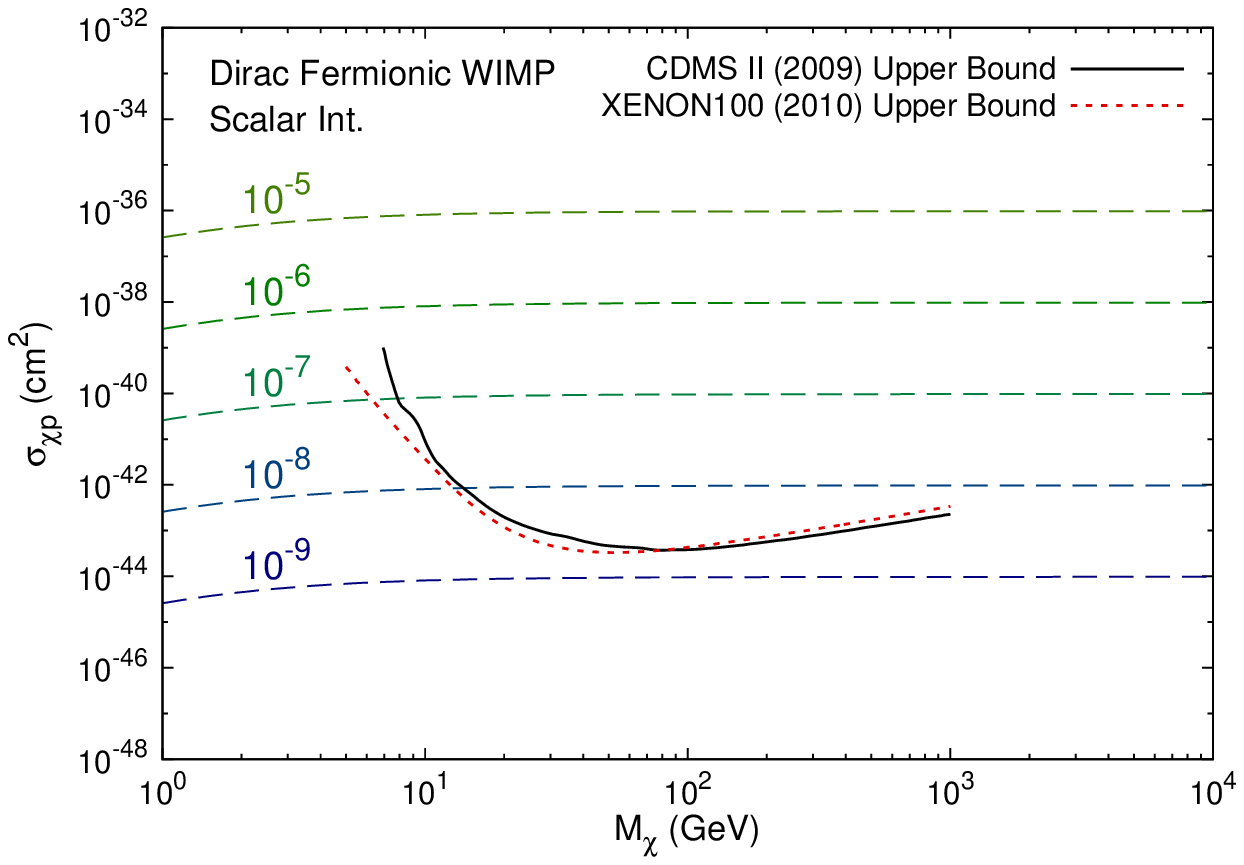}%
\hspace{0.008\textwidth}%
\includegraphics[width=0.49\textwidth]{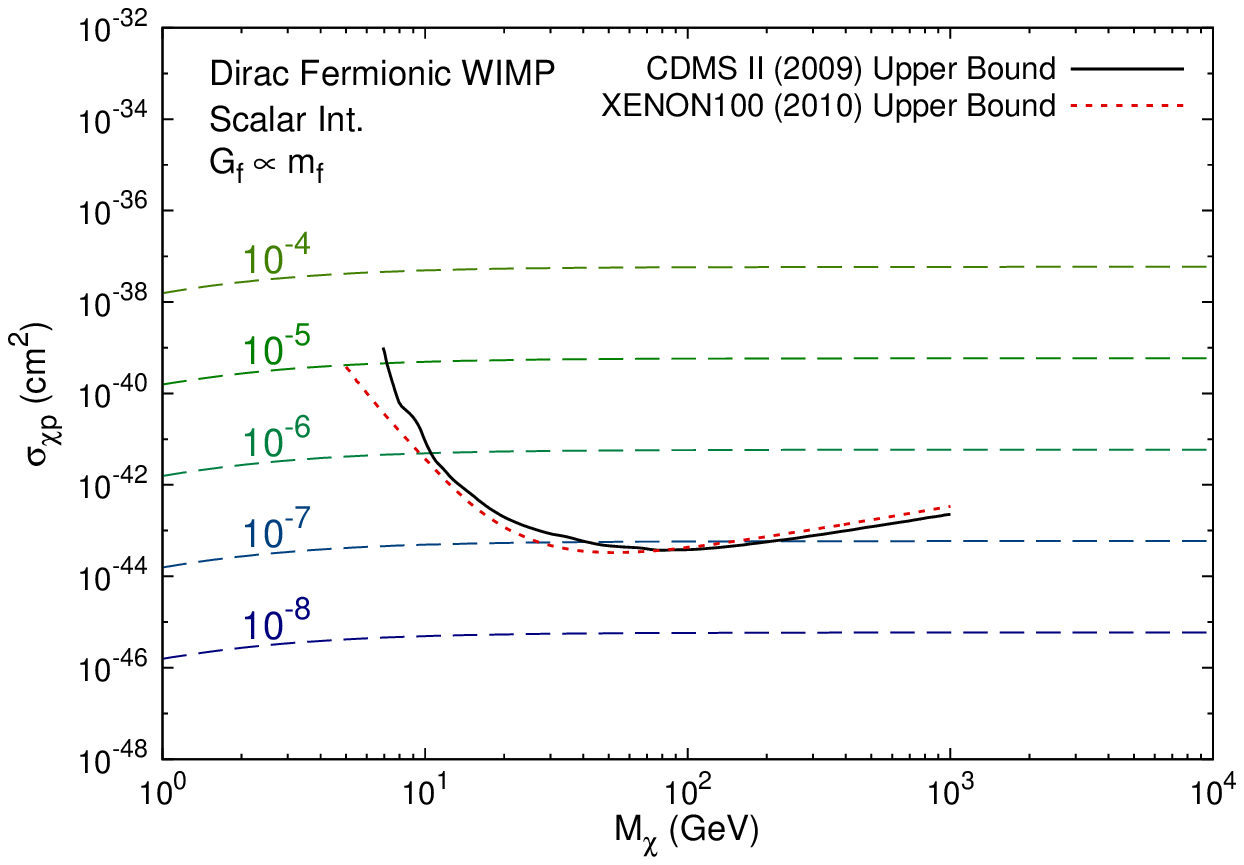}%
\\
\includegraphics[width=0.49\textwidth]{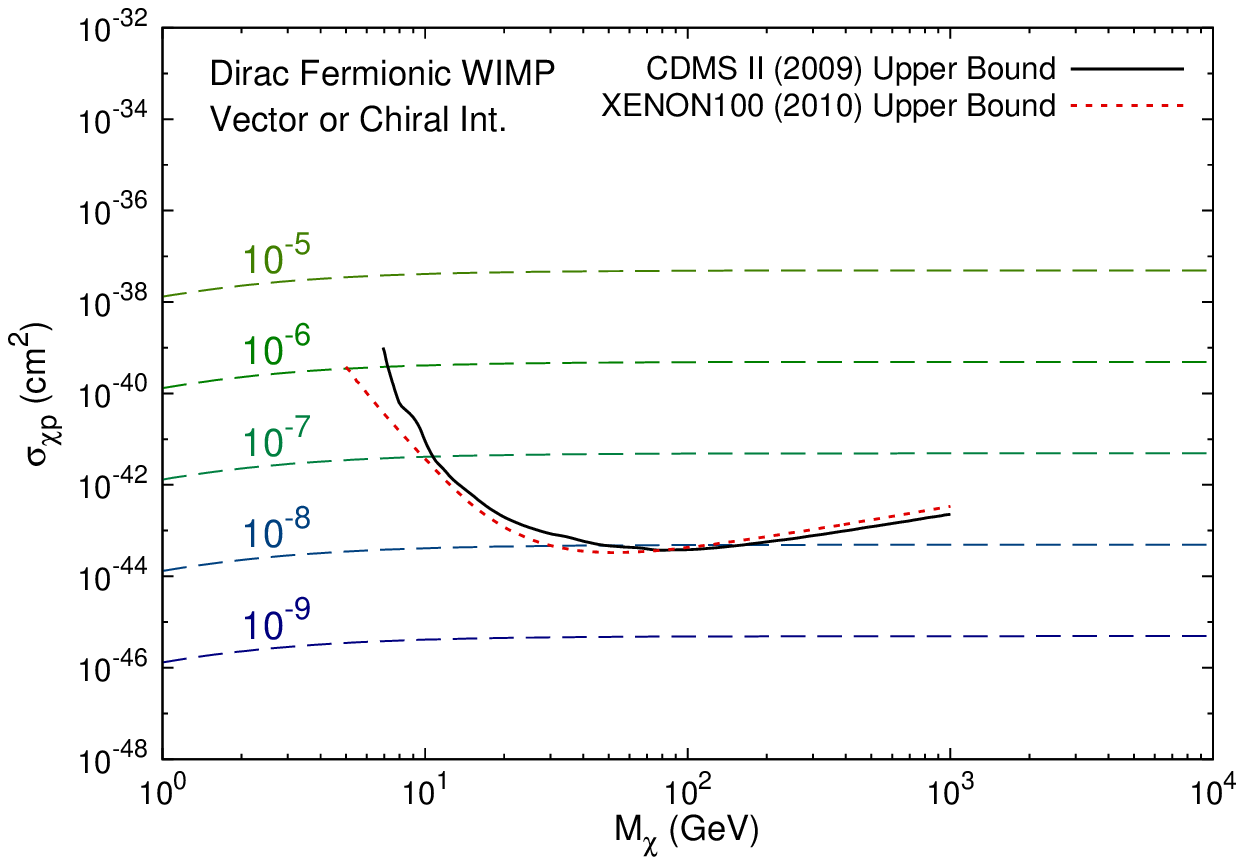}%
\hspace{0.008\textwidth}%
\includegraphics[width=0.49\textwidth]{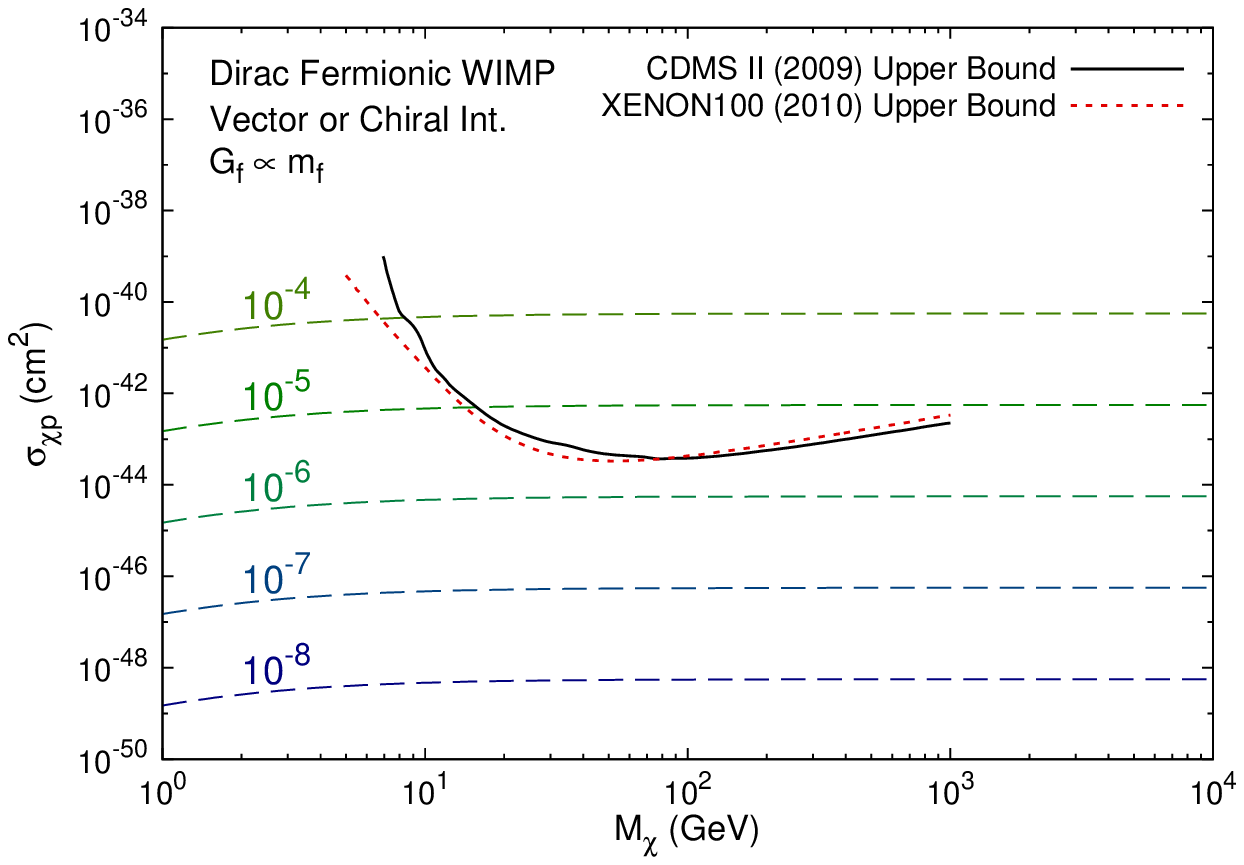}
\caption{The spin-independent (SI) WIMP-proton cross sections (dashed lines) for Dirac fermionic WIMPs with scalar, vector and chiral interactions. The results for chiral interactions are approximately identical to those for vector interactions. In the upper left and lower left frames, the results are shown for scalar and vector interactions for the case of universal couplings $G_f = 10^{-9}$, $10^{-8}$, $10^{-7}$, $10^{-6}$ and $10^{-5}~\mathrm{GeV}^{-2}$. In the remaining two frames, the results are shown for the case when the coupling constants proportion to the fermion mass $m_f$, $G_f \times (1~\mathrm{GeV} / m_f) = 10^{-8}$, $10^{-7}$, $10^{-6}$, $10^{-5}$ and $10^{-4}~\mathrm{GeV}^{-2}$. Also shown as solid and dotted curves are the current upper bounds from the experiments of CDMS II (2009) \cite{Ahmed:2009zw} and XENON100 (2010) \cite{Aprile:2010um}.
\label{fig:dirac:scat:SI}}
\end{figure}
\begin{figure}[!htbp]
\centering
\includegraphics[width=0.49\textwidth]{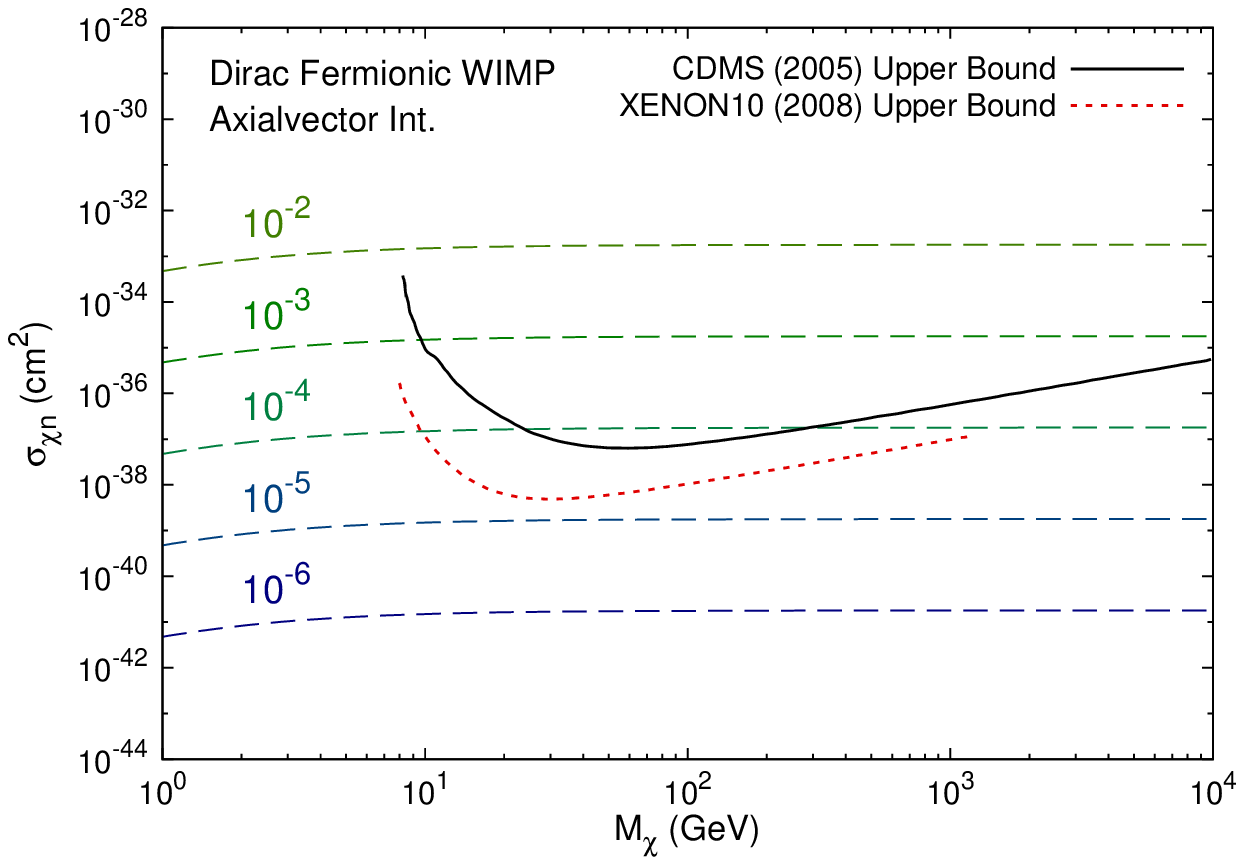}%
\hspace{0.008\textwidth}%
\includegraphics[width=0.49\textwidth]{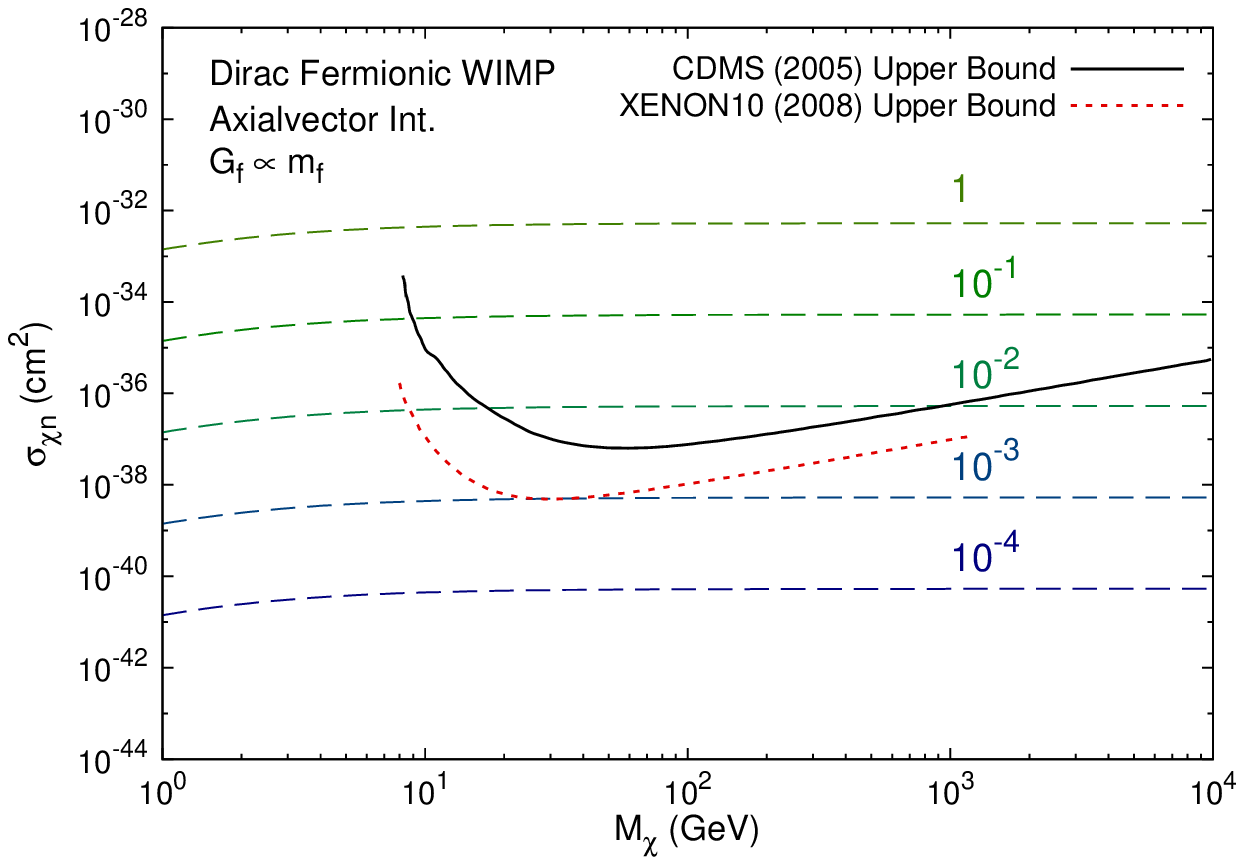}%
\\
\includegraphics[width=0.49\textwidth]{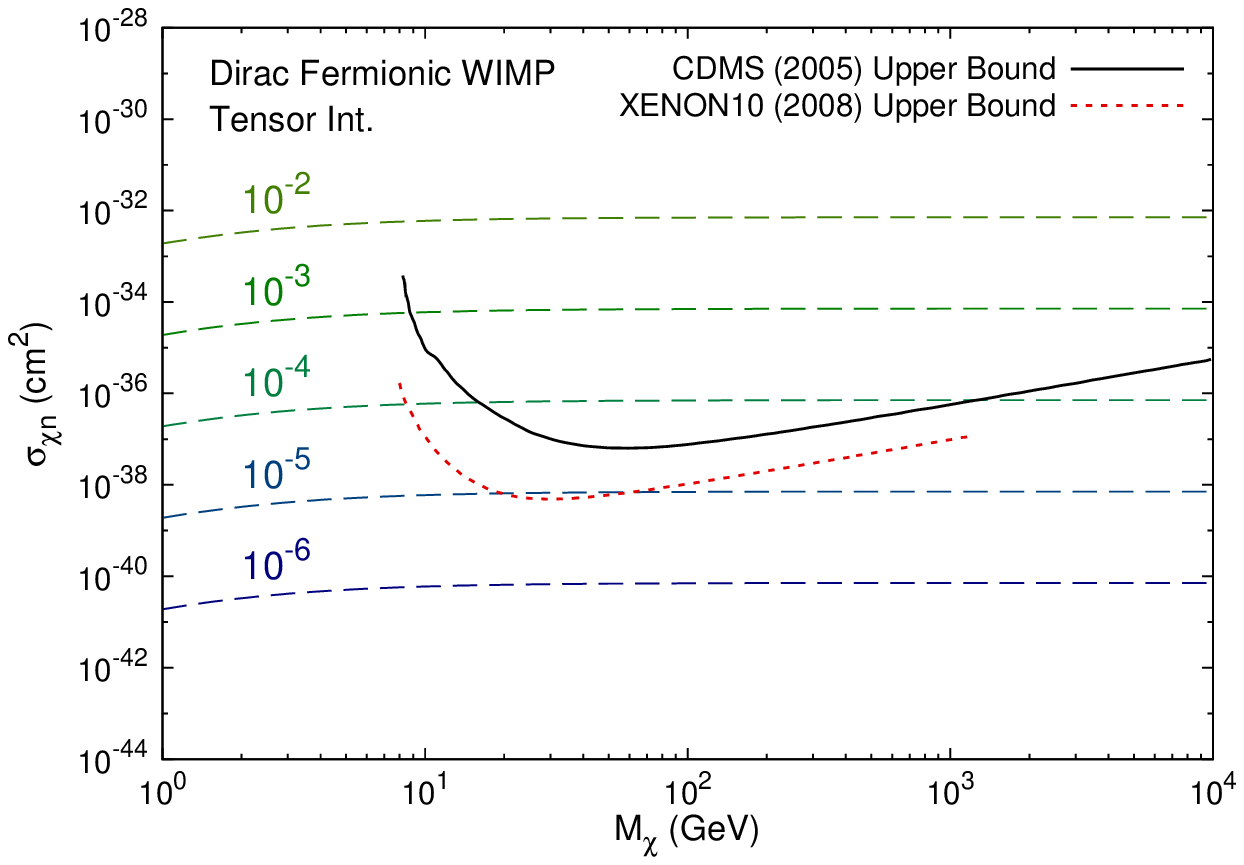}%
\hspace{0.008\textwidth}%
\includegraphics[width=0.49\textwidth]{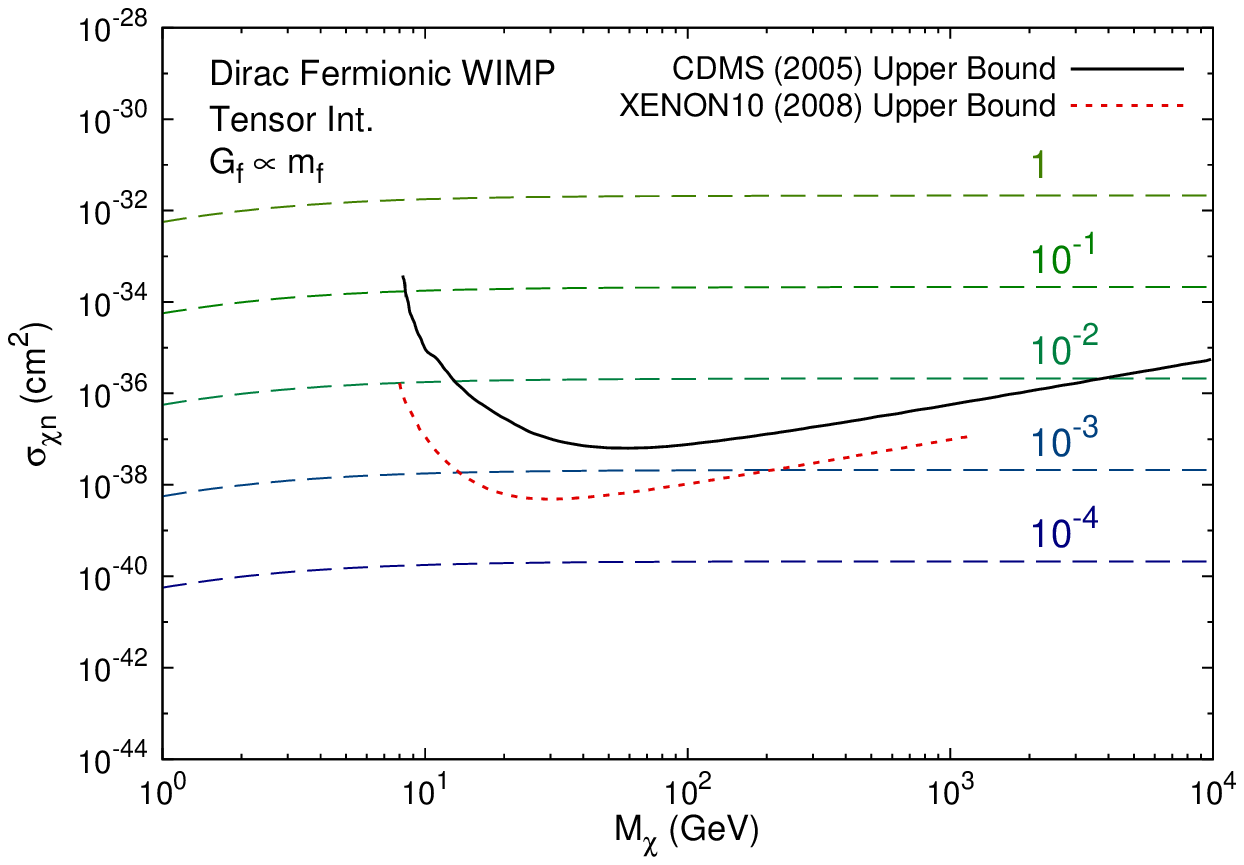}
\caption{The spin-dependent (SD) WIMP-neutron cross sections (dashed lines) for Dirac fermionic WIMPs with axialvector and tensor interactions. In the upper left and lower left frames, the results are shown for axialvector and tensor interactions for the case of universal couplings $G_f = 10^{-6}$, $10^{-5}$, $10^{-4}$, $10^{-3}$ and $10^{-2}~\mathrm{GeV}^{-2}$. In the remaining two frames, the results are shown for the case when the coupling constants proportion to the fermion mass $m_f$, $G_f \times (1~\mathrm{GeV} / m_f) = 10^{-4}$, $10^{-3}$, $10^{-2}$, $10^{-1}$ and $1~\mathrm{GeV}^{-2}$. Also shown as solid and dotted curves are the upper bounds from the experiments of CDMS (2005) \cite{Akerib:2005za} and XENON10 (2008) \cite{Angle:2008we}. \label{fig:dirac:scat:SD}}
\end{figure}

Eqs. \eqref{cross-sec-chi-N-S}, \eqref{cross-sec-chi-N-V} -- \eqref{cross-sec-chi-N-C} predict the normalized WIMP-nucleon cross sections $\sigma_{\chi N}$ for the different interactions as  functions of the WIMP mass $M_\chi$ and the effective couplings to nucleons $G_N$, which further depend on the more fundamental effective couplings to quarks $G_q$. To illustrate the results and compare with the experimental bounds, we draw the curves of $\sigma_{\chi N}$ vs. $M_\chi$ for fixed coupling constants in these effective models.

In Fig.~\ref{fig:dirac:scat:SI}, we show the SI cross sections for
a Dirac WIMP elastically scattering with a proton in the effective
models of scalar, vector and chiral interactions, respectively.
The results for chiral interactions are approximately identical to
those for vector interactions due to
Eqs.~\eqref{cross-sec-chi-N-V} and \eqref{cross-sec-chi-N-C}, so
they are shown in the same frame. We still consider the two types
of coupling constants. In the upper left and lower left frames of
Fig.~\ref{fig:dirac:scat:SI}, the results are shown for scalar and
vector interactions for the case of universal couplings $G_f =
10^{-9}$, $10^{-8}$, $10^{-7}$, $10^{-6}$ and
$10^{-5}~\mathrm{GeV}^{-2}$. In the remaining two frames, the
results are shown for the case when the coupling constants
proportional to the fermion mass $m_f$, $G_f \times
(1~\mathrm{GeV} / m_f) = 10^{-8}$, $10^{-7}$, $10^{-6}$, $10^{-5}$
and $10^{-4}~\mathrm{GeV}^{-2}$. We also show in
Fig.~\ref{fig:dirac:scat:SI} the current upper limits for the SI
WIMP-nucleon elastic scattering from the experiments of CDMS II
(2009) \cite{Ahmed:2009zw} and XENON100 (2010)
\cite{Aprile:2010um} for comparison.

In Fig.~\ref{fig:dirac:scat:SD}, we show the SD cross sections for
a  Dirac WIMP elastiscally scattering with a neutron in the
effective models of axialvector and tensor interactions,
respectively. In the upper left and lower left frames of
Fig.~\ref{fig:dirac:scat:SD}, the results are shown for
axialvector and tensor interactions for the case of universal
couplings $G_f = 10^{-6}$, $10^{-5}$, $10^{-4}$, $10^{-3}$ and
$10^{-2}~\mathrm{GeV}^{-2}$. In the remaining two frames, the
results are shown for the case when the coupling constants
proportion to the fermion mass $m_f$, $G_f \times (1~\mathrm{GeV}
/ m_f) = 10^{-4}$, $10^{-3}$, $10^{-2}$, $10^{-1}$ and
$1~\mathrm{GeV}^{-2}$. Also shown in Fig.~\ref{fig:dirac:scat:SD}
are the upper bounds for the SD WIMP-neutron elastic scattering
from the experiments of CDMS (2005) \cite{Akerib:2005za} and
XENON10 (2008) \cite{Angle:2008we}.

From Figs.~\ref{fig:dirac:scat:SI} and \ref{fig:dirac:scat:SD}, we
see that the experimental constraints for the SI interactions (i.e.,
scalar or vector interactions) are much more stringent than those
for the SD interactions (i.e., axialvector or tensor interactions).
In addition, the experimental constraints in the case of universal
couplings are more stringent than those in the case of
Yukawa-like couplings ($G_f\propto
m_f$) for both the SI and SD interactions. Due to the factor $4$
difference in the WIMP-nucleon cross sections between us and that in
\cite{Beltran:2008xg} our predicted WIMP-proton SI cross sections
(dashed lines) in Figs.~\ref{fig:dirac:scat:SI} are lower than them.

We summarize that the scalar, vector and chiral interactions are
mainly constrained by the experimental SI upper limits of
WIMP-nucleus elastic scattering, while the axialvector and tensor
interactions are constrained by the SD upper limits of
WIMP-nucleus elastic scattering. However, the other 6 types of
interactions, $\mathcal{L}_{\mathrm{P}}$,
$\mathcal{L}_{\mathrm{SP}}$, $\mathcal{L}_{\mathrm{PS}}$,
$\mathcal{L}_{\mathrm{VA}}$, $\mathcal{L}_{\mathrm{AV}}$ and
$\mathcal{L}_{\tilde{\mathrm{T}}}$, cannot be constrained by
direct detection experiments.

\section{Indirect Detection\label{sec-indirect}}

In addition to the direct search method for the WIMP DM at
underground laboratories, an indirect detection method is used to
look for the DM annihilation or decay products which include
neutrinos, gamma rays, positrons and antiprotons. These particles
can be detected by  cosmic ray experiments. For the charged
particles they are deflected by the Galactic magnetic field and
interact with the interstellar medium when they propagate in the
Galaxy. Therefore we have to study the propagation process of the
charged particles to compare predictions with observations.

The propagation of cosmic rays in the Galaxy can be described by
\cite{Strong:1998pw}
\begin{eqnarray}
\frac{\partial \psi}{\partial t}= Q(\vec{r},p)
+\nabla\cdot(D_{xx}\nabla\psi-\vec{V}_c\psi) +\frac{\partial}{\partial p}\left[
{{p^2}{D_{pp}}{\partial\over{\partial p}}\bigg({{1\over{{p^2}}}\psi}\bigg)}\right]
-{\partial\over{\partial p}}\left[{\dot p\psi  - {p \over 3}( {\nabla  \cdot
{\vec{V}}_c})\psi } \right] - {1 \over {{\tau _f}}}\psi  - {1
\over {{\tau _r}}}\psi \label{propeq}
\end{eqnarray}
where $\psi = \psi ( {\vec{r}},p,t) $ is the number density of
cosmic-ray particles per unit momentum interval, $Q(\vec{r},p)$ is
the source term, $D_{xx}$ is the spatial diffusion coefficient,
$\vec{V}_c$ is the convection velocity, $D_{pp}$ is the diffusion
coefficient in momentum space describing the reacceleration
process, $\dot p \equiv dp/dt$ is the momentum loss rate, $\tau_f$
and $\tau_r$ are the time scales for fragmentation and the
radioactive decay, respectively. The most accurate method to treat
the propagation is to solve Eq. \eqref{propeq} by a numerical code,
GALPROP \cite{Strong:1998pw,Strong:1999sv}. For given source
functions and boundary conditions, GALPROP can solve the equations
for various cosmic-ray species and give reasonable fit to many
cosmic ray data.

Antiprotons are rarely produced in usual astrophysical processes,
and the observed antiproton-to-proton flux ratio $\bar{p}/p$ is small, typically $\mathcal{O}(10^{-5})\sim\mathcal{O}(10^{-4})$,
from 100 MeV to 100 GeV in kinetic energy. The annihilation of
WIMPs, however, produces protons and antiprotons in equal numbers.
Thus the $\bar p / p$ spectrum may be sensitive to the annihilation
of WIMPs. In this work we use the antiproton-to-proton flux ratio $\bar p /
p$ measured by the satellite-borne experiment PAMELA
\cite{Adriani:2010rc} to constrain the effective models.

The source term of antiprotons contributed by the annihilation of
Dirac fermionic WIMPs is given by
\begin{equation}
Q_{\mathrm{ann}}\left( {{\mathbf{x}},E} \right) =
\frac{{\left\langle {\sigma_{\mathrm{ann}} v}
\right\rangle_\text{tot} }} {{4M_\chi
^2}}\left[\sum\limits_q{B_q\left(\frac{{dN_{\bar p}}} {{dE_{\bar
p}}}\right)_q}\right]{\rho ^2}( {\mathbf{x}} ), \label{dm_anni_pbar}
\end{equation}
with
\begin{equation}
{\left\langle {{\sigma _{{\mathrm{ann}}}}v} \right\rangle
_{{\text{tot}}}} = \sum\limits_{q} {{{\left\langle {{\sigma
_{{\mathrm{ann}}}}v} \right\rangle }_q}} ,\quad \text{and}\quad
{B_q} = \frac{{{{\left\langle {{\sigma _{{\mathrm{ann}}}}v}
\right\rangle }_q}}} {{{{\left\langle {{\sigma _{{\mathrm{ann}}}}v}
\right\rangle }_{{\text{tot}}}}}},\quad q = u,d,s,c,b,t,
\label{bran_ratio}
\end{equation}
where $\rho \left(\mathbf{x}\right)$ is the DM mass density
distribution of the Galaxy, $\left({dN_{\bar p}}/{dE_{\bar
p}}\right)_q$ is the number per unit energy interval of the
antiprotons produced by the annihilation of a pair of WIMPs in the
$q \bar q$ channel, and $B_q$ is the branching ratio of the $q
\bar q$ channel. The source term of protons is similar to
Eq.\eqref{dm_anni_pbar}. We use the Monte Carlo program PYTHIA
\cite{Sjostrand:2006za} to simulate the particle production of the
WIMP annihilation processes, and pick out the events in which the
final states are $\bar p$ to build the $\left({dN_{\bar
p}}/{dE_{\bar p}}\right)_q$ spectrum in each quark-antiquark
channel. Due to the different quark masses, the $\left({dN_{\bar
p}}/{dE_{\bar p}}\right)_q$ spectra in the six $q \bar q$ channels
are slightly different.

The NFW profile \cite{Navarro:1996gj} is taken to describe the DM
mass density distribution of the Galaxy:
\begin{equation}
\rho \left( r \right) = \dfrac{{{\rho _s}}} {{\left( {r/{r_s}}
\right){{\left( {1 + r/{r_s}} \right)}^2}}},
\end{equation}
where $\rho_s$ is the characteristic density, and $r_s $ is the
scale radius. We choose $\rho_s =
0.334~\mathrm{GeV}/\mathrm{cm}^3$ and $r_s = 20~\mathrm{kpc}$.
These values guarantee the local DM density $\rho \left(
8.33~\mathrm{kpc} \right) = 0.4~\mathrm{GeV}/\mathrm{cm}^3$, which
is consistent with the recent results, such as
\cite{Catena:2009mf} and \cite{Salucci:2010qr}. The CDM particles
in the Galaxy should obey the Maxwell-Boltzmann velocity
distribution $f\left( {{v_0}} \right) = {\left( {M_\chi/{2\pi k_B
T}} \right)^{3/2}}\exp \left( { - {M_\chi v_0^2}/{2k_B T}}
\right)$. Their velocity dispersion $\bar v \equiv \sqrt
{\left\langle {v_0^2} \right\rangle }$ is chosen to be the
canonical value $270 ~ {\text{km/s}}$ \cite{Jungman:1995df}. Since
the DM particles in the Galaxy nowadays are extremely nonrelativistic, $v_\mathrm{M{\o}l}$ almost equals
$v_\mathrm{rel}$ by neglecting the $\mathcal{O}(v^2/c^2)$ term, and we need not to distinguish them. And $v_\mathrm{rel}$ has the same value in
different frames with the extremely nonrelativistic limit. Thus, as a very good approximation, we expand $\left\langle\sigma_\mathrm{ann}v
\right\rangle$ to be $\left\langle\sigma_\mathrm{ann}v \right\rangle \simeq a + b\left\langle v^2\right\rangle$ with the coefficients $a$ and $b$ obtained in Eqs.~\eqref{sigma_v-S}--\eqref{sigma_v-Ch}. The relation between $\left\langle v^2\right\rangle$ and $\bar v$ is given by
$\left\langle {{v^2}} \right\rangle = \left\langle
{{\mathbf{v}}_1^2 - 2{{\mathbf{v}}_1} \cdot {{\mathbf{v}}_2} +
{\mathbf{v}}_2^2} \right\rangle = 2\left\langle {v_0^2}
\right\rangle  = 2{\bar v^2}$.

\begin{figure}[!htbp]
\centering
\includegraphics[width=0.49\textwidth]{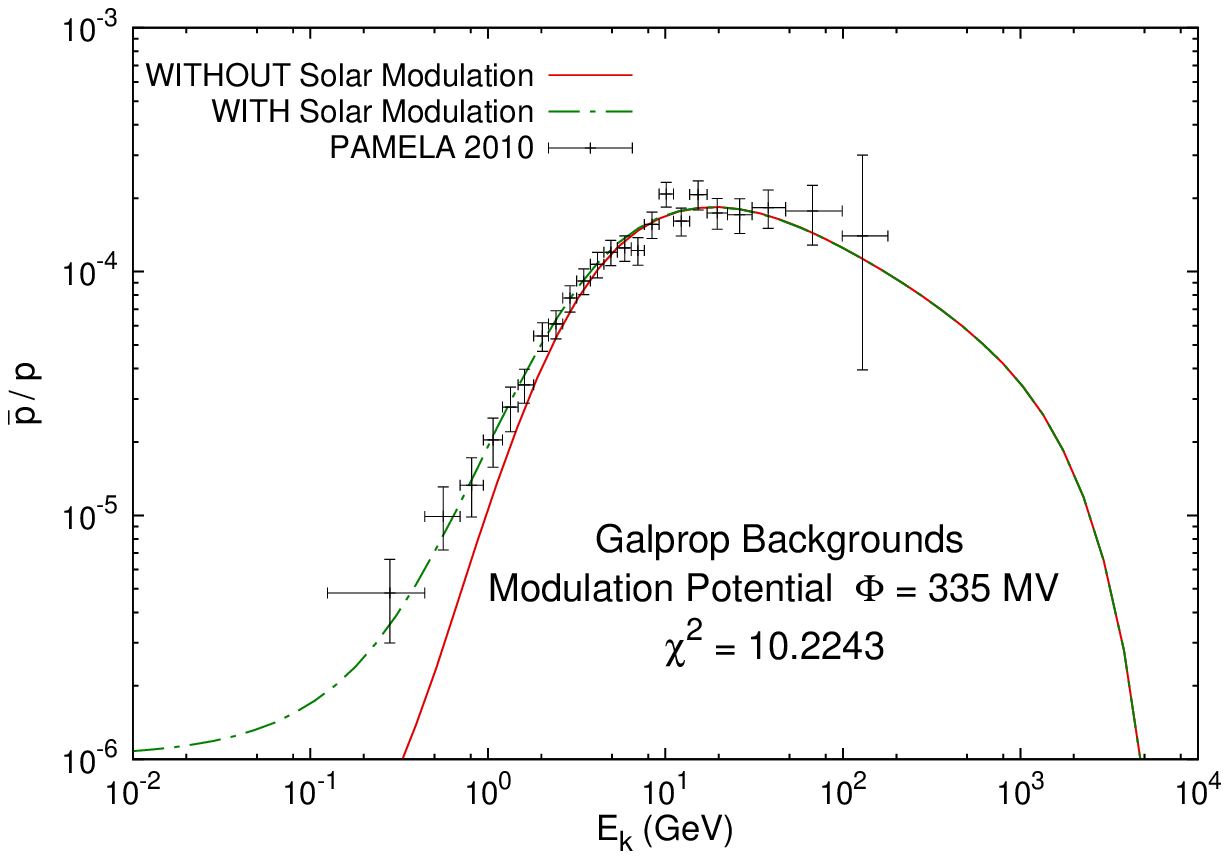}%
\hspace{0.008\textwidth}
\includegraphics[width=0.49\textwidth]{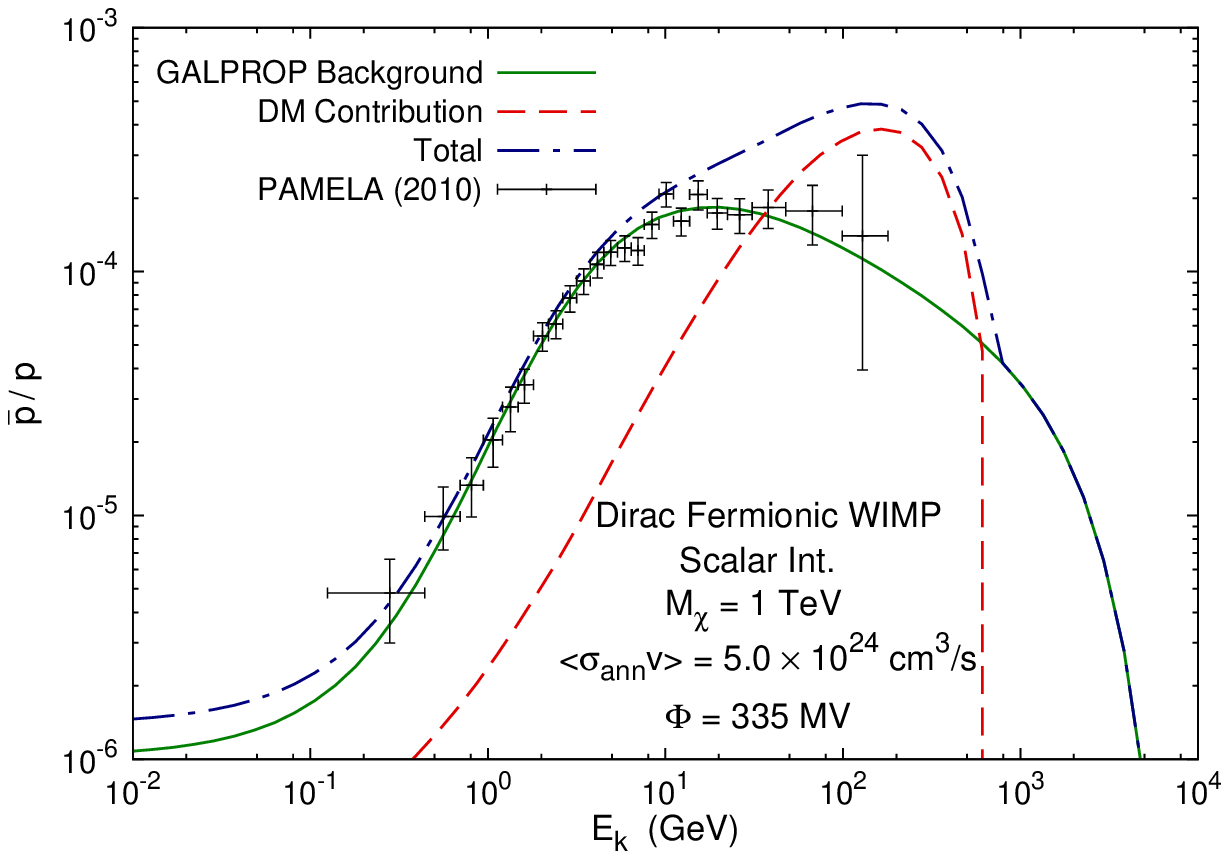}
\caption{The $\bar p / p$ spectrum calculated by GALPROP and that
measured by PAMELA \cite{Adriani:2010rc} (error bars). In the left
frame, we show the GALPROP predicted $\bar p / p$ spectrum without the DM
contribution, where the Solar modulation potential $\Phi$ is taken to be 335 MV to fit the PAMELA data. In the right frame, we show the predicted total $\bar p / p$ spectrum  (GALPROP background + DM
contribution) in the effective model of scalar
interactions as an example. $M_\chi = 1~\mathrm{TeV}$ and $\left\langle {{\sigma
_{{\mathrm{ann}}}}v} \right\rangle _\mathrm{tot} = 5.0 \times
10^{24} ~\mathrm{cm}^3/\mathrm{s} $ are set here.} \label{fig:pbarp}
\end{figure}
\begin{figure}[!htbp]
\centering
\includegraphics[width=0.88\textwidth]{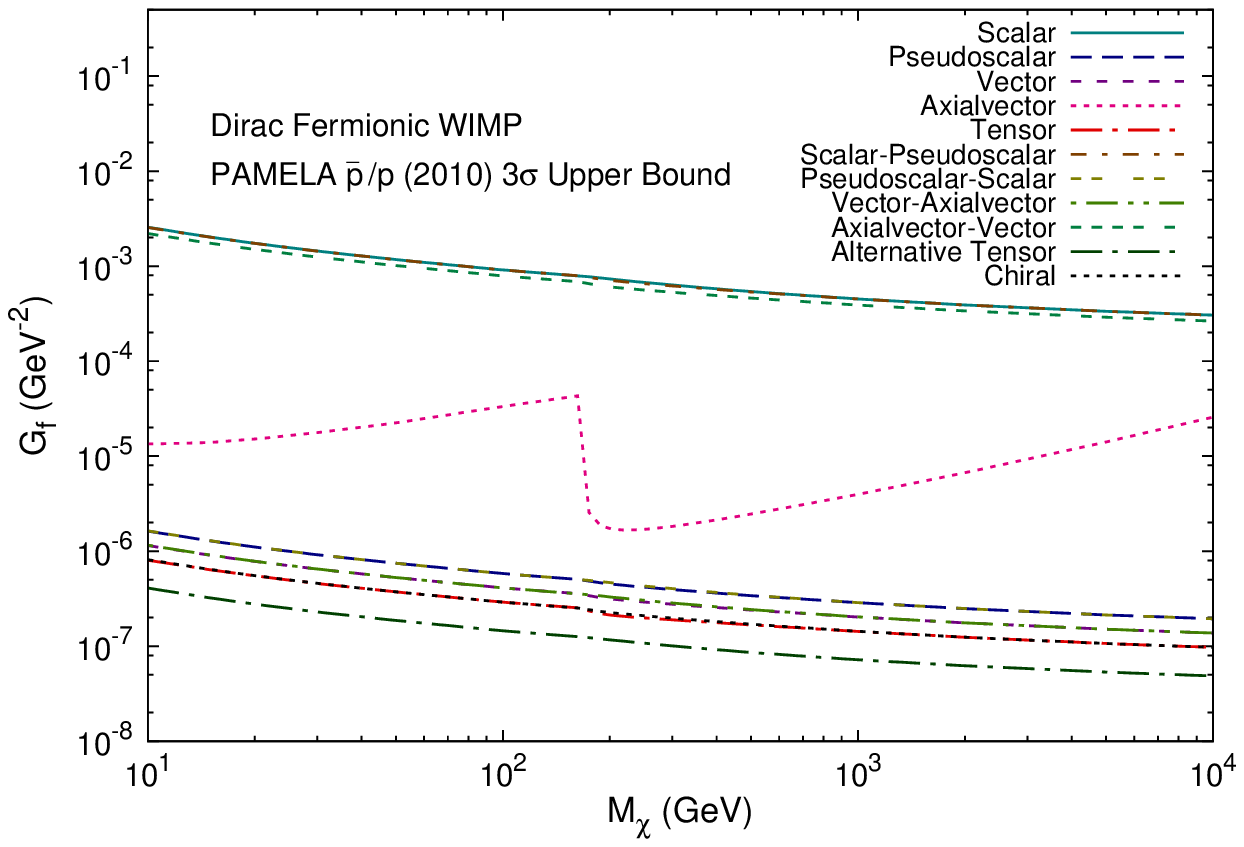}%
\\
\includegraphics[width=0.88\textwidth]{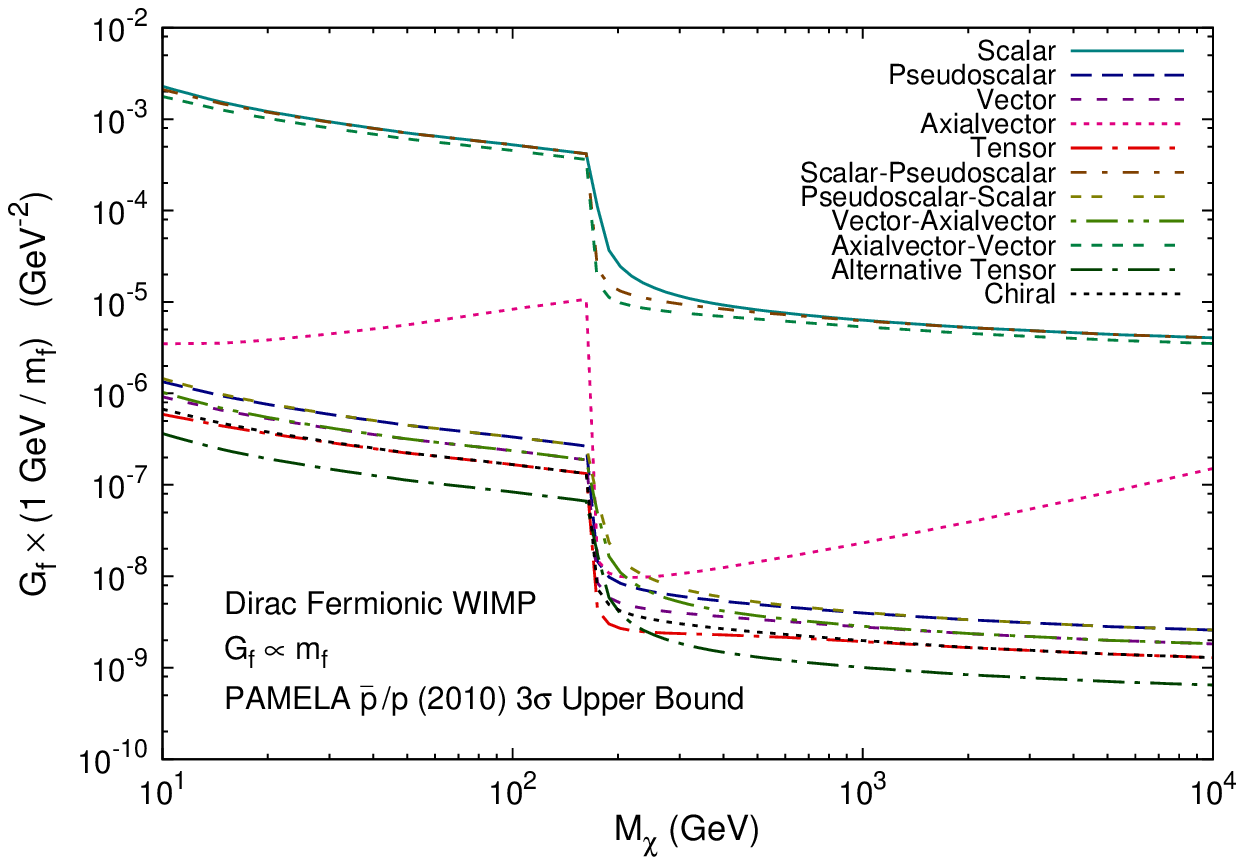}%
\caption{The $3\sigma$ upper bounds on the coupling constants $G_f$ from the PAMELA $\bar p / p$ spectrum \cite{Adriani:2010rc} in each effective model
of 4-fermion interaction operators. In the upper frame, results are given for the case when the effective couplings to all the standard model fermions are equal (universal couplings). In the lower frame, results are shown for the case when the coupling constants are proportional to the fermion mass $m_f$.  \label{fig-antiproton}}
\end{figure}

In the calculation of the $\bar p / p$ spectrum with GALPROP, we
adopt the Galaxy propagation model with diffusion and convection,
and set the half-height of the Galaxy propagation halo to be $z_h =
4~\mathrm{kpc}$. Although the GALPROP expected $\bar p / p$ spectrum
without the DM contribution can fit the PAMELA result
\cite{Adriani:2010rc} well, as shown in the left frame of
Fig.~\ref{fig:pbarp}, the present data cannot rule out the
possibility that the $\bar p / p$ spectrum might receive a small
portion of contribution from the DM annihilation. If we add the DM
contribution to the $\bar p / p$ spectrum, however, the total
spectrum (GALPROP background + DM contribution) may deviate from the
PAMELA result, as shown in the right frame of Fig.~\ref{fig:pbarp}.
For the universal couplings and the
Yukawa-like couplings ($G_f\propto
m_f$) in our numerical calculation, the DM contribution is actually
determined by two parameters, $\left\langle {{\sigma
_{{\mathrm{ann}}}}v} \right\rangle _\mathrm{tot}$ and $M_\chi$. For
fixed $M_\chi$, the smaller $\left\langle {{\sigma
_{{\mathrm{ann}}}}v} \right\rangle _\mathrm{tot}$ is, the smaller
the DM contribution will be. We must let $\left\langle {{\sigma
_{{\mathrm{ann}}}}v} \right\rangle _\mathrm{tot}$ be small enough to
keep the total $\bar p / p$ spectrum within an acceptable deviation
range of the PAMELA result when we take into account the DM
contribution.

Now we calculate the $\chi ^2$ value to set upper bound on the DM
coupling constants from the $\bar{p}/p$ data. Since $\left\langle
{{\sigma _{{\mathrm{ann}}}}v} \right\rangle _\mathrm{tot}$
monotonously depends on the coupling constants $G_f$, we can derive
the 3$\sigma$ upper bounds on $G_f$ for fixed $M_\chi$ in the cases
of universal couplings and
Yukawa-like couplings ($G_f\propto
m_f$) for each type of effective interactions, as shown in
Fig.~\ref{fig-antiproton}. Here the solar modulation potential is
set to be $\Phi = 335~\mathrm{MV}$, which gives minimal $\chi ^2$
for the background only.

We see that several pairs of upper bound curves in
Fig.~\ref{fig-antiproton} are nearly identical, which is similar to
the situation of Fig.~\ref{fig:dirac:rd_coupling}. The reason is the
same as explain before. As pointed out in Sec.~\ref{sec-relic}, the
quantity $\sigma_{\mathrm{ann}} v$ in each pair differ only by terms
of $\mathcal{O}(v^2)$ and/or terms of $m_f^2/M_\chi^2$. Using the
abbreviated notation defined in Sec.~\ref{sec-relic}, we may call
the pairs of nearly identical upper bound curves by S$\;\simeq\;$SP,
P$\;\simeq\;$PS, V$\;\simeq\;$VA, and T$\;\simeq\;$C. These 4 pairs
are the same as those in Fig.~\ref{fig:dirac:rd_coupling} in
Sec.~\ref{sec-relic}. The exception in this situation is that the
upper bound curves for axialvector (A) and axialvector-vector (AV)
interactions are rather different. It may comes from the fact that
$\sigma_{AV,\,\mathrm{ann}}v$ is of order $\mathcal{O}(v^2)$, while
$\sigma_{A,\,\mathrm{ann}}v$ is of order $\mathcal{O}(v^0)$ and
proportional to $m_f^2$. The difference of the curves for A and AV
is somewhat like the difference of the cases of universal couplings
and Yukawa-like couplings
($G_f\propto m_f$) for the same effective interaction. In addition,
we note that in Fig.~\ref{fig-antiproton} the upper bound curves for
scalar (S), scalar-pseudoscalar (SP) and axialvector-vector (AV)
interactions lie well above the other curves, for their
$\sigma_{\mathrm{ann}} v$ are of order $\mathcal{O}(v^2)$. And we
note that there are downward bends in the upper bound curves at
about $M_\chi\sim m_t=171.2$~GeV account for the $\chi~
\bar{\chi}\to t~\bar{t}$ threshold effect.

\section{Validity region of effective models and combined constraints\label{sec-combine}}

In this section, let us discuss the validity region where the method
of effective theory can be used. For a generic 4-fermion interaction
operator
$\frac{G_f}{\sqrt{2}}\,\bar{\chi}\,\Gamma_1\chi\bar{f}\,\Gamma_2 f$,
the mass dimension of the coupling $G_f$ is $-2$. Since we have used
the two types of coupling constants, i.e., the universal couplings
and the Yukawa-like couplings
($G_f\propto m_f$), in the numerical calculation throughout the last
three sections, let us consider them case by case:
\begin{itemize}
  \item For the universal couplings, we can write the coupling as
  $\frac{G_f}{\sqrt{2}}=\frac{\alpha}{\Lambda^2}$, where $\Lambda$ is the cutoff energy scale and $\alpha$ is the coupling of the fundamental theory beyond
  $\Lambda$, which may be of order 1. The transfer momentum of the annihilation process $\chi~ \bar{\chi}\to f~\bar{f}$ must be well below the cutoff,
  that is, $2M_\chi\ll \Lambda$, so that the effective theory can be used. On the other hand, a weakly coupled UV completion of the effective theory usually
  requires $\alpha < 4\pi$ such that the perturbation can be adopted \cite{Goodman:2010ku,Bi:2009am}. From the above 3 relations, we obtain
\begin{equation}
G_f\ll \frac{\sqrt{2}\pi}{M_\chi^2} \label{constraint-effective-theory-Univ}
\end{equation}
  \item For the Yukawa-like couplings ($G_f\propto m_f$), we have $\frac{G_f}{\sqrt{2}}=\frac{\alpha\, m_f}{\Lambda^3}$, where $\Lambda$ is a cutoff and $\alpha$ is of order 1. Likewise, we still have $2M_\chi\ll \Lambda$ and $\alpha < 4\pi$. From the above 3 relations, we obtain
\begin{equation}
\frac{G_f}{m_f}\ll \frac{\pi}{\sqrt{2}M_\chi^3}\;\;
.\label{constraint-effective-theory-Yukawa}
\end{equation}
\end{itemize}

\begin{figure}[!htbp]
\centering
\includegraphics[width=0.49\textwidth]{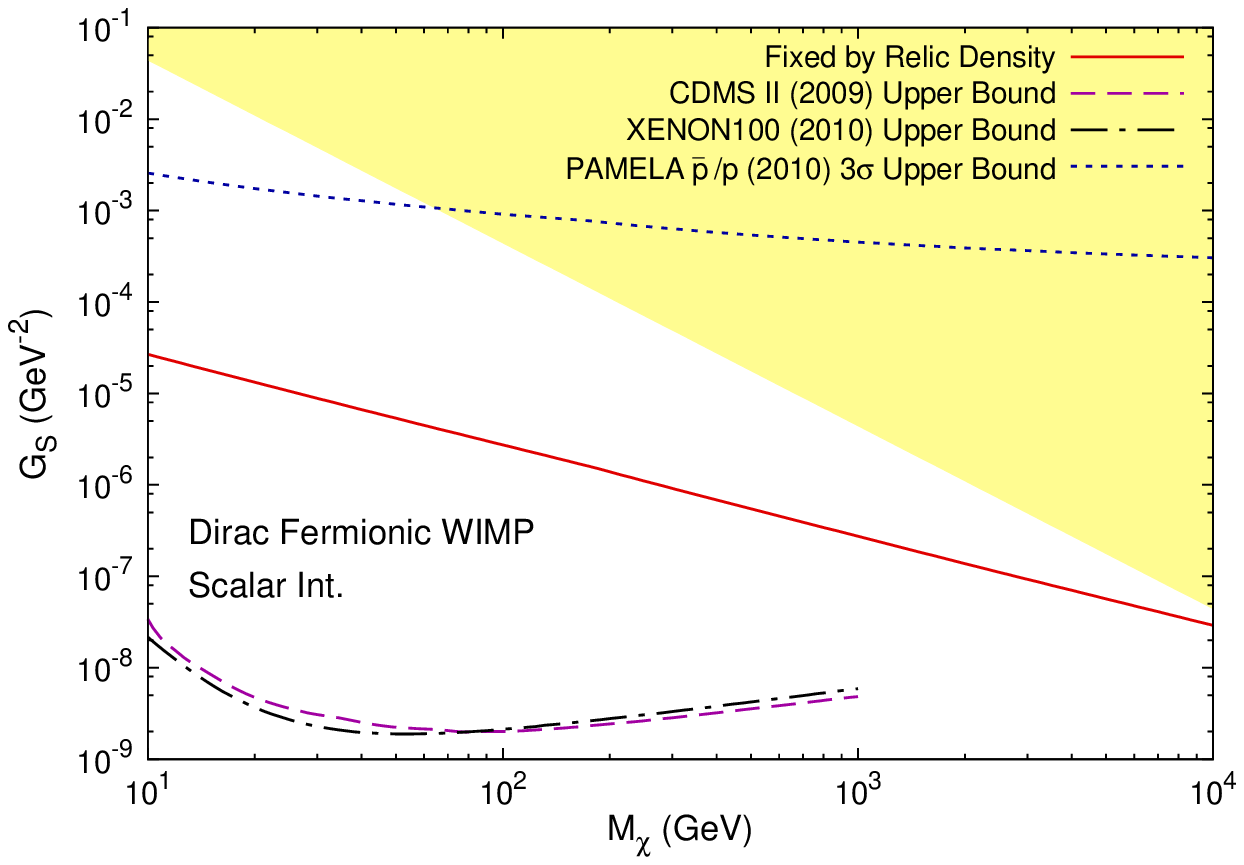}%
\hspace{0.008\textwidth}
\includegraphics[width=0.49\textwidth]{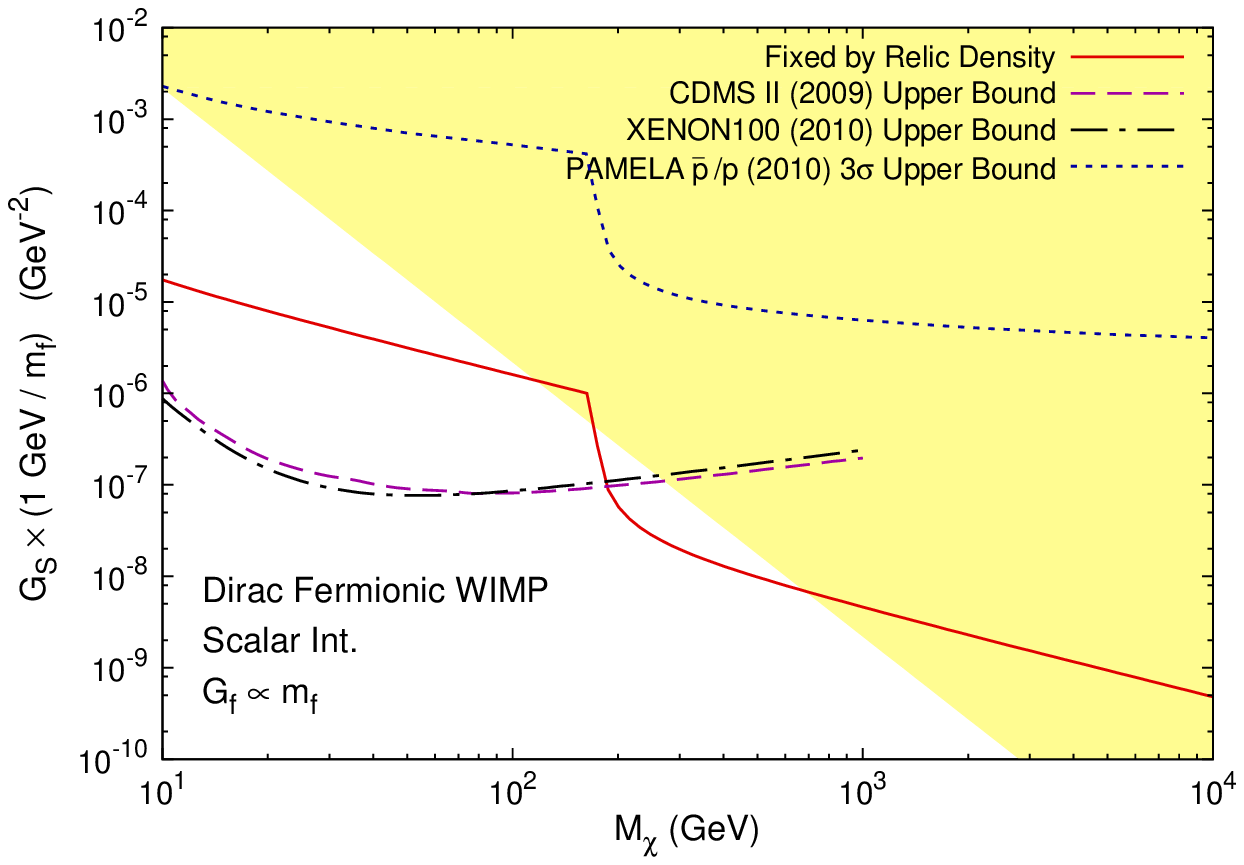}
\caption{Combined constraints on coupling constants $G_f$ of Dirac
fermionic WIMPs with scalar (S) interactions from relic density,
direct detection experiments of CDMS II and XENON100, PAMELA $\bar p
/ p$ ratio, and validity of effective theory. The yellow region
denotes invalid parameter space of effective field theory. The left
frame is shown for the case of universal couplings, while the right
frame for the case of Yukawa-like
couplings  ($G_f\propto m_f$). The constraints for
scalar-pseudoscalar (SP) interactions from relic density, PAMELA
$\bar p / p$ ratio, and validity of effective theory are nearly the
same as above, but SP interactions are insensitive to direct
detection experiments. \label{fig-combined-S}}
\end{figure}

\begin{figure}[!htbp]
\centering
\includegraphics[width=0.49\textwidth]{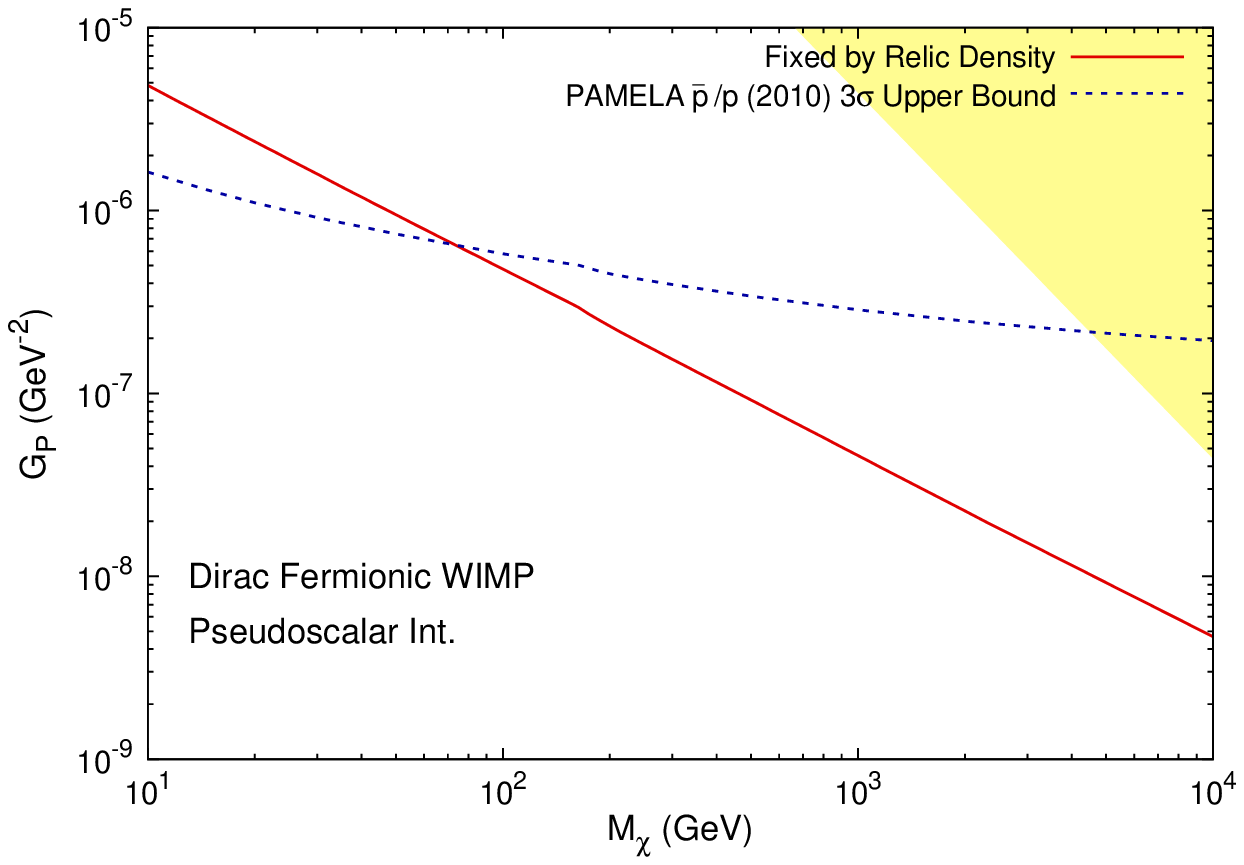}%
\hspace{0.008\textwidth}
\includegraphics[width=0.49\textwidth]{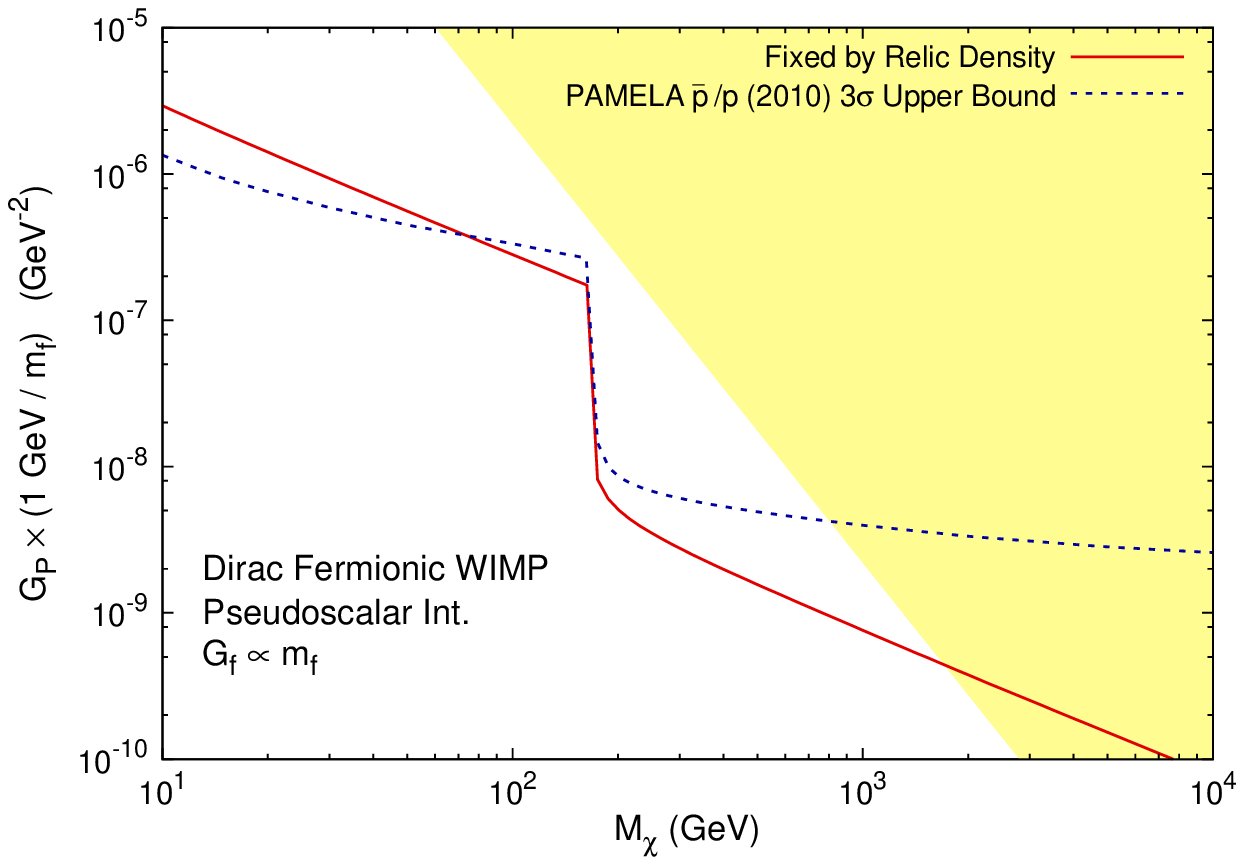}
\caption{Combined constraints on coupling constants $G_f$ of Dirac fermionic WIMPs with pseudoscalar (P) interactions from relic density, PAMELA $\bar p / p$ ratio, and validity of effective theory. The constraints for pseudoscalar-scalar (PS) interactions from relic density, PAMELA $\bar p / p$ ratio, and validity of effective theory are nearly the same as above. \label{fig-combined-P}}
\end{figure}

\begin{figure}[!htbp]
\centering
\includegraphics[width=0.49\textwidth]{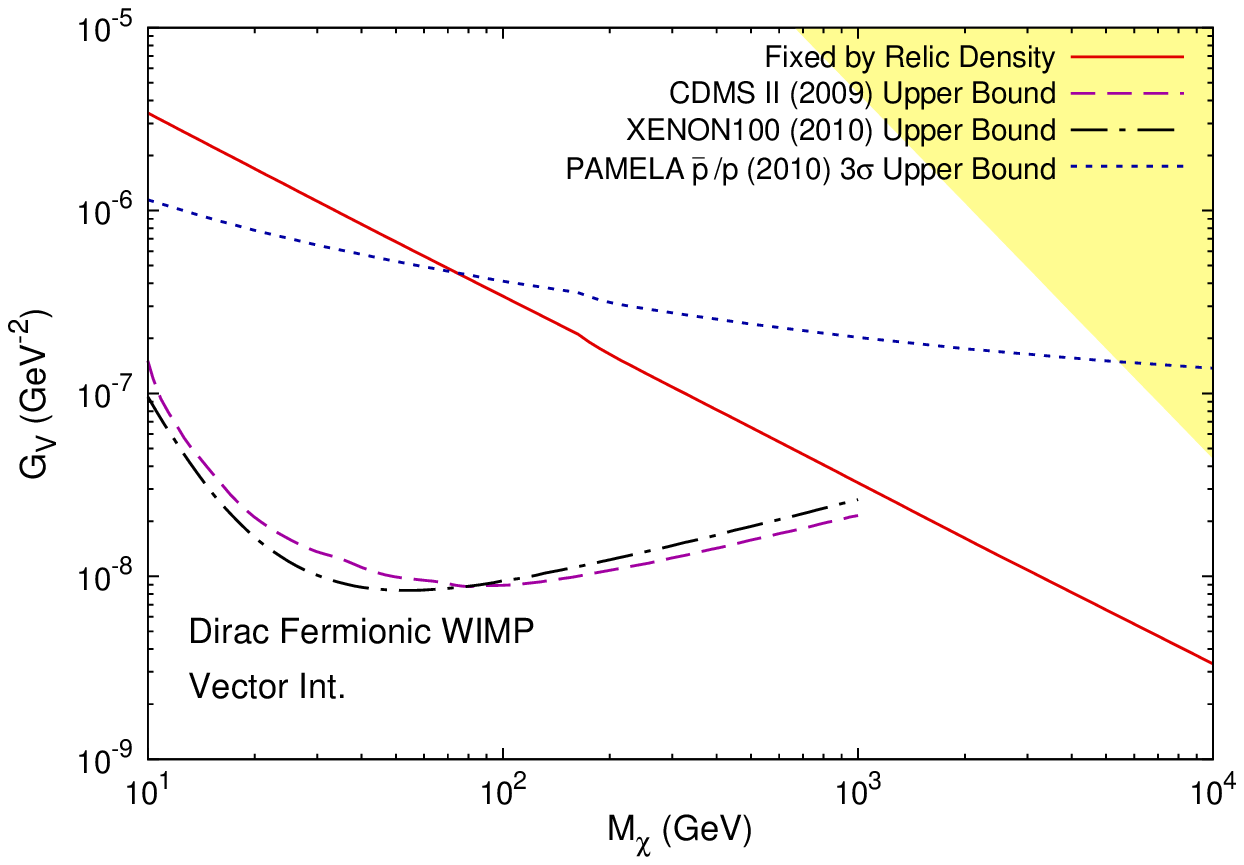}%
\hspace{0.008\textwidth}
\includegraphics[width=0.49\textwidth]{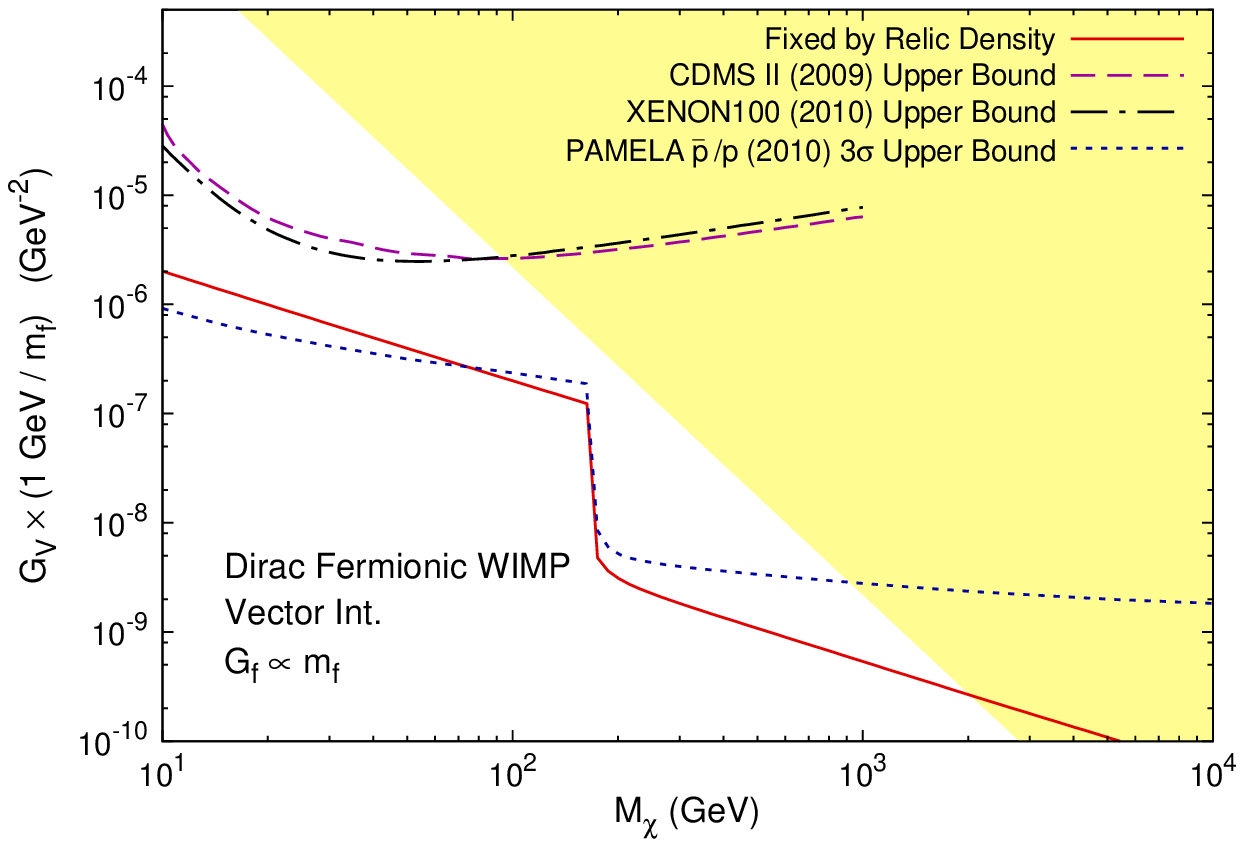}
\caption{Combined constraints on coupling constants $G_f$ of Dirac fermionic WIMPs with vector (V) interactions from relic density, direct detection experiments of CDMS II and XENON100, PAMELA $\bar p / p$ ratio, and validity of effective theory. The constraints for vector-axialvector (VA) interactions from relic density, PAMELA $\bar p / p$ ratio, and validity of effective theory are nearly the same as above, but VA interactions are insensitive to direct detection experiments. \label{fig-combined-V}}
\end{figure}

\begin{figure}[!htbp]
\centering
\includegraphics[width=0.49\textwidth]{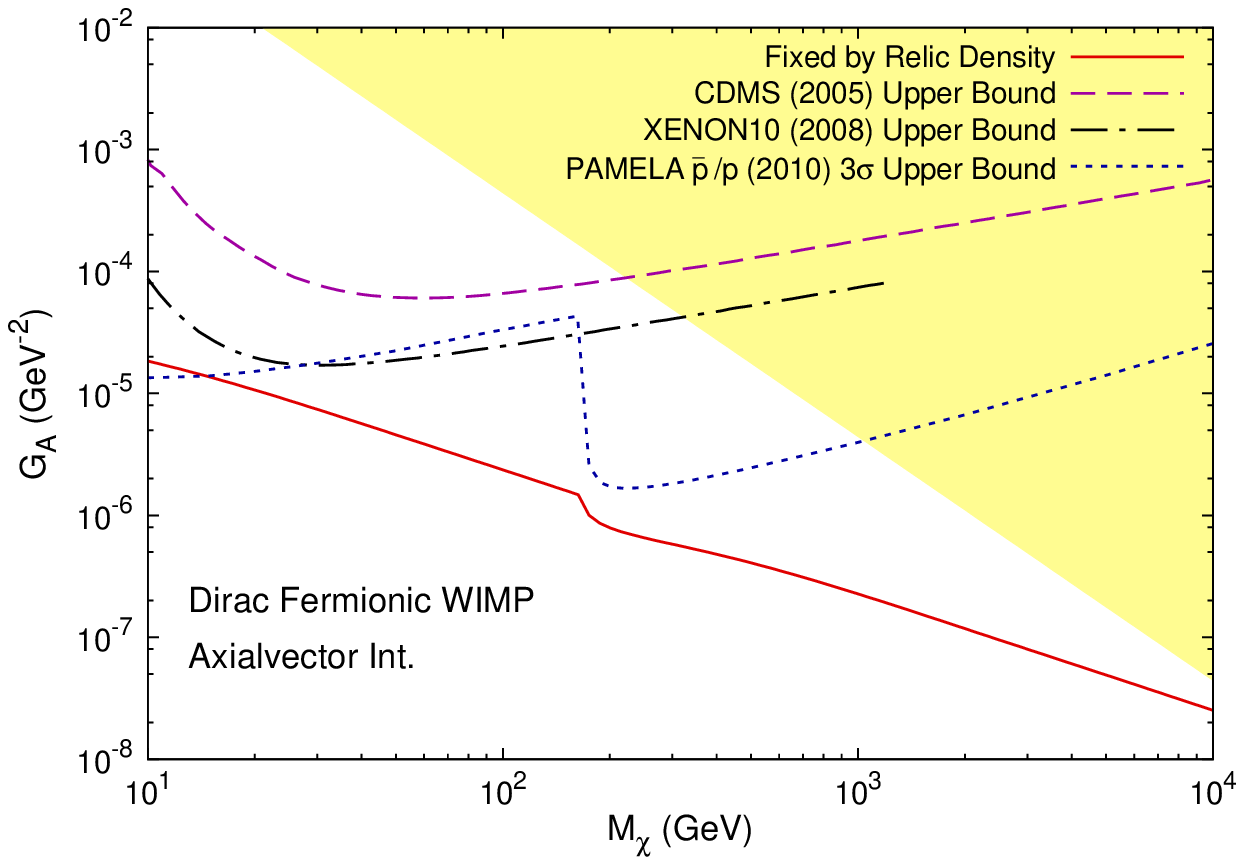}%
\hspace{0.008\textwidth}
\includegraphics[width=0.49\textwidth]{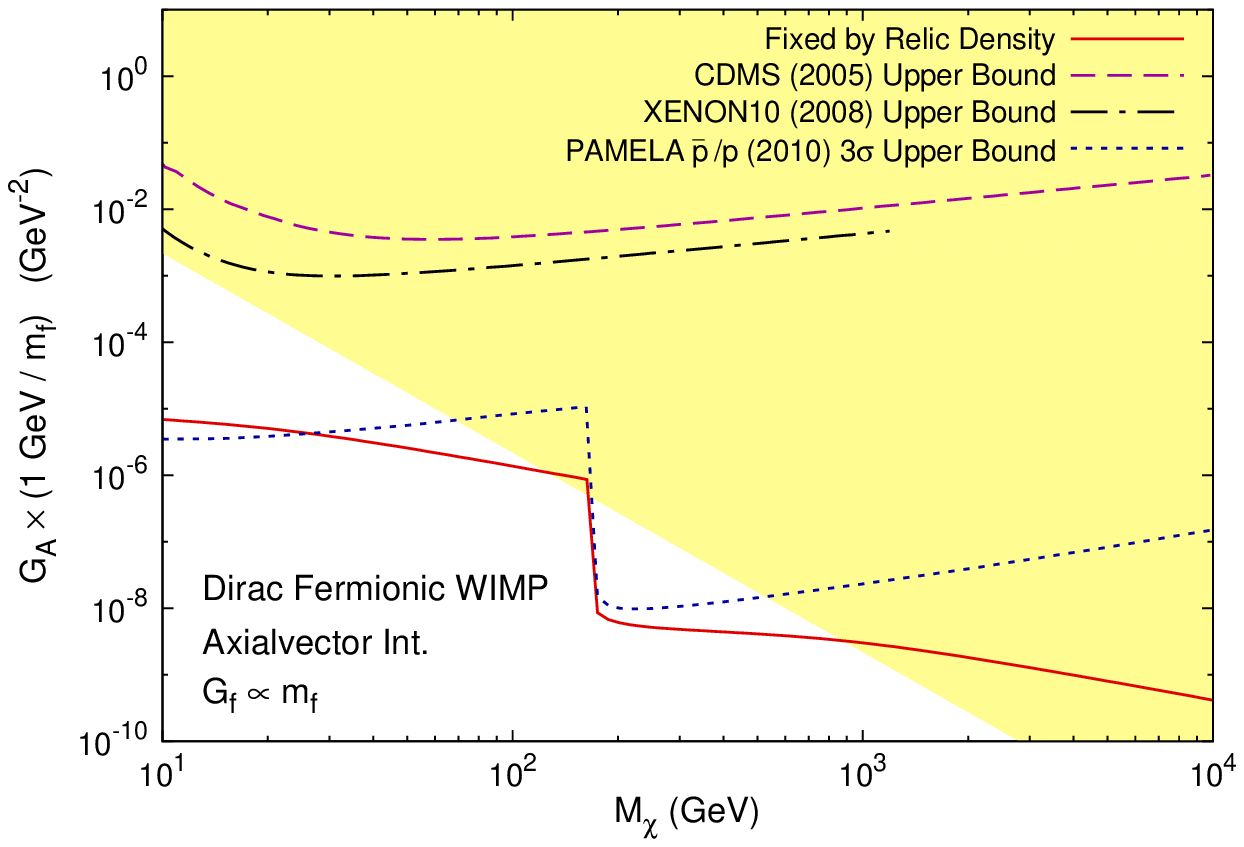}
\caption{Combined constraints on coupling constants $G_f$ of Dirac fermionic WIMPs with axialvector (A) interactions from relic density, direct detection experiments of CDMS and XENON10, PAMELA $\bar p / p$ ratio, and validity of effective theory. \label{fig-combined-A}}
\end{figure}

\begin{figure}[!htbp]
\centering
\includegraphics[width=0.49\textwidth]{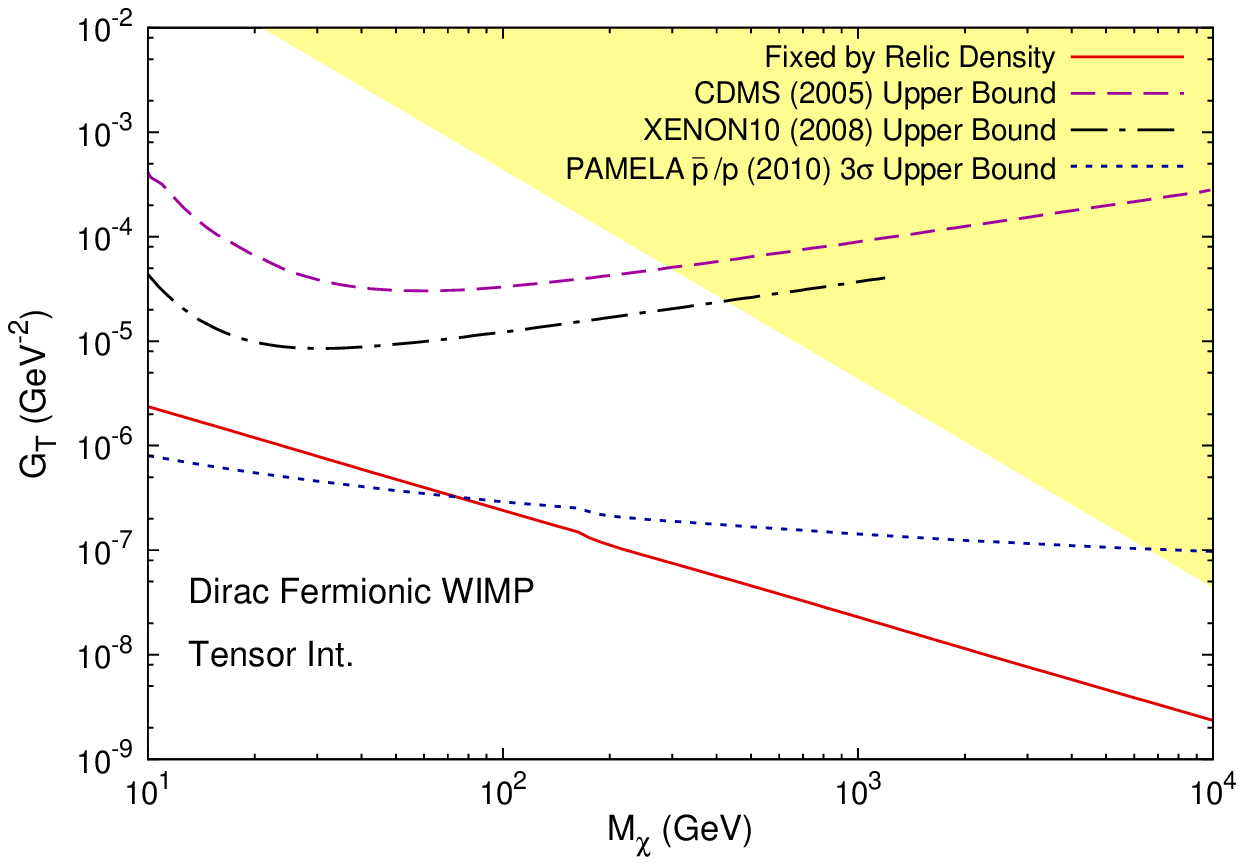}%
\hspace{0.008\textwidth}
\includegraphics[width=0.49\textwidth]{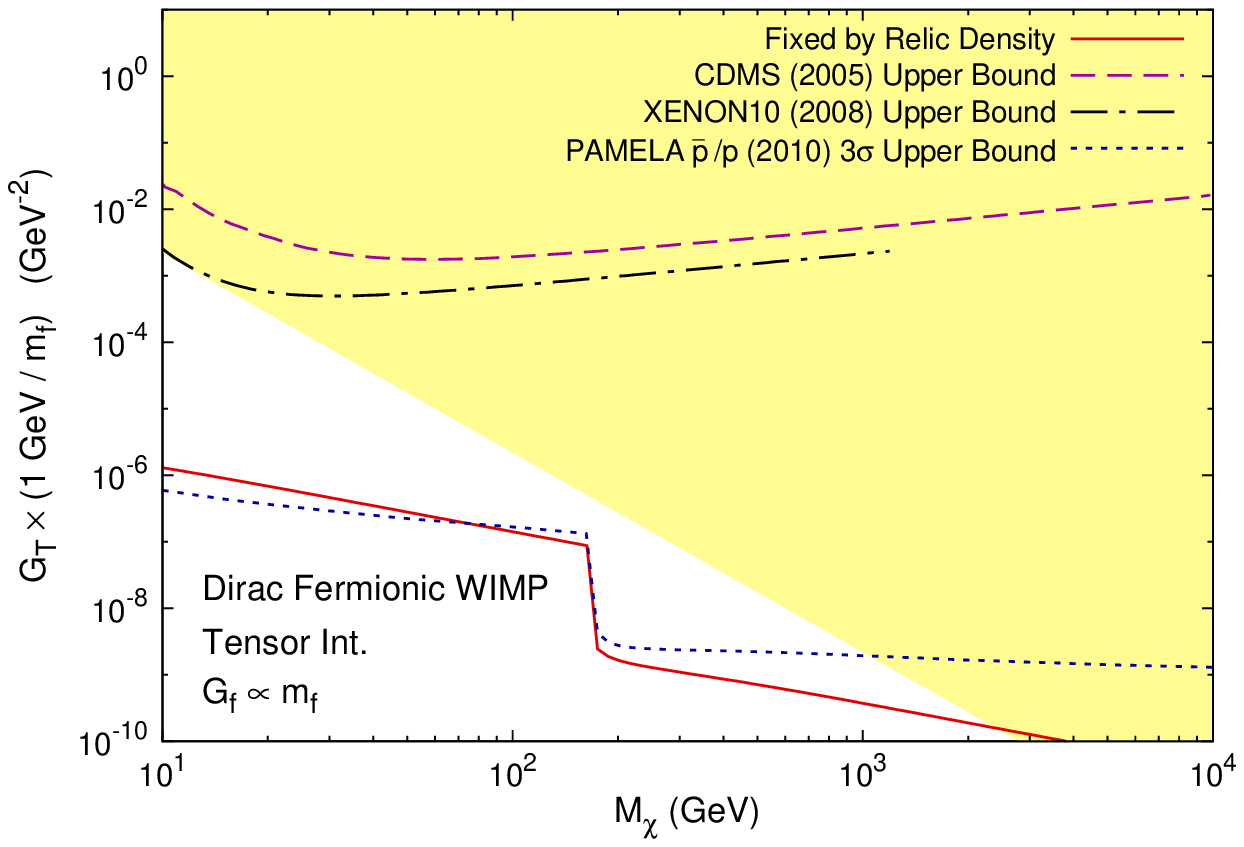}
\caption{Combined constraints on coupling constants $G_f$ of Dirac fermionic WIMPs with tensor (T) interactions from relic density, direct detection experiments of CDMS and XENON10, PAMELA $\bar p / p$ ratio, and validity of effective theory. \label{fig-combined-T}}
\end{figure}

\begin{figure}[!htbp]
\centering
\includegraphics[width=0.49\textwidth]{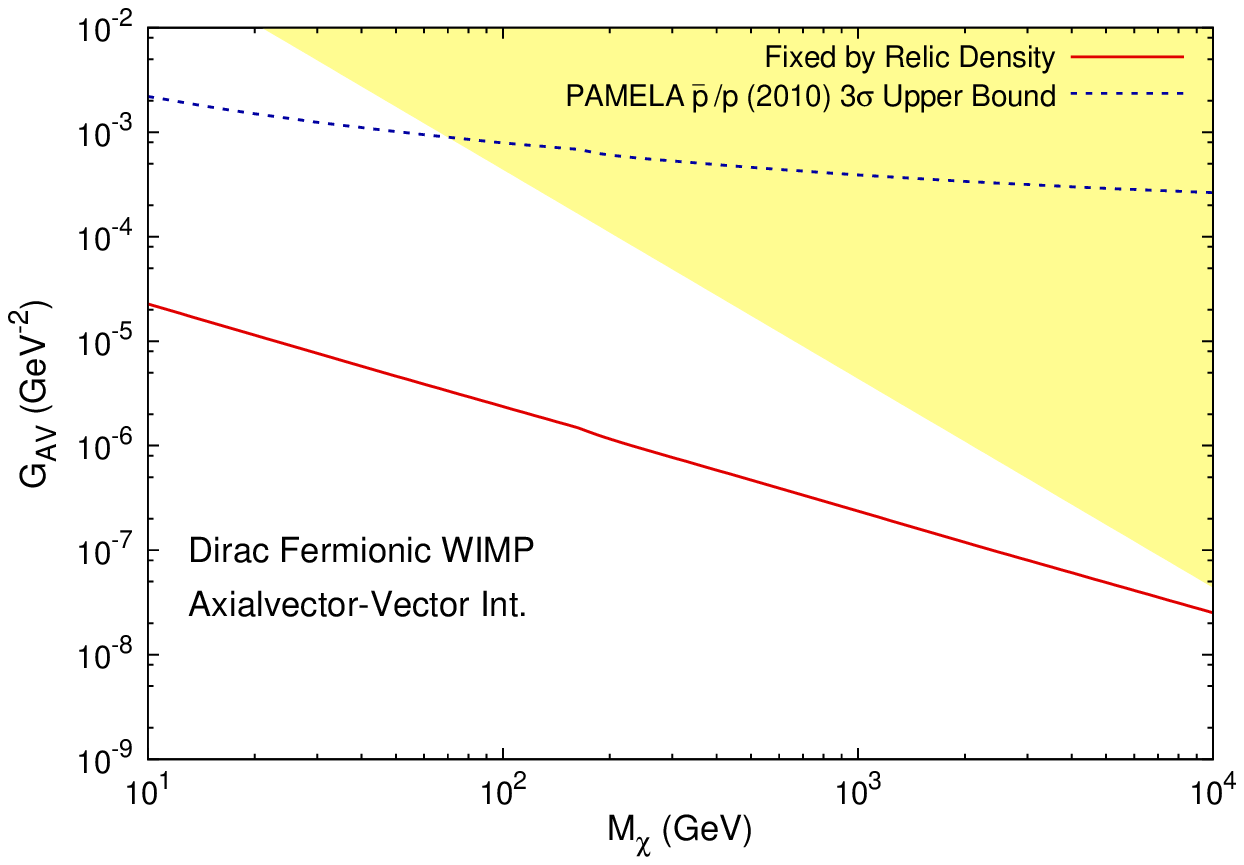}%
\hspace{0.008\textwidth}
\includegraphics[width=0.49\textwidth]{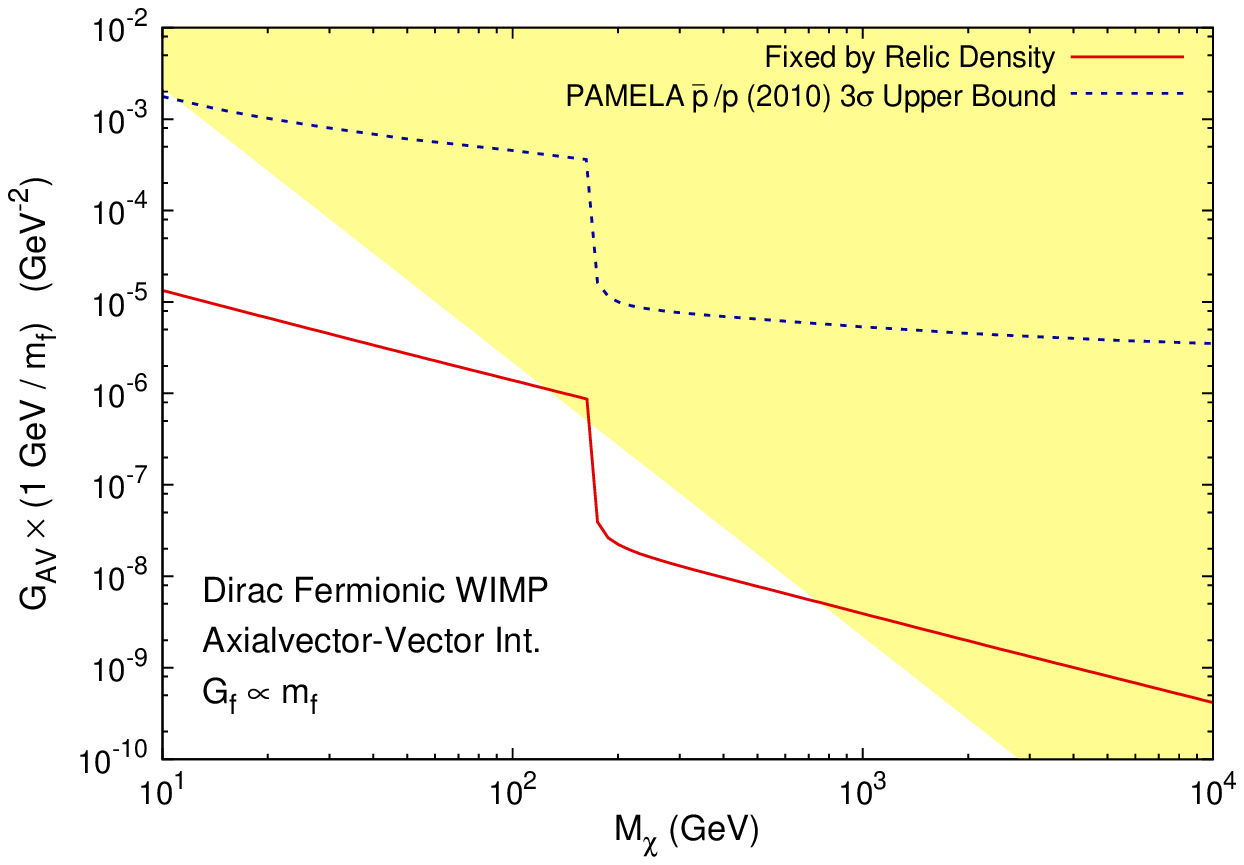}
\caption{Combined constraints on coupling constants $G_f$ of Dirac fermionic WIMPs with axialvector-vector (AV) interactions from relic density, PAMELA $\bar p / p$ ratio, and validity of effective theory. \label{fig-combined-AV}}
\end{figure}

\begin{figure}[!htbp]
\centering
\includegraphics[width=0.49\textwidth]{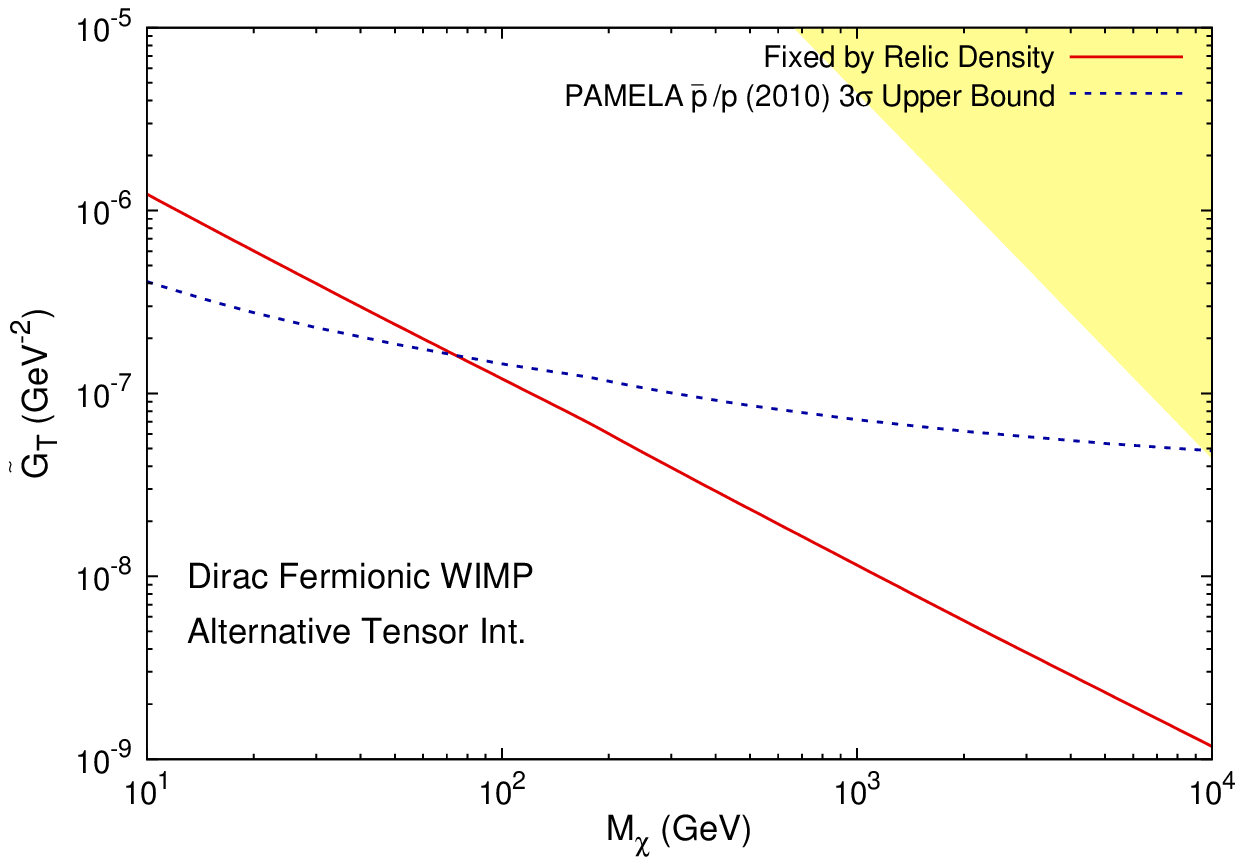}%
\hspace{0.008\textwidth}
\includegraphics[width=0.49\textwidth]{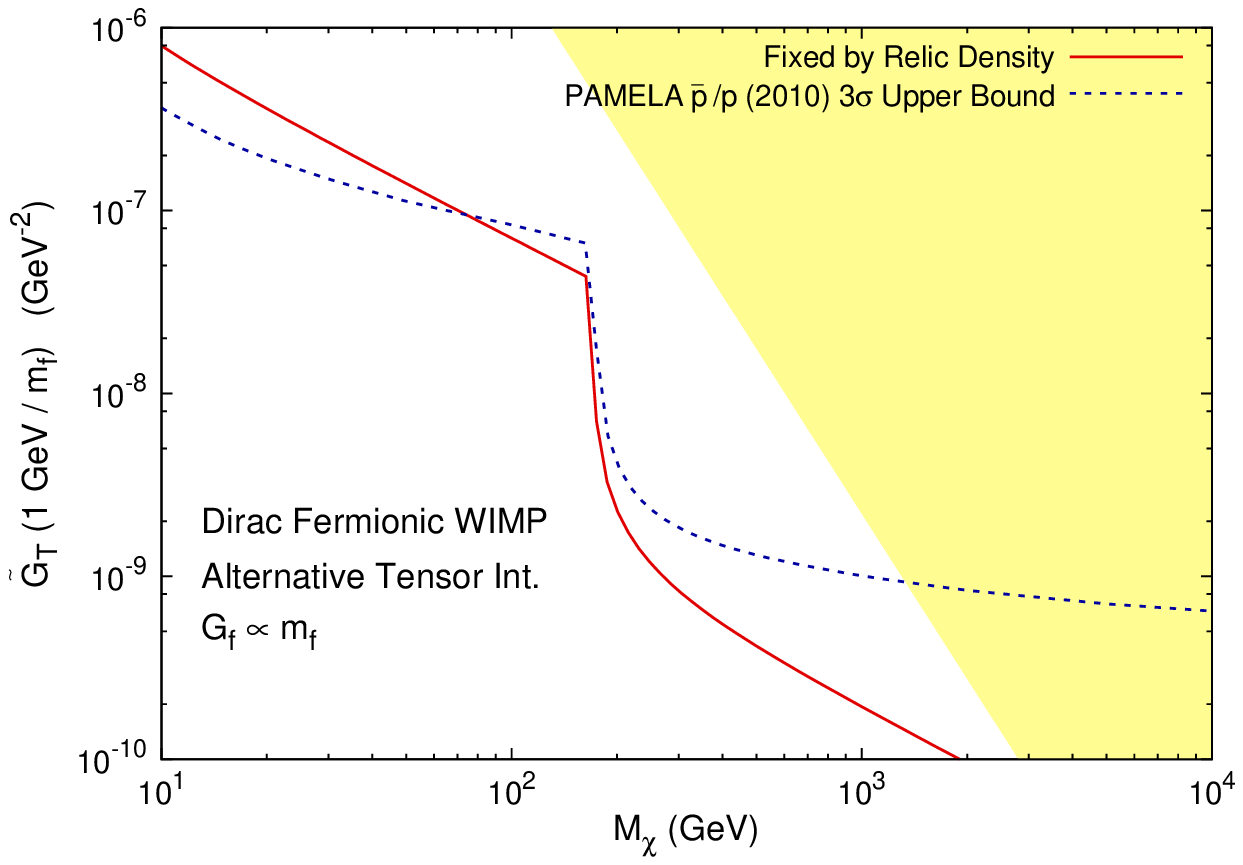}
\caption{Combined constraints on coupling constants $G_f$ of Dirac fermionic WIMPs with alternative tensor ($\tilde{\mathrm{T}}$) interactions from relic density, PAMELA $\bar p / p$ ratio, and validity of effective theory. \label{fig-combined-T-tild}}
\end{figure}

\begin{figure}[!htbp]
\centering
\includegraphics[width=0.49\textwidth]{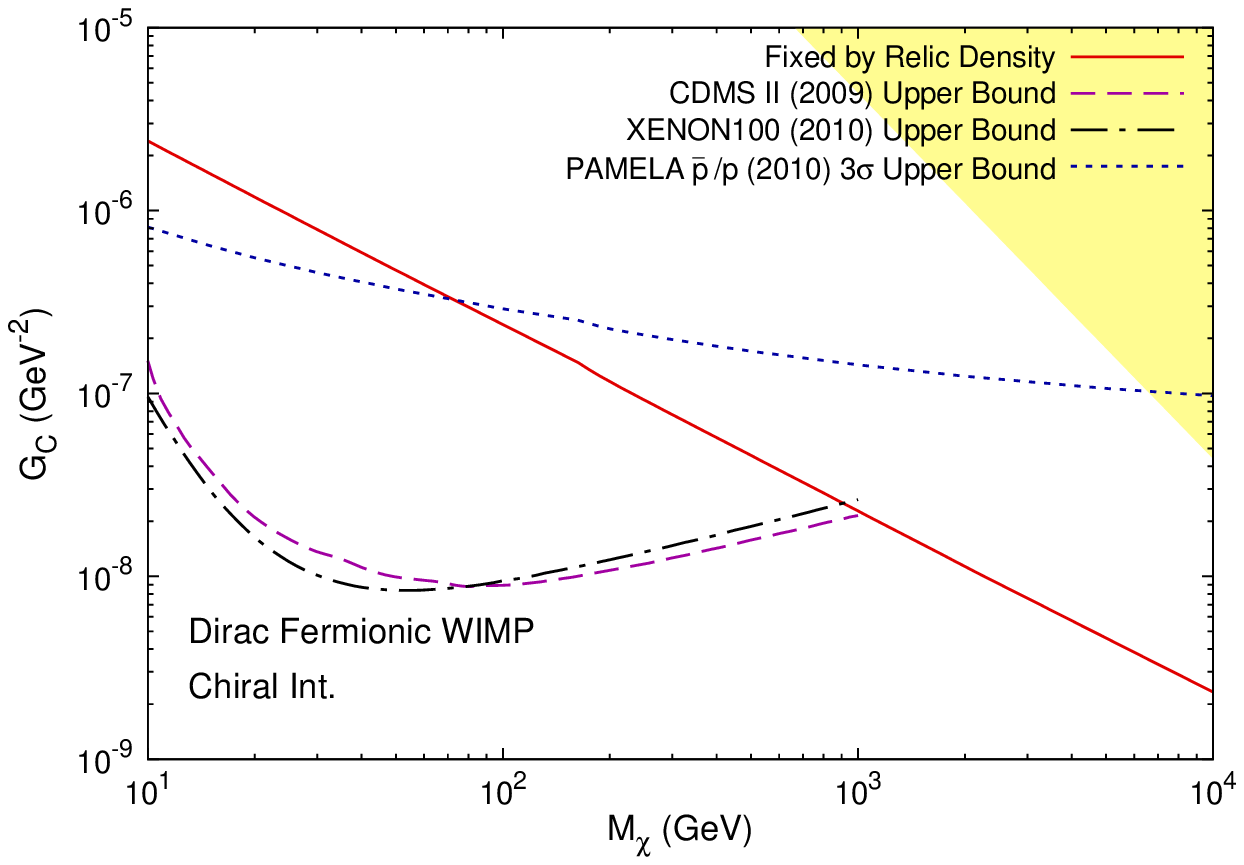}%
\hspace{0.008\textwidth}
\includegraphics[width=0.49\textwidth]{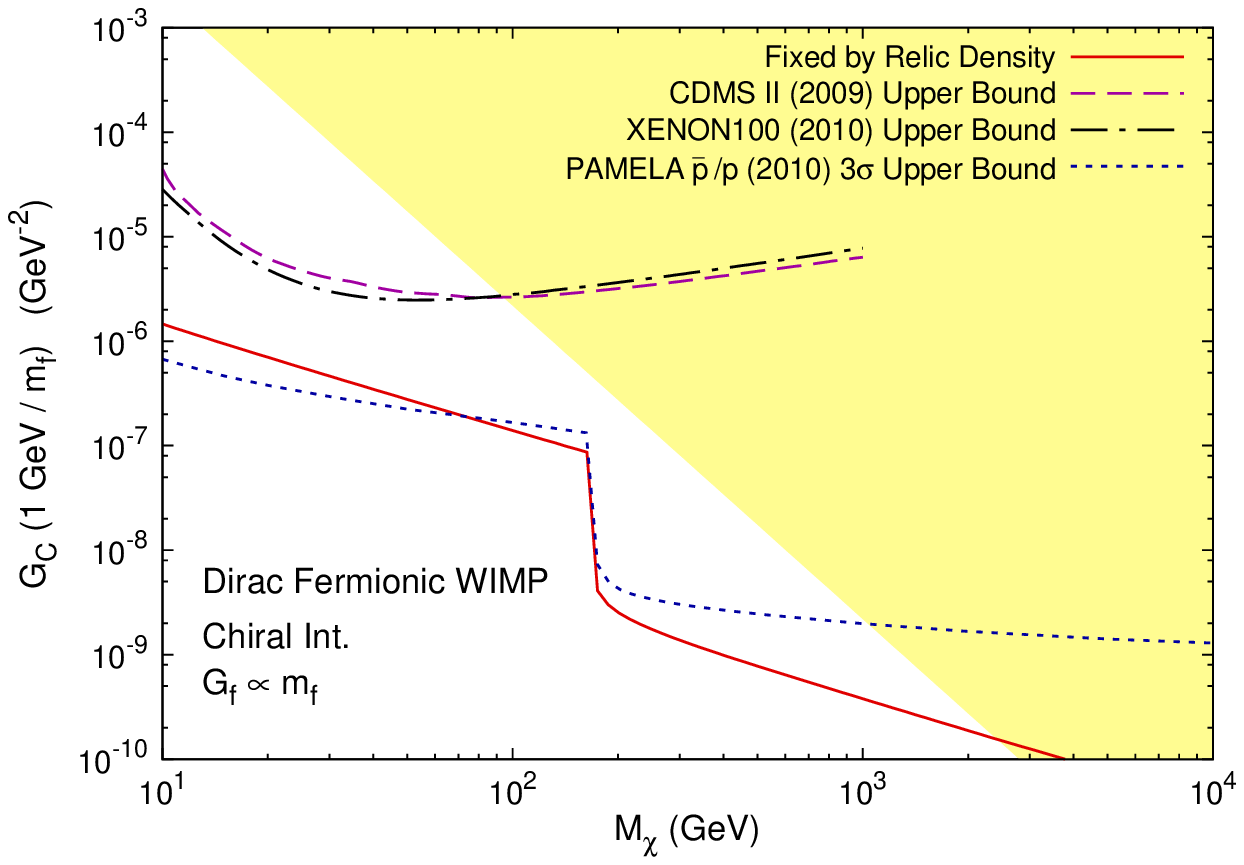}
\caption{Combined constraints on coupling constants $G_f$ of Dirac fermionic WIMPs with chiral (C) interactions from relic density, direct detection experiments of CDMS II and XENON100, PAMELA $\bar p / p$ ratio, and validity of effective theory. \label{fig-combined-Chiral}}
\end{figure}

Eqs.~\eqref{constraint-effective-theory-Univ} and
\eqref{constraint-effective-theory-Yukawa} can be used to set the
valid regions of the effective theories. Considering these validity
conditions altogether with the other phenomenological constraints,
we get the combined constraints on the effective models. In
Figs.~\ref{fig-combined-S} -- \ref{fig-combined-Chiral}, we show the
combined constraints on the coupling constants $G_f$ of Dirac
fermionic WIMPs with scalar (S), pseudoscalar (P), vector (V),
axialvector (A), tensor (T), axialvector-vector (AV), alternative
tensor ($\tilde{\mathrm{T}}$), and chiral (C) interactions,
respectively. The constraints for the pseudoscalar-scalar (PS)
interactions are nearly the same as those for the P interactions.
The constraints for the scalar-pseudoscalar (SP) and
vector-axialvector (VA) interactions are nearly the same as those
for the S and V interactions, respectively, except for that SP and
VA interactions are insensitive to direct detection experiments.
Note that only the effective models of the S, V, A, T, and C
interactions suffer constraints from direct detection experiments.
The SI constraints on the S, V and C interactions are much more
stringent than the SD constraints on the A and T interactions.

From these figures, we can get interesting results. For the scalar
and vector interactions that induce SI scattering with nuclei, the
constraints from direct detection can be much stronger than the other
constraints for the universal couplings, but weaker if the
coupling is Yukawa-like. This is easily understood by considering
the relation of DM-nucleon coupling constants and the DM-quark
coupling constants given in Sec. IV. But for the axialvector and
tensor operators that induce the SD interaction with nuclei we
note that the direct detection constraints are much weaker than the
indirect $\bar{p}/p$ and relic density constraints. Further, we
also note that in several cases the $\bar{p}/p$ constraints can be
stronger than the relic density constraints for DM mass lighter
than 100 GeV. If the direct or indirect detection constraints are
stronger than relic density, the constraints on the effective
coupling are so weak that the thermal production may overclose the
universe. Therefore the DM models in such cases should be
excluded, or else some exotic entropy generation process should occur
after DM freeze out.


\section{Conclusion\label{sec-con}}

\begin{table}[!htbp]
\begin{center}
\belowcaptionskip=0.2cm \caption{A summary for Dirac fermionic WIMPs
with various effective interactions. The excluded regions of
$M_\chi$ given by direct and indirect experiments are indicated.}
\label{tab:dirac_sum}
\renewcommand{\arraystretch}{1.3}
\small
\begin{tabular}{ccc}
\hline \hline \multicolumn{3}{c}{Universal coupling} \\
Interaction & Direct detection & PAMELA $\bar p / p$ \\
\hline
Scalar & Excluded $M_\chi \simeq 10~\mathrm{GeV} - \mathrm{above}~1~\mathrm{TeV}$ & Not sensitive \\
Pseudoscalar & Not sensitive & Excluded $M_\chi \simeq 10 - 70~\mathrm{GeV}$ \\
Vector & Excluded $M_\chi \simeq 10~\mathrm{GeV} - 1~\mathrm{TeV}$ & Excluded $M_\chi \simeq 10 - 70~\mathrm{GeV}$ \\
Axialvector & Not sensitive & Excluded $M_\chi \simeq 10 - 14~\mathrm{GeV}$ \\
Tensor & Not sensitive & Excluded $M_\chi \simeq 10 - 70~\mathrm{GeV}$ \\
Scalar-pseudoscalar & Not sensitive & Not sensitive \\
Pseudoscalar-scalar & Not sensitive & Excluded $M_\chi \simeq 10 - 70~\mathrm{GeV}$ \\
Vector-axialvector & Not sensitive & Excluded $M_\chi \simeq 10 - 70~\mathrm{GeV}$ \\
Axialvector-vector & Not sensitive & Not sensitive \\
Alternative tensor & Not sensitive & Excluded $M_\chi \simeq 10 - 70~\mathrm{GeV}$ \\
Chiral & Excluded $M_\chi \simeq 10~\mathrm{GeV} - 1~\mathrm{TeV}$ & Excluded $M_\chi \simeq 10 - 70~\mathrm{GeV}$ \\
\hline \hline \multicolumn{3}{c}{$G_f \propto m_f$} \\
Interaction & Direct detection & PAMELA $\bar p / p$ \\
\hline
Scalar & Excluded $M_\chi \simeq 10 - 185~\mathrm{GeV}$ & Not sensitive \\
Pseudoscalar & Not sensitive & Excluded $M_\chi \simeq 10 - 70~\mathrm{GeV}$ \\
Vector & Not sensitive & Excluded $M_\chi \simeq 10 - 70~\mathrm{GeV}$ \\
Axialvector & Not sensitive & Excluded $M_\chi \simeq 10 - 25~\mathrm{GeV}$ \\
Tensor & Not sensitive & Excluded $M_\chi \simeq 10 - 70~\mathrm{GeV}$ \\
Scalar-pseudoscalar & Not sensitive & Not sensitive \\
Pseudoscalar-scalar & Not sensitive & Excluded $M_\chi \simeq 10 - 70~\mathrm{GeV}$ \\
Vector-axialvector & Not sensitive & Excluded $M_\chi \simeq 10 - 70~\mathrm{GeV}$ \\
Axialvector-vector & Not sensitive & Not sensitive \\
Alternative tensor & Not sensitive & Excluded $M_\chi \simeq 10 - 70~\mathrm{GeV}$ \\
Chiral & Not sensitive & Excluded $M_\chi \simeq 10 - 68~\mathrm{GeV}$ \\
\hline \hline
\end{tabular}
\end{center}
\end{table}

In this work we give a general analysis of the 4-fermion
interaction between the DM and the standard model particles. We
have considered the most general form of the 4-fermion operators
and corrected some errors in the previous works. We find that the
constraints from DM relic density, DM direct detection and
indirect detection of $\bar{p}/p$ data are complementary to each
other. Generally, the SI constraints are the most stringent while
the SD constraints are quite weak. For light DM ($\lesssim 70$
GeV) the $\bar{p}/p$ data give very strong constraints on the
interaction.

Assuming that one operator dominates the effective interaction between DM and the
SM fermions, we find that some cases get so strong constraints that the
universe will be overclosed by DM thermal production. In such
cases the DM models are actually excluded assuming a standard
cosmology. As a summary, in Tab.~\ref{tab:dirac_sum}, we indicate
the excluded regions of $M_\chi$ given by direct and indirect
experiments for Dirac fermionic WIMPs with various effective
interactions. We find that recent direct detection experiments
only exclude some regions of $M_\chi$ for the scalar, vector and
chiral interactions with universal couplings, and for the scalar
interaction with $G_f \propto m_f$. The PAMELA $\bar p / p$
spectrum, however, excludes some small $M_\chi$ regions ($\lesssim
70~\mathrm{GeV}$) for most of the effective interactions.

\begin{acknowledgments}
This work is supported by the 973 project under Grant No. 2010CB833000,
the National Natural Science Foundation
of China (NSFC) under Grant Nos. 10773011, 11005163 and 11075169, the
Specialized Research Fund for the Doctoral Program of Higher
Education (SRFDP) under Grant No. 200805581030, the Fundamental
Research Funds for the Central Universities, and Sun Yat-Sen
University Science Foundation.
\end{acknowledgments}

\end{document}